\documentclass[english,structabstract]{aa}
\usepackage{mathptmx}
\usepackage[T1]{fontenc}
\usepackage[utf8]{inputenc}
\usepackage{url}
\usepackage{amssymb}
\usepackage{graphicx}

\makeatletter
\bibpunct{(}{)}{;}{a}{}{,} 
\newcommand{\teff}{T_{\mathrm{eff}}}
\newcommand{\logg}{\log g}
\newcommand{\feh}{\left[\mathrm{Fe}/\mathrm{H}\right]}
\newcommand{\ltaur}{\log\tau_{\mathrm{Ross}}}
\newcommand{\taur}{\tau_{\mathrm{Ross}}}

\newcommand{\ssbot}{s_{\mathrm{bot}}}
\newcommand{\ssmin}{s_{\mathrm{min}}}
\newcommand{\dss}{\Delta s}

\newcommand{\vzrmsp}{v_{z,\mathrm{rms}}^{\mathrm{peak}}}

\newcommand{\nsad}{\vec{\nabla}_{\mathrm{sad}}}

\newcommand{\hav}{\left\langle \mathrm{3D}\right\rangle}
\newcommand{\havz}{\hav_z}
\newcommand{\havr}{\hav_{\mathrm{Ross}}}

\usepackage{hyperref}
\usepackage{txfonts}
\titlerunning{The Stagger-grid -- I. Methods and general properties}
\authorrunning{Magic et al.}

\makeatother

\usepackage{babel}
\begin{document}

\title{The \textsc{Stagger}-grid: A Grid of 3D Stellar Atmosphere Models}

\subtitle{I. Methods and General Properties}

\author{Z. Magic\inst{1,2}, R. Collet\inst{2,3,1}, M. Asplund\inst{2,1},
R. Trampedach\inst{4}, W. Hayek\inst{1,2}, A. Chiavassa\inst{5},
R. F. Stein\inst{6} \and  {\AA}. Nordlund\inst{3}}

\institute{Max-Planck-Institut f{\"u}r Astrophysik, Karl-Schwarzschild-Str.
1, 85741 Garching, Germany \\
\email{magic@mpa-garching.mpg.de} \and  Research School of Astronomy
\& Astrophysics, Cotter Road, Weston ACT 2611, Australia\and  StarPlan,
Natural History Museum of Denmark/Niels Bohr Institute, {\O}ster
Voldgade 5-7, DK--1350 Copenhagen, Denmark\and  JILA, University
of Colorado and National Institute of Standards and Technology, 440
UCB, Boulder, CO 80309, USA\and  Laboratoire Lagrange, UMR 7293,
CNRS, Observatoire de la C{\^o}te d’Azur, Universit{\'e} de Nice
Sophia-Antipolis, Nice, France\and  Department of Physics \& Astronomy,
Michigan State University, East Lansing, MI 48824, USA}

\offprints{magic@mpa-garching.mpg.de}

\date{Received ...; Accepted...}

\abstract{}{We present the \textsc{Stagger}-grid, a comprehensive grid of
time-dependent, three-dimensional (3D), hydrodynamic model atmospheres
for late-type stars with realistic treatment of radiative transfer,
covering a wide range in stellar parameters. This grid of 3D models
is intended for various applications besides studies of stellar convection
and atmospheres \emph{per se}, including stellar parameter determination,
stellar spectroscopy and abundance analysis, asteroseismology, calibration
of stellar evolution models, interferometry, and extrasolar planet
search. In this introductory paper, we describe the methods we applied
for the computation of the grid and discuss the general properties
of the 3D models as well as of their temporal and spatial averages
(here denoted $\left\langle 3\mathrm{D}\right\rangle $ models).}{All
our models were generated with the \textsc{Stagger}-code, using realistic
input physics for the equation of state (EOS) and for continuous and
line opacities. Our $\sim220$ grid models range in effective temperature,
$T_{\mathrm{eff}}$, from $4000$ to $7000\,\mathrm{K}$ in steps
of $500\,\mathrm{K}$, in surface gravity, $\log g$, from $1.5$
to $5.0$ in steps of 0.5 dex, and metallicity, $\left[\mathrm{Fe}/\mathrm{H}\right]$,
from $-4.0$ to $+0.5$ in steps of $0.5$ and $1.0\,\mathrm{dex}$.}{We
find a tight scaling relation between the vertical velocity and the
surface entropy jump, which itself correlates with the constant entropy
value of the adiabatic convection zone. The range in intensity contrast
is enhanced at lower metallicity. The granule size correlates closely
with the pressure scale height sampled at the depth of maximum velocity.
We compare the $\left\langle 3\mathrm{D}\right\rangle $ models with
currently widely applied one-dimensional (1D) atmosphere models, as
well as with theoretical 1D hydrostatic models generated with the
same EOS and opacity tables as the 3D models, in order to isolate
the effects of using self-consistent and hydrodynamic modeling of
convection, rather than the classical mixing length theory (MLT) approach.
For the first time, we are able to quantify systematically over a
broad range of stellar parameters the uncertainties of 1D models arising
from the simplified treatment of physics, in particular convective
energy transport. In agreement with previous findings, we find that
the differences can be rather significant, especially for metal-poor
stars. } {}


\maketitle

\section{Introduction}

The primary source of information for stellar objects is the light
they emit, which carries information about the physical conditions
at its origin. However, in order to interpret the information correctly,
one first needs either theoretical or semi-empirical models of the
atmospheric layers at the surface of stars from where the stellar
radiation escapes. Therefore, models of stellar atmospheres are essential
for much of contemporary astronomy. 

In the case of late-type stars, the theoretical modeling of stellar
atmospheres is complicated by the presence of convective motions and
turbulent flows as well as of magnetic fields in their envelopes \citep[see review by][and references therein]{Nordlund:2009p4109}.
In particular, convection can significantly affect both the atmospheric
stratification and emergent spectral energy distribution in these
stars. Hence, in order to correctly represent the temperature stratifications
in the outer layers of stars, from where the stellar light escapes,
it is vital to accurately account for the interaction between radiative
and convective energy transport at the optical surface .\\

The first realistic grids of line-blanketed atmosphere models for
late-type stars appeared with the publication of \textsc{MARCS} \citep{Gustafsson:1975p3743,Gustafsson:2008p3814}
and \textsc{ATLAS} models \citep{Kurucz:1979p4707,Castelli:2004p4949}.
Subsequently, other one-dimensional (1D) atmosphere codes, e.g. \textsc{PHOENIX}
\citep{Hauschildt:1999p4899} and\textsc{ MAFAGS} \citep{Grupp:2004p14918},
were developed to model the atmospheres of stars. In general, these
theoretical 1D atmosphere models assume hydrostatic equilibrium, flux
constancy, and local thermodynamic equilibrium (LTE). For the modeling
of convective energy transport, they commonly employ the mixing-length
theory \citep[MLT, see ][]{BohmVitense:1958p4822}, which is characterized
by several free parameters, the most commonly known being the mixing-length
$l_{m}$, or equivalently, the parameter $\alpha_{\mathrm{MLT}}=l_{m}/H_{P}$.
Alternatively, some relatives thereof are available, such as the full
turbulence spectrum (FTS) theory by \citet{Canuto:1991p6553}, which
itself also has a free parameter. The values of these free parameters
are not known from first principles and need to be calibrated based
on observations or simulations. The mixing-length theory has in total
four free parameters \citep[see][]{BohmVitense:1958p4822,Henyey:1965p15592,Mihalas:1970p21310}.
These free parameters can be calibrated based on their effect on synthetic
spectra, but usually only $\alpha_{\mathrm{MLT}}$ is calibrated based
on the reproduction of selected lines \citep{Fuhrmann:1993p15161,Barklem:2002p20825,Smalley:2002A&A...395..601S}.
Moreover, the free mixing length is calibrated in stellar evolutionary
calculations by matching the observed luminosity and radius of the
Sun at its current age \citep[e.g.][]{Magic:2010p13816}. To construct
simple yet realistic 1D models of convection is rather difficult,
in particular convective overshooting beyond the classical Schwarzschild
instability criterion is normally not considered in 1D atmospheric
modeling. Attempts have been made at including its effects in 1D model
atmospheres albeit with only limited success \citep{Castelli:1997p4955}.\\

The first numerical 1D model stellar atmosphere codes usually assumed
a plane-parallel geometry for the atmospheric stratification. This
was later improved upon by changing to a spherical symmetry, leading
to lower temperatures in the upper layers, in particular for giant
stars, due to the dilution of the radiation field with increasing
radial distance, which can cover a significant fraction of the stellar
radius at low $\logg$ \citep[see][]{Gustafsson:2008p3814}. Initially
line blanketing was included by means of opacity distribution functions
\citep[ODFs,][]{Gustafsson:1975p3743} with a few hundred ODFs covering
the entire spectrum, eventually replaced by opacity sampling (OS)
including thousands of wavelength points \citep{Johnson:1976p21373}.
Nowadays, thousands of ODFs or hundreds of thousands of OS wavelengths
are used. Despite such high resolution in wavelength, the computational
costs for 1D atmosphere models are currently quite small, at least
for LTE models. Large, homogeneous grids of atmospheres with up to
$\sim10^{5}$ models exist \citep{Gustafsson:2008p3814,Cassisi:2004p1158,Hauschildt:1999p4899},
covering a wide range of stellar atmosphere parameters ($T_{\mathrm{eff}}$,
$\log g$, and $\left[\mathrm{Fe}/\mathrm{H}\right]$).\\

Even though the 1D atmosphere models are based on numerous simplifications,
they have demonstrated high predictive capabilities owing to major
improvements in the atomic and molecular data (e.g. line lists by
\citet{Kurucz:1993p21505} or VALD by \citet{Piskunov:1995p20700}).
Also, the continuum opacity sources and the EOS have undergone similar
developments. Thanks to these, 1D atmosphere models are in many respects
very successful in comparisons with observations and are widely applied
in astronomy today.\\

Another approach, almost exclusively used for solar atmosphere modeling,
is the use of semi-empirical models. In these models, the temperature
stratification is inferred from observations (e.g. from lines forming
at different heights or continuum center-to-limb variations). Often-used
semi-empirical 1D solar atmosphere models are the \citet{Holweger:1974p5049},
\textsc{VAL3C} \citep{Vernazza:1976p21109}, \citet{Maltby:1986p21111}
and \textsc{MISS} \citep{AllendePrieto:2001p21112} models. A similar
approach can be used to integrate spatially resolved observations
and thus infer the three-dimensional (3D) atmosphere structures using
inversion techniques \citep{RuizCobo:1992p21128,SocasNavarro:2011p5800}.
Semi-empirical modeling is rarely attempted for other stars, although
exceptions exist \citep[e.g.][]{AllendePrieto:2000p21415}.\\

Constructing more realistic models requires one to go beyond the 1D
framework and model convection without relying on MLT. Stellar convection
is an inherently 3D, time-dependent, non-local, and turbulent phenomenon.
Therefore, one cannot expect 1D models to reproduce all observed properties
accurately, even with access to free parameters to tweak. The next
natural step is to abandon some of these crude simplifications by
constructing realistic 3D atmosphere models of solar convection. Early
hydrodynamic simulations \citep{Nordlund:1982p6697,Nordlund:1990p6720,Steffen:1989p18861}
revealed that stellar surface convection operates in a distinctly
different fashion from the MLT picture. Instead of the homogeneous
convective elements, they displayed highly asymmetrical motions with
slow broad steady upflows interspersed with fast narrow turbulent
downdrafts, sometimes even supersonic (e.g. \citealt[hereafter SN98]{Stein:1998p3801};
\citealt{Asplund:2000p20875,Nordlund:2009p4109,Carlsson:2004p12217,Ludwig:1999p7606}).
The advent of 3D simulations, which are constructed from first principles,
has enabled astronomers to predict various observables such as solar
granulation properties and spectral line profiles astonishingly well.
More recent solar 3D simulations are remarkably good at reproducing
the observed center-to-limb variation \citep[e.g.][]{Pereira:2009p17415,Asplund:2009p3308,Ludwig:2006p7592}.\\

3D atmosphere models are by design free from the adjustable parameters
of MLT and other parameters such as micro- and macro-turbulence that
have hampered stellar spectroscopy for many decades. Instead, in 3D
simulations, convection emerges naturally, by solving the time-dependent
hydrodynamic equations for mass-, momentum- and energy-conservation,
coupled with the 3D radiative transfer equation in order to account
correctly for the interaction between the radiation field and the
plasma. Also, the non-thermal macroscopic velocity fields associated
with convective motions are rendered realistically, and various natural
kinetic consequences such as overshooting and excitation of waves
emerge from the simulations, without the need for further ad hoc modeling
or additional free parameters. The inhomogeneities in the convective
motions arise spontaneously and self-organize naturally to form a
distinct flow pattern that exhibits the characteristic granulation
at the surface. Furthermore, additional spectral observables such
as limb-darkening and detailed spectral line shapes, including asymmetries
and shifts, are also modeled unprecedentedly accurately with 3D models
for the Sun \citep{Nordlund:2009p4109,Pereira:2009p17405}\\

For metal-poor late-type stars it has been shown \citep{Asplund:1999p11771,Collet:2006p10854,Collet:2007p5617}
that the assumption of pure radiative equilibrium in the convectively
stable photospheric layers of classical hydrostatic models is generally
insufficient. In particular in the upper photosphere, the thermal
balance is instead primarily regulated by radiative heating due to
spectral line re-absorption of the continuum-radiation from below
and adiabatic cooling due to the expansion of upflowing gas. In metal-poor
stars, the balance between heating by radiation and cooling by mechanical
expansion of the gas occurs at lower temperatures because of the weakness
and scarcity of spectral lines at low metallicities. By contrast,
1D MLT models have no velocity fields outside their convection zones,
and are therefore in pure radiative equilibrium. The temperature stratification
there is therefore regulated solely by radiative heating and cooling,
thus neglecting altogether the adiabatic cooling component. This results
in an overestimation of the temperatures by up to $\sim1\,000\,\mathrm{K}$
in 1D models at very low metallicities, which can potentially lead
to severe systematic errors in abundance determinations based on 1D
models \citep[see][]{Asplund:1999p11771,Asplund:2001p21515,Ludwig:2010p4652,Collet:2009p4419,GonzalezHernandez:2010p21570}.
These shortcomings of 1D models are manifested as inconsistencies
in the analysis of observed spectra, such as abundance trends with
excitation potential of the lines (e.g. analysis of NH lines in the
very metal-poor star HE1327-2326, by \citet{Frebel:2008p21240} and
discrepant abundances between atomic and molecular lines involving
the same elements \citep[e.g.][]{Nissen:2002p21232}. For further
discussion, we refer to a review of possible impacts of 3D models
on stellar abundance analysis by \citet{Asplund:2005p7802}.\\

Additionally, there are discrepancies between observations and predictions
from 1D models of the solar structure in the context of helioseismology,
which point to mistakes in the outer layers of theoretical 1D stellar-structure
models, and which are usually referred to as \textit{surface effects}
\citep{Rosenthal:1999p9777}. With classical 1D stellar structures,
higher frequency p-modes of the Sun are systematically shifted due
to discrepancies at the upper turning points of the modes, which occur
in the superadiabatic peak at the top of the convection envelope.
\citet{Rosenthal:1999p9777} found better agreement of stellar structures
with helioseismic observations, when including the mean stratification
of solar 3D models at the top, since the turbulent pressure, usually
neglected in 1D models, extends the resonant cavity. Also, it was
found that with 3D solar models the predicted p-mode excitation rates
are much closer to helioseismic observations\citep{Nordlund:2001p6371,Stein:2001p10845}.
\citet{Ludwig:2009p10872} compared the power spectra of the photometric
micro-variability induced by granulation and found good agreement
between the theoretical predictions of 3D solar models and observations
with SOHO.\\

With the aid of 3D simulations, stellar radii have been derived for
a number of red giants from interferometric observations \citet{Chiavassa:2010p6257,Chiavassa:2012p22493}.
The determined stellar radii are slightly larger than estimated with
the use of 1D models, which has an impact on the zero point of the
effective temperature scale derived by interferometry. Furthermore,
\citet{Chiavassa:2012p22493} showed that for interferometric techniques
a detailed knowledge of the granulation pattern of planet-hosting
stars is crucial for the detection and characterization of exoplanets.\\

Several 3D magnetohydrodynamics codes with realistic treatment of
radiative transfer have been developed and applied to the modeling
of stellar surface convection. Here, we make use of the \textsc{Stagger}-code,
which is developed specifically to run efficiently on the massively
parallel machines available today (Nordlund \& Galsgaard 1995%
\footnote{\url{http://www.astro.ku.dk/~kg/Papers/MHD_code.ps.gz}%
}; \citealt{Kritsuk:2011p10673}). The \textsc{Bifrost}-code is an
Oslo derivative of the \textsc{Stagger}-code \citep{Gudiksen:2011p17512},
tailored for simulations of the solar photosphere and chromosphere,
and therefore including true scattering \citep{Hayek:2010p4435}.
Other widely used codes are \textsc{CO$^{5}$BOLD} \citep{Freytag:2012p23073},
\textsc{MURaM} \citep{Vogler:2005p11515} and \textsc{ANTARES} \citep{Muthsam:2010p15549},
which have been independently developed in the last decades. \citet{Beeck:2012p13340}
compared solar models from three of the above 3D stellar atmosphere
codes (\textsc{Stagger}, \textsc{CO$^{5}$BOLD} and \textsc{MURaM}),
and showed that the models are overall very similar, despite the distinct
numerical approaches. Most of the available 3D stellar convection
codes are now highly parallelized, which when coupled with the computational
power available today makes it feasible to construct grids of 3D convection
simulations within a reasonable time-scale. Grids of 2D and 3D atmosphere
models already exist \citep{Ludwig:1999p7606,Ludwig:2009p4627,Trampedach:2007p5614,Trampedach:2013arXiv1303.1780T,Tanner:2013arXiv1302.5707T}.
Clearly, the age of 3D atmosphere modeling has arrived, partly driven
by the rising demand created by improved high-resolution spectroscopic
and asteroseismic observations.\\

In this paper, we present a new large grid of 3D model atmospheres
for late-type stars, covering an extensive range of effective temperatures,
surface gravities, and metallicities. In Section \ref{sec:Methods}
we describe the methods that we followed in order to compute the 3D
model atmospheres, with emphasis on the convection code we used (Sect.
\ref{sub:The-Stagger-code}) and on the tools we developed for scaling
the grid models and post-process the results of the numerical calculations
(Sect. \ref{sub:Scaling-and-relaxing}). In Section \ref{sec:Results},
we present an overview of general properties (Sect. \ref{sub:Global-properties})
of our simulations, and discuss the temporally and spatially averaged
atmospheres (Sect. \ref{sub:The-mean-atmosphere}) from the 3D model
atmospheres. We also compare our results with theoretical 1D models
in Sect. \ref{sub:Comparison-with-1D} corresponding to the same stellar
parameters%
\footnote{We refer always to stellar \textit{atmospheric} parameters.%
}. These have been computed with a specifically newly developed 1D
code that employs exactly the same EOS and opacities as the 3D simulations.
Also, we compare our 3D model atmospheres with 1D models from grids
widely adopted by the astronomical community (\textsc{MARCS} and \textsc{ATLAS}).
Finally, in Section \ref{sec:Conclusions}, we summarize our findings
and outline a roadmap for our future ambitions on the many possible
applications of the \textsc{Stagger}-grid.

\section{Methods\label{sec:Methods}}

\subsection{The \textsc{Stagger}-code\label{sub:The-Stagger-code}}

The 3D model atmospheres presented here were constructed with a custom
version of the \textsc{Stagger}-code, a state-of-the-art, multipurpose,
radiative-magnetohydrodynamics (R-MHD) code originally developed by
Nordlund \& Galsgaard (1995), and continuously improved over the years
by its user community. In pure radiation-hydrodynamics mode, the \textsc{Stagger}-code
solves the time-dependent hydrodynamic equations for the conservation
of mass (Eq. \ref{eq:mass}), momentum (Eq. \ref{eq:momentum}), and
energy (Eq. \ref{eq:energy}) in a compressible flow 
\begin{eqnarray}
\partial_{t}\rho & = & -\vec{\nabla}\cdot(\rho\vec{v}),\label{eq:mass}\\
\partial_{t}\rho\vec{v} & = & -\vec{\nabla}\cdot(\rho\vec{v}\vec{v}+\underline{\underline{\tau}})-\vec{\nabla}p+\rho\vec{g},\label{eq:momentum}\\
\partial_{t}e & = & -\vec{\nabla}\cdot(e\vec{v})-p\vec{\nabla}\cdot\vec{v}+q_{\mathrm{rad}}+q_{\mathrm{visc}},\label{eq:energy}
\end{eqnarray}
coupled to the radiation field via the heating and cooling (per unit
volume) term
\begin{eqnarray}
q_{\mathrm{rad}} & = & 4\pi\rho\int_{\lambda}\kappa_{\lambda}\left(J_{\lambda}-S_{\lambda}\right)\, d\lambda,\label{eq:radiative_heating}
\end{eqnarray}
which is computed from the solution of the radiative transfer equation
\begin{eqnarray}
\hat{n}\cdot\vec{\nabla}I_{\lambda} & = & \varrho\kappa_{\lambda}(S_{\lambda}-I_{\lambda})\label{eq:radiative_transfer_first}
\end{eqnarray}
to account properly for the energy exchange between matter and radiation
(here $\rho$ denotes the density, $\vec{v}$ the velocity field,
$p$ the thermodynamic pressure, $e$ the internal energy per unit
volume%
\footnote{In the following, we will indicate the internal energy per unit mass
with $\varepsilon=e/\rho$ .%
}, $g$ the gravity, $\underline{\underline{\tau}}$ the viscous stress
tensor, $q_{\mathrm{visc}}=\sum_{ij}\tau_{ij}\partial v_{i}/\partial r_{j}$
the viscous dissipation rate, $\kappa_{\lambda}$ the monochromatic
opacity, $I_{\lambda}$ the monochromatic intensity, $J_{\lambda}=1/4\pi\int_{\Omega}I_{\lambda}d\Omega$
the monochromatic mean intensity averaged over the entire solid angle,
and $S_{\lambda}$ the source function). We have ignored magnetic
fields in the present grid of 3D convection simulations, however,
we will study their effects in a future work.

\subsubsection{Details on the numerics\label{sub:numerics}}

The \textsc{Stagger}-code uses a sixth-order explicit finite-difference
scheme for numerical derivatives and the corresponding fifth-order
interpolation scheme. The solution of the hydrodynamic equations is
advanced in time using an explicit third-order Runge-Kutta integration
method \citep{Williamson:1980p24417}. The code operates on a staggered,
Eulerian, rectangular mesh: the thermodynamic variables, density and
internal energy per volume, are cell-centered, while momentum components
are defined at cell faces. Also, in the MHD mode, the components of
the magnetic field $B$ (electric field $E$) are defined at the cell
faces (edges). This configuration allows for a flux-conservative formulation
of the magnetohydrodynamic equations, at the same time ensuring that
the magnetic field remains divergence-free. The solution of the discretized
equations is stabilized by hyper-viscosity which aims at minimizing
the impact of numerical diffusion on the simulated flow, while providing
the necessary diffusion for large-eddy simulations with finite-difference
schemes (see also Nordlund \& Galsgaard 1995 for further details).
The values of the numerical viscosity parameters%
\footnote{The actual values we used are $n_{1}=0.005$ and $n_{2}=0.8$ \citep[see Eq. 9 in][]{Kritsuk:2011p10673}.%
} are empirically tuned for the solar surface-convection simulation:
they are set large enough to stabilize the numerical solution of the
hydrodynamic equations and, at the same time, kept small enough to
reduce their smoothing of the flow’s structures. The same optimized
values of the parameters are then applied to all other simulations
in the grid.

The version of the \textsc{Stagger}-code we used for this work is
fully MPI-parallel. The parallelization scales well with the number
of cores. For this project, the simulations were typically run on
$64$ cores.

\subsubsection{Geometrical properties\label{sub:Geometrical-properties}}

The setup of the simulations is of the so-called \textit{box-in-a-star}
type: the domain of the simulations is limited to a small representative
volume located around the stellar photosphere and including the top
portion of the stellar convective envelope. The boundary conditions
of the simulation box are periodic in the horizontal directions and
open vertically. Gravity is assumed to be constant over the whole
extent of the box, neglecting sphericity effects. However, since the
size of the simulation domains correspond to only a fraction of the
total radii of the stars ($0.4\,\%$ and $\sim10\,\%$ of the stellar
radius for the solar simulation and for a typical $\log g=1.5$ red
giant simulation, respectively) such effects can be regarded as small
for the purposes of the current grid of models. Also, for simplicity,
the effects of stellar rotation and associated Coriolis forces are
neglected in the present simulation setup, as it would add two more
dimensions to the grid.

At the bottom, the inflowing material has a constant value of specific
entropy per unit mass, which ultimately determines the emerging effective
temperature. While the domains of our simulations cover only a small
fraction of the convective zone, the box-in-a-star setup is still
valid because the bulk up-flows at the bottom boundary of the simulations
carry essentially the same entropy value as in deeper layers and are
mostly unaffected by entrainment with cooler downflows. At the beginning
of each simulation, the entropy of the inflowing gas at the bottom
is adjusted in order to yield the desired $\teff$ and, after that,
is kept unchanged during the entire run (see Sect. \ref{sub:Scaling-and-relaxing}).
Furthermore, pressure is assumed to be constant over the whole bottom
layer. 

The physical dimensions in the horizontal directions are chosen to
be large enough to cover an area corresponding to about ten granular
cells. The vertical dimensions are extended enough for the simulations
to cover at least the range of $-5.0<\log\tau_{\mathrm{Ross}}<+6.0$
in terms of Rosseland optical depth (in fact they range on average
from $-7.3<\log\tau_{\mathrm{Ross}}<+7.5$), which typically corresponds
to approximately six orders of magnitude in pressure (about 14 pressure
scale heights). All of the simulations have a mesh resolution of $240^{3}$,
since a resolution of about $200^{3}-250^{3}$ was found to be adequate
\citep[see][]{Asplund:2000p20875}. Five layers at the bottom and
the top are reserved for the so-called \textit{ghost-zones}: these
extra layers serve to enforce boundary conditions for the high-order
derivatives in the vertical direction. The spacing between cells in
the horizontal direction ($\Delta x,\Delta y$) is constant, ranging
from about 6 km in dwarfs to about 25 Mm in giants, while it varies
smoothly with depth in the vertical direction, in order to resolve
the steep temperature gradients near the optical surface. These are
the layers from where the continuum radiation escapes; they are characterized
by a sharp transition between stellar interior and outer layers in
terms of thermodynamic quantities such as temperature, internal energy,
and entropy that marks the beginning of the photosphere. Also, the
steepest temperature gradients are found in the superadiabatic region
just below the optical surface ($0.0<\log\tau_{\mathrm{Ross}}<2.0$).
Therefore, it is very important that the thin transition layer around
the optical surface is well-resolved in order to ensure an accurate
modeling of the radiative transfer and to avoid spurious numerical
artifacts from insufficient spatial resolution.

\subsubsection{Equation of state\label{sub:Equation-of-state}}

We use the realistic equation of state (EOS) by \citet{Mihalas:1988p20892},
which explicitly treats excitation to all bound states of all ionization
stages, of all included elements. We have custom computed tables for
a mix of the 17 most abundant elements ($\mathrm{H},\mathrm{He},\mathrm{C},\mathrm{N},\mathrm{O},\mathrm{Ne},\mathrm{Na},\mathrm{Mg},\mathrm{Al},\mathrm{Si},\mathrm{S},\mathrm{Ar},\mathrm{K},\mathrm{Ca},\mathrm{Cr},\mathrm{Fe}$
and $\mathrm{Ni}$). The only molecules that are included in the EOS
are $\mathrm{H}_{2}$ and $\mathrm{H}_{2}^{+}$, and they are treated
on equal footing with the atoms and ions. For the solar abundances
we employed the latest chemical composition by \citet{Asplund:2009p3308},
which is based on a solar simulation performed with the same code
and atomic physics as presented here. Our choice for the EOS, is supported
by \citet{diMauro:2002p21228} who showed that solar models based
on the EOS by \citet{Mihalas:1988p20892} show better agreement with
helioseismology in the outer $20\,\mathrm{Mm}$ ($\ge0.97\, R_{\odot}$),
compared to models based on the OPAL-EOS. We inverted the \citet{Mihalas:1988p20892}
EOS tables, hence the temperatures and the thermodynamic pressures
are tabulated as a function of density and internal energy. This inversion
exploits the analytical derivatives provided in the EOS tables to
minimize losses in accuracy. These analytical derivatives are also
used in the bi-cubic spline interpolation in the inverted tables.

\subsubsection{Opacity\label{sub:Opacity}}

We use the continuum absorption and scattering coefficients listed
in detail and with references by \citet{Hayek:2010p4435}. These include
the sophisticated calculations by \citet{Nahar:2004p23948}%
\footnote{\url{http://www.astronomy.ohio-state.edu/~nahar/}.%
} for the first three ions of all metals we include, except for $\mathrm{K}$
and $\mathrm{Cr}$. These calculations are improvements over those
forming the basis for the OP opacities%
\footnote{\url{http://cdsweb.u-strasbg.fr/topbase/}.%
} \citep{Badnell:2005p23952}. The line opacity is supplied by the
opacity sampling (OS) data that was also used for the newest MARCS
grid of stellar atmospheres \citet{Gustafsson:2008p3814}, which are
in turn based on the VALD-2 database%
\footnote{\url{http://www.astro.uu.se/~vald/php/vald/}.%
} \citep{Stempels:2001p23953} of atomic and molecular lines.

\subsubsection{Radiative transfer\label{sub:Radiative-transfer}}

The radiative heating and cooling rate (Eq. \ref{eq:radiative_heating})
is evaluated by solving the radiative transfer equation 
\begin{equation}
\frac{dI_{\lambda}}{d\tau_{\lambda}}=I_{\lambda}-S_{\lambda},\label{eq:radiative_transfer}
\end{equation}
where $\tau_{\lambda}=\int\rho\kappa_{\lambda}ds$ denotes the monochromatic
optical depth along a given direction $s$, with a method similar
to that by \citet{Feautrier:1964p21596}. The equation is solved at
each time step and grid point on long characteristics, along the vertical
direction and along eight additional inclined angles (two $\mu=\cos\theta$
and four $\varphi$-angles) by tilting the (domain-decomposed) 3D
cube. Given the opacity $\kappa_{\lambda}$ and the source function
$S_{\lambda}$, the monochromatic intensity $I_{\lambda}$ can be
obtained by solving Eq. (\ref{eq:radiative_transfer}) and the radiative
heating and cooling rate computed by integrating $\rho\kappa_{\lambda}(I_{\lambda}-S_{\lambda})$
over solid angle and wavelength. We use the Radau quadrature to determine
the optimal ray directions to approximate the angular integral in
the calculation of the radiative heating and cooling rate as a weighted
sum. For the radiative transfer calculations, we employ opacities
as described above (Sect. \ref{sub:Opacity}).

Computing the full monochromatic solution to the radiative transfer
equation in 3D at each time step is extremely expensive. The cost
of the radiative transfer calculations however can be reduced enormously
by opting instead for an approximated solution based on the \textit{opacity
binning} or \textit{multi-group} method \citep{Nordlund:1982p6697,Skartlien:2000p9857}.
Following this method, we sort all sampled wavelength points into
different bins based on the spectral range they belong to and on their
associated \emph{opacity strength} or, better, their \emph{formation
depth}, i.e. the Rosseland optical depth $\tau_{\mathrm{Ross}}\left(\tau_{\lambda}=1\right)$,
where the monochromatic optical depth equals unity. In this way, wavelength
points characterized by similar formation heights and belonging to
the same spectral interval are grouped together (see Fig. \ref{fig:opacity-binning}).
For each simulation, we use the 1D temporal and spatial mean stratification
to estimate the formation heights of the various wavelengths and sort
the wavelengths into the different opacity bins. The bin selection
and wavelength sorting process is performed twice during the simulation's
relaxation phase after updating the individual mean stratifications,
but is kept unchanged during the production runs that make up the
time-series presented in this work.

To each bin, we assign a mean opacity $\kappa_{i}$ which accounts
for the contribution from both continuum and line opacities. To compute
the mean opacities, we differentiate between diffusion and free-streaming
limits, i.e. between the optical thick and optical thin regimes, below
and above the photospheric transition zone, respectively. The mean
bin-opacity $\kappa_{i}$ is calculated as a Rosseland-like average
\begin{equation}
\kappa_{\mathrm{Ross},i}=\left.\int_{\lambda\left(i\right)}\frac{dB_{\lambda}}{dT}d\lambda\middle/\int_{\lambda\left(i\right)}\frac{1}{\kappa_{\lambda}}\frac{dB_{\lambda}}{dT}d\lambda\right.\label{eq:rosseland_opacity}
\end{equation}
in the optical thick regime, and as a mean-intensity-weighted mean
opacity 
\begin{equation}
\kappa_{J,i}=\left.\int_{\lambda\left(i\right)}\kappa_{\lambda}J_{\lambda}d\lambda\middle/\int_{\lambda\left(i\right)}J_{\lambda}d\lambda\right.\label{eq:intensity-weighted-mean-opacity}
\end{equation}
in the optical thin regime, where $\lambda\left(i\right)$ is the
set of wavelength points assigned to bin $i$. For bin $i$, the transition
from one regime to the other around that bins optical surface is achieved
by means of an exponential bridging of the two averages: 
\begin{equation}
\kappa_{i}=e^{-2\tau_{\mathrm{Ross},i}}\kappa_{J,i}+\left(1-e^{-2\tau_{\mathrm{Ross},i}}\right)\kappa_{\mathrm{Ross},i}.
\end{equation}

All simulations presented here have been run with the radiative transfer
in the strict LTE approximation, i.e. under the assumption that he
monochromatic source function $S_{\lambda}$ (in Eq. \ref{eq:radiative_heating})
is the Planck function at the local gas temperature, i.e. $S_{\lambda}\left(T\right)=B_{\lambda}\left(T\right)$.
For each bin $i$, we compute an integrated source function by summing
up the contributions from all wavelength points in the bin;

\begin{equation}
S_{i}=B_{i}=\int_{\lambda(i)}B_{\lambda}d\lambda
\end{equation}

\citet{Collet:2011p6147} showed that, with this opacity binning implementation,
the approximation of strict LTE results in a temperature stratification
very similar to the case, where scattering is properly treated, as
long as the contribution of scattering to the extinction is \textit{excluded}
when averaging the mean opacities $\kappa_{J,i}$ (Eq. \ref{eq:intensity-weighted-mean-opacity})
in the optically thin layers (\textquotedbl{}streaming-regime\textquotedbl{}),
but \emph{include} it as true absorption when averaging the mean opacities
$\kappa_{\mathrm{Ross},i}$ (Eq. \ref{eq:rosseland_opacity}) in the
optically thick layers (\textquotedbl{}diffusion approximation regime.\textquotedbl{}).
They also showed that including scattering as true absorption leads
to erroneous atmosphere structures due to overestimated radiative
heating in the optically thin layers. However, these findings have
so far being verified only a small sample of stellar parameters, therefore
we cannot rule out that scattering needs to be accounted for properly
in certain cases. Nonetheless, evaluating the radiative transfer in
strict LTE greatly eases the computational burden compared to the
case, where the contribution of scattering is included to the total
extinction \citep{Hayek:2010p4435}.

The radiative transfer equation is solved for the individual opacity
bins (Eq. \ref{eq:radiative_transfer_first}) for all layers that
have $\max\left(\tau_{\mathrm{Ross}}\right)<300$, while in the deeper
layers, we use instead the diffusion approximation, which is fulfilled
to a high degree at such depths. With the opacity binning approximation,
the radiative heating rate term (Eq. \ref{eq:radiative_heating})
takes then the form 
\begin{eqnarray}
q_{\mathrm{rad}} & = & 4\pi\rho\sum_{i}\kappa_{i}\left(J_{i}-B_{i}\right)\label{eq:rad_trans}
\end{eqnarray}
where $J_{i}$ is the mean intensity computed from the solution of
the radiative transfer equation for bin $i$. 

For the relaxation phase of the simulation runs we considered six
bins, while for the final models we used twelve opacity bins. We have
developed an algorithm for the bin selection, which will be explained
further below (see Sect. \ref{sub:Selection-opacity-bins}). Towards
lower surface gravities ($\log g\lesssim2.0$) and higher effective
temperatures, numerical artifacts in the radiative transfer can occasionally
develop and manifest as a Moir{\'e} pattern in the integrated outgoing
intensities due to very steep temperature gradients in the photosphere.
For those situations, we solve the radiative transfer equation on
an adaptive mesh with finer vertical resolution, which is dynamically
optimized to resolve regions where temperature gradients are steeper.
The radiative heating and cooling rates computed on the adaptive mesh
are then interpolated back to the coarser hydrodynamic depth scale
under the consideration of energy conservation.

\subsection{The \textsc{Stagger}-grid models\label{sub:The-Stagger-grid-setup}}

\begin{figure}
\includegraphics[width=88mm]{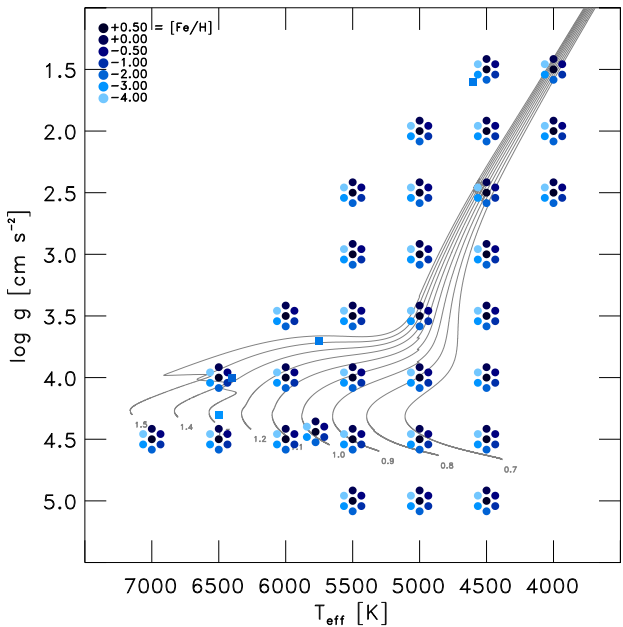}

\caption{Kiel diagram ($T_{\mathrm{eff}}-\log g$ diagram) showing the targeted
\textsc{Stagger}-grid parameters for the 217 models, comprising seven
different metallicities (colored circles). Four additional standard
stars (see text) are also indicated (squares). In the background,
the evolutionary tracks for stellar masses from $0.7$ to $1.5\, M_{\odot}$
and for solar metallicity are shown (thin grey lines).}
\label{fig:stagger-grid}
\end{figure}
The \textsc{Stagger}-grid covers a broad range in stellar parameters
with 217 models in total. The range in effective temperature is from
$T_{\mathrm{eff}}=4000\,\mathrm{K}$ to $7000\,\mathrm{K}$ in steps
of $500\,\mathrm{K}$, while the gravity ranges from $\log g=1.5$
to $5.0$ in steps of $0.5$. The grid also covers a broad range in
metallicity starting from $\left[\mathrm{Fe}/\mathrm{H}\right]=-4.0$
to $+0.5$ in steps of $1.0$ below $-1.0$, and steps of $0.5$ above
that%
\footnote{We use the bracket notation $\left[X/\mathrm{H}\right]=\log\left(N_{X}/N_{\mathrm{H}}\right)_{\star}-\log\left(N_{X}/N_{\mathrm{H}}\right)_{\odot}$
as a measure of the relative stellar to solar abundance of element
$X$ with respect to hydrogen.%
}. We decided to apply the same parameters $T_{\mathrm{eff}}$ and
$\log g$ for all metallicities, in order to facilitate the interpolation
of (averaged) models within a regular grid in stellar parameters.
In addition, the grid also includes the Sun with its non-solar metallicity
analogs, and four additional standard stars, namely HD 84937, HD 140283,
HD 122563 and G 64-12 that are presented in \citet{Bergemann:2012p20128}.
For metal-poor chemical compositions with $\left[\mathrm{Fe}/\mathrm{H}\right]\le-1.0$
we applied an $\alpha$-enhancement of $\left[\alpha/\mathrm{Fe}\right]=+0.4\,\mathrm{dex}$,
in order to account for the enrichment by core-collapse supernovae
\citep{Ruchti:2010p18149}.\\

In Figure \ref{fig:stagger-grid}, we present an overview of our simulations
in stellar parameter space. Therein, we also show evolutionary tracks
\citep{Weiss:2008p163} for stars with masses from $0.7$ to $1.5\, M_{\odot}$
and solar metallicity, in order to justify our choice of targeted
stellar parameters. Hence, the grid covers the evolutionary phases
from the main-sequence (MS) over the turnoff (TO) up to the red-giant
branch (RGB) for low-mass stars. In addition, the RGB part of the
diagram in practice also covers stars with higher masses, since these
are characterized by similar stellar atmospheric parameters.

\subsection{Scaling and relaxing 3D models\label{sub:Scaling-and-relaxing}}

Generating large numbers of 1D atmosphere models is relatively cheap
in terms of computational costs, but the same is not true for 3D models.
Based on our experiences from previous simulations of individual stars,
we designed a standard work-flow of procedures for generating our
grid. More specifically, we developed a large set of IDL-tools incorporating
the various necessary steps for generating new 3D models, which we
then applied equally to all simulations. The steps are: 
\begin{itemize}
\item Scale the starting model from an existing, relaxed 3D simulation,
and perform an initial run with six opacity bins, so that the model
can adjust to the new stellar parameters. 
\item Check the temporal variation of $T_{\mathrm{eff}}$ and estimate the
number of convective cells. If necessary, adjust the horizontal sizes,
in order to ensure that the simulation box is large enough to enclose
at least ten granules. 
\item If the optical surface has shifted upwards during the relaxation,
add new layers at the top of it to ensure that $\left\langle \log\tau_{\mathrm{Ross}}\right\rangle _{\mathrm{top}}<-6.0$. 
\item Determine the period $\pi_{0}$ of the radial p-mode with the largest
amplitude, then damp these modes with an artificial exponential-friction
term with period $\pi_{0}$ in the momentum equation (Eq. \ref{eq:momentum}).
\item Let the natural oscillation mode of the simulation emerge again by
decreasing the damping stepwise before switching it off completely. 
\item Re-compute the opacity tables with 12 bins for the relaxed simulation. 
\item Evolve the simulations for at least $\sim7$ periods of the fundamental
p-mode, roughly corresponding to $\sim2$ convective turnover times,
typically, a few thousand time-steps, of which 100 -- 150 snapshots
equally spaced were stored and used for analysis. 
\end{itemize}
During these steps the main quantities of interest are the time evolution
of effective temperature, p-mode oscillations, and drifts in the values
of the mean energy per unit mass and of the mean density at the bottom
boundary, which indicate the level of relaxation. When the drifts
in these above properties stop, we regard the simulation as relaxed.
If these conditions were not fulfilled, we continued running the model,
to give the simulation more time to properly adjust towards its new
quasi-stationary equilibrium state. Also, when the resulting effective
temperature of an otherwise relaxed simulation deviated more than
100 K from the targeted $T_{\mathrm{eff}}$, we re-scaled the simulation
to the targeted value of $\teff$ and started over from the top of
our list of relaxation steps.\\

The interplay between EOS, opacities, radiative transfer and convection
can shift the new location of the photosphere, when the initial guess
made by our scaling procedure slightly misses it. This is the case
for a few red giant models leading to upwards-shifts of the optical
surface and of the entire upper atmosphere during the adjustment phase
after the scaling, with the average Rosseland optical depth ending
up to be larger than required, i.e. $\left\langle \log\tau_{\mathrm{Ross}}\right\rangle _{\mathrm{top}}\ge-6.0$.
In order to rectify this, we extended those simulations at the top
by adding extra layers on the top, until the top layers fulfilled
our requirements of $\left\langle \log\tau_{\mathrm{Ross}}\right\rangle _{\mathrm{top}}<-6.0$.

\subsubsection{Scaling the initial models\label{sub:Scaling-models}\label{sub:Granule-counting}}

To start a new simulation, we scale an existing one with parameters
close to the targeted ones, preferably proceeding along lines of constant
entropy of the inflowing gas at the bottom in stellar parameter space
(see Fig. \ref{fig:const-ssbot-lines-1}). In this way, we find that
the relaxation process is much faster. In order to generate an initial
model for a set of targeted parameters, we scale temperature, density,
and pressure with depth-dependent scaling ratios derived from two
1D models, with parameters corresponding to the current and intended
3D model \citep{Ludwig:2009p4627}. For this, we used specifically
computed 1D envelope models (\textsc{MARCS} or our own 1D models,
see Sect. \ref{sub:1D-models}), which extend to $\log\tau_{\mathrm{Ross}}>4.0$.
The reference depth-scale for all models in the scaling process is
the Rosseland optical depth above the photosphere and gas pressure
normalized to the gas pressure at the optical surface below it ($\log\tau_{\mathrm{Ross}}>0.0$).\\

\begin{figure}
\includegraphics[width=88mm]{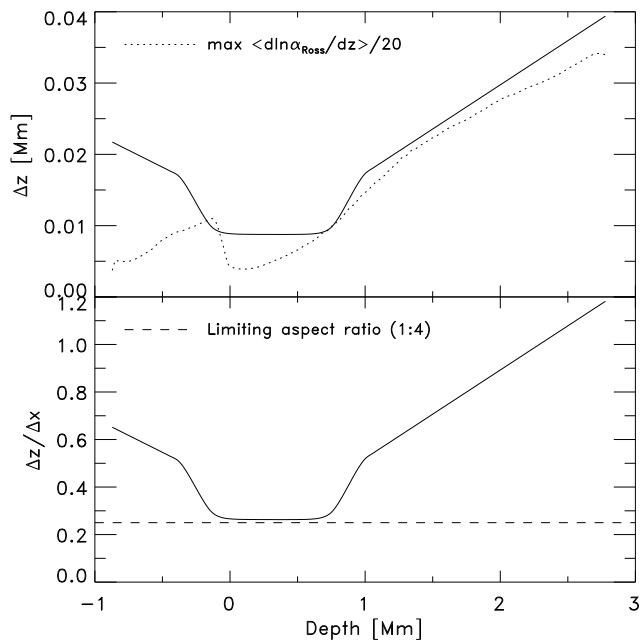}

\caption{In the top panel, we display the non-equidistant vertical spacing
$\Delta z$ of the depth scale as a function of geometrical depth
in our solar model (solid line). The $z$-scale is optimized to resolve
the flows and thermal structure, which naturally results in the highest
spatial resolution around the photosphere. Furthermore, we also show
the maximum of the vertical gradient of the absorption coefficient
$\max\left\langle d\ln\alpha_{\mathrm{Ross}}/dz\right\rangle $ as
a function of depth (dotted line). In the bottom panel, we show the
aspect ratio $\Delta z/\Delta x$ (solid line) and we also indicated
its lower allowed limit with $1:4$ (dashed line). The actual vertical-to-horizontal
aspect ratio ranges from $0.26$ at the photosphere to $1.18$ at
the bottom of the simulation domain.}
\label{fig:depth-scaling}
\end{figure}
After the initial scaling, we construct the geometrical depth scale
$z$ for the new simulation by enforcing the same (quasi-)hydrostatic-equilibrium
condition as in the starting simulation, but with the newly scaled
pressure and density. The resulting new $z$-scale is usually not
smooth, therefore we generate a new $z$-scale, which is optimized
to resolve the region with the steepest (temperature) gradients, as
shown in Fig. \ref{fig:depth-scaling}. The density-, energy-, and
velocity cubes are then interpolated to this new geometrical depth
scale. The new $z$-scale is constructed using the variation with
depth of the (smoothed) maximum of the derivative of the Rosseland
absorption coefficient, $\max\left\langle d\ln\alpha_{\mathrm{Ross}}/dz\right\rangle $,
as a guide. The basic idea behind this approach is to vertically distribute
the mesh points as evenly as possible on the optical-depth scale.
With such an optimized $z$-scale we can efficiently resolve the same
features with fewer grid-points, compared to an equidistant vertical
mesh. Furthermore, a limiting vertical-to-horizontal aspect ratio
($\Delta z/\Delta x$ and $\Delta z/\Delta y$) of $1:4$ over the
whole vertical extent is enforced. We find that this value represents
in practice an optimal lower limit to the aspect ratio, with respect
to numerical stability and accuracy of the solution of the radiative
transfer equation along inclined rays. Finally, the position of the
zero-point in the depth scale is adjusted to coincide with the position
of the mean optical surface, i.e. $\left\langle \taur\right\rangle _{\left(z=0\right)}=1$.\\

At fixed surface gravity and metallicity, the mean diameter of the
granules, which is used for determining the horizontal extent of the
simulation, increases with higher effective temperature (see Figs.
\ref{fig:granule-size-1} and Sect. \ref{sub:Granule-size}). The
number of granules present in the simulation box is retrieved with
the aid of the \textsc{contour} routine in IDL. Based on the map of
the temperature below the surface (the vertical velocity would serve
equally well), a contour chart of the significantly hotter granules
is extracted, from which the number of granules is counted. Concerning
the temporal resolution of the simulation sequences of the final production
runs, the frequency, at which snapshots are stored, is based on the
sound-crossing time of one pressure scale height, $H_{P}$, in the
photosphere, i.e. 
\begin{eqnarray}
\Delta t_{\mathrm{snap}} & = & \left\langle H_{P}/c_{s}\right\rangle _{\left(\tau=2/3\right)}
\end{eqnarray}
(see Cols. 14 and 15 in Table \ref{tab:global_properties}). With
the help of functional fits of the dependence of granule sizes and
sound-crossing time scales on stellar parameters, the horizontal sizes
of the simulation boxes and the snapshot sampling times can be estimated
rather accurately in advance (see App. \ref{sub:Functional-fits}).

\subsubsection{Selection of the opacity bins\label{sub:Selection-opacity-bins}}

\begin{figure}
\includegraphics[width=88mm]{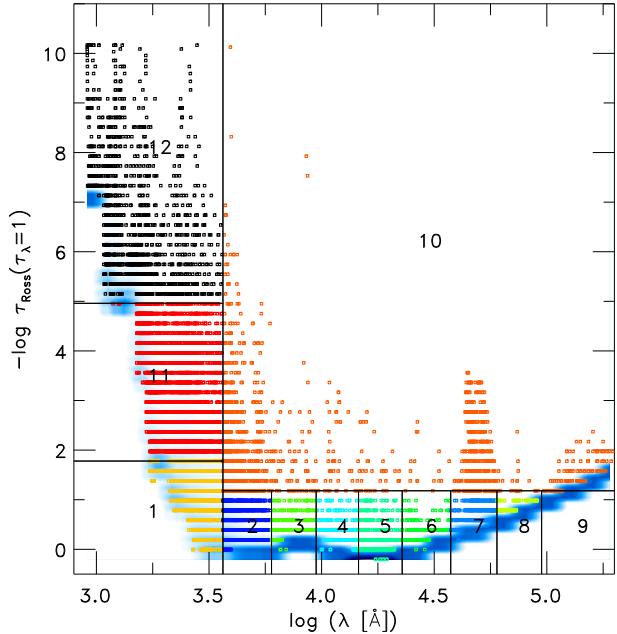} 

\caption{We show the twelve opacity bins selected for the solar simulation
by plotting the opacity strength (or, more precisely, the formation
height) against wavelength for all sampled wavelength points. The
individual bin elements are indicated by colored symbols. For clarity,
we plotted only a subset of the wavelength points considered for the
opacity binning procedure. In the background, we included the smoothed
histogram of the opacity strength distribution (blue contour). This
shows how the majority of $\lambda$-points are mostly concentrated
close to the continuum-forming layers and only a smaller fraction
contributes to lines. }
\label{fig:opacity-binning}
\end{figure}

As we mentioned earlier, in Sect. \ref{sub:Radiative-transfer}, the
purpose of the opacity-binning approximation is to reproduce the radiative
heating and cooling rates as accurately as possible with a small number
of opacity-bins, in order to reduce the computational burden. For
the assignment of wavelength points to bins, we first compute the
opacity strengths for all of the $\gtrsim10^{5}$ wavelength points
in the opacity-sampling (OS) data from the \textsc{MARCS} package
\citep{Gustafsson:2008p3814}. The histogram of their distribution
as a function of wavelength (see Fig. \ref{fig:opacity-binning})
exhibits a characteristic \textquotedbl{}L\textquotedbl{}-shape. Shorter
wavelengths (UV) require more bins to resolve the wide range in opacity
strength, while the lower part of the L-shaped distribution at longer
wavelengths (optical and IR) calls for a better resolution in terms
of wavelength. Therefore, we initially make a division in wavelength
at $\lambda_{X}$, between the UV and the optical/IR (see boundaries
of bin 1, 11 and 12 in Fig. \ref{fig:opacity-binning}) and comprising
approximately an equal number of wavelength points. These two regions
are then in turn subdivided evenly into opacity bins according to
the number of $\lambda$-points. By trial and error, we found that
a binning scheme with three bins in the $\lambda<\lambda_{X}$ region
and eight bins for $\lambda>\lambda_{X}$, one of which being a large
one and comprising the stronger lines in the optical and IR (bin number
10 in Fig. \ref{fig:opacity-binning}) gives a good representation
of the monochromatic radiative heating and cooling. We iterate the
bin selection with slight differences (e.g., one additional division
in opacity strength for the 8 bins in the lower part of the optical
and IR) and by small adjustments, and choose the bin selection with
the smallest relative difference between the total heating rates computed
with opacity binning $q_{\mathrm{bin}}$ and the full monochromatic
solution $q_{\lambda}$ for the average stratification of the 3D simulation,
i.e. 
\begin{equation}
\max\left[\delta q_{\mathrm{bin}}\right]=\frac{\max\left|q_{\mathrm{bin}}-q_{\lambda}\right|}{\max\left|q_{\lambda}\right|}.
\end{equation}

We found that the individual selection of some of the bins displays
a highly non-linear response to small changes. In most cases an even
distribution was favored by the minimization. Naturally, our method
will typically find only a local minimum due to the small sample of
iterations instead of a true global minimum. However, our method is
a fast, repeatable, and automatic selection of the opacity bins, which
minimizes the human effort significantly, while at the same time yielding
very satisfactory results. Moreover, the possible deviation from the
global minimum due to our automated bin selection and its resulting
uncertainties are anyways smaller than the overall uncertainties associated
with the opacity binning method.\\

\begin{figure}
\includegraphics[width=88mm]{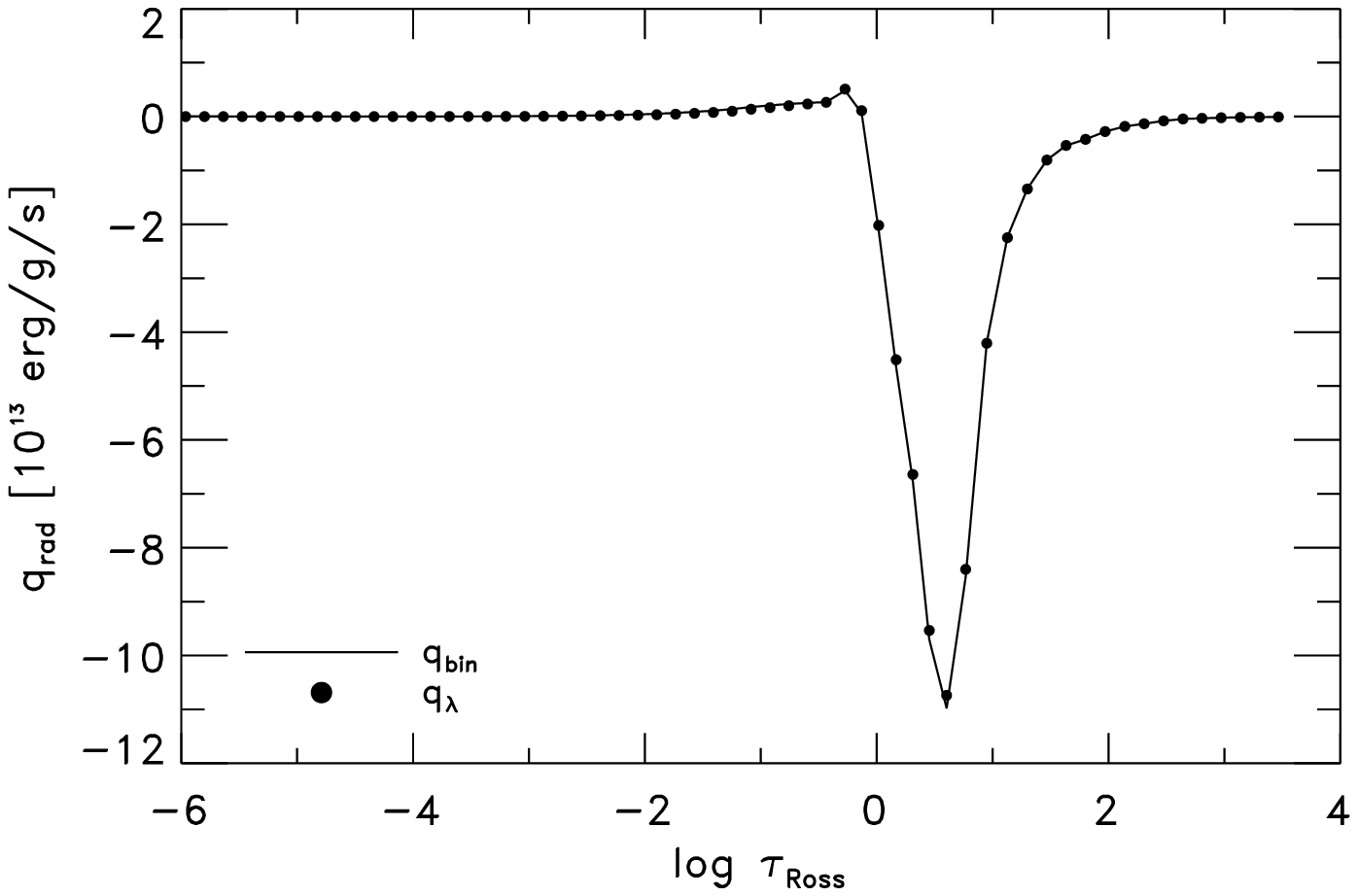}

\includegraphics[width=88mm]{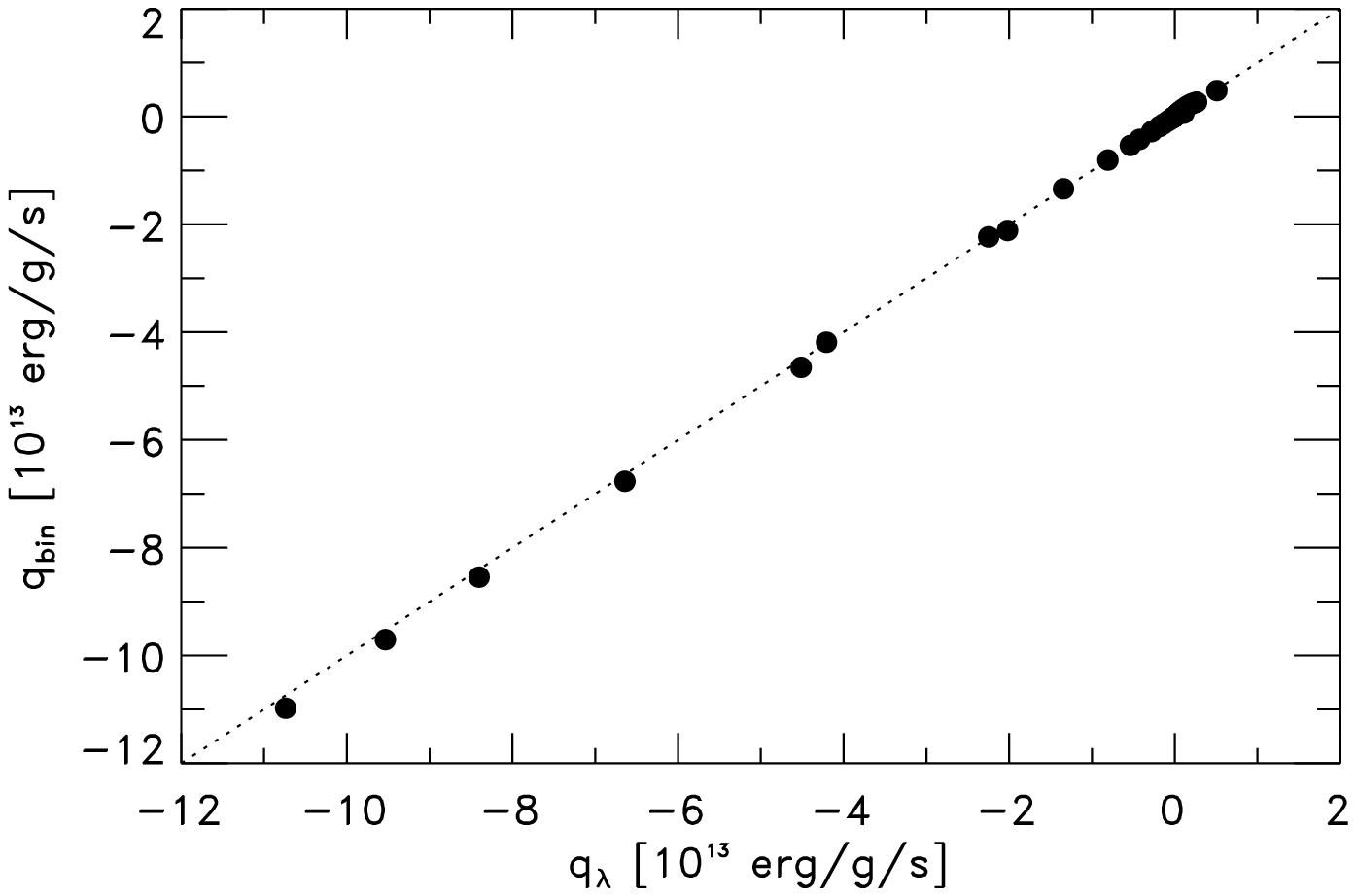}

\caption{Comparison of the radiative heating and cooling resulting from monochromatic
computations $q_{\lambda}$ (filled dots) and the opacity binning
method $q_{\mathrm{bin}}$ (solid line) for the solar model mean stratification.
In the top panel we show both $q_{\mathrm{rad}}$ vs. optical depth,
while in the bottom panel, we compare the two against each other. }
\label{fig:comp_qqmono_vs_bin}
\end{figure}

In Fig. \ref{fig:comp_qqmono_vs_bin}, we compare the resulting radiative
heating and cooling rates from the monochromatic calculation against
those from the opacity binning solution for the mean stratification
of our solar model. The radiative heating and cooling rates from the
simplified opacity binning appear rather similar to those from the
monochromatic solution, thereby supporting our approach. For the solar
model, our algorithm finds a bin selection that is just slightly less
accurate ($\max\left[\delta q_{\mathrm{bin}}\right]=2.78\,\%$) than
an optimized manual bin selection ($1.86\,\%$). Incidentally, with
six bins, we get $\max\left[\delta q_{\mathrm{bin}}\right]=3.54\,\%$.
We obtain an average $\max\left[\delta q_{\mathrm{bin}}\right]$ for
all the grid models of $\overline{\max\left[\delta q_{\mathrm{bin}}\right]}=2.38\,\%$,
while with six bins we get $\overline{\max\delta q_{\mathrm{bin}}}=3.0\,\%$.
We find that $\max\left[\delta q_{\mathrm{bin}}\right]$ increases
slightly with $\teff$ and $\feh$. We note that the opacity binning
method with its small number of bins states an approximation for the
radiative transfer, therefore, despite the small values for $\max\left[\delta q_{\mathrm{bin}}\right]$
further improvement is necessary.

\section{Results\label{sec:Results}}

The spatially and temporally averaged mean 3D stratifications (hereafter
$\left\langle 3\mathrm{D}\right\rangle $) from all of our 3D models
will be available online. The methods we applied to average our models
are explicitly described in a separate paper. We provide the models
in our own format, but also in various commonly used formats suited
for standard 1D spectrum synthesis codes such as \textsc{MOOG} \citep{Sneden:1973p21509},
\textsc{SYNTHE} \citep{Kurucz:1993p21505} and \textsc{Turbospectrum}
\citep{Plez:2008p20773,deLaverny:2012p23075}, together with routines
to interpolate the $\left\langle 3\mathrm{D}\right\rangle $ models
to arbitrary stellar parameters. In this paper the discussion will
be confined to global properties and mean stratifications only. More
extensive discussions and presentations of the wealth of details present
in the data of our 3D RHD models will be performed systematically
in subsequent papers.

\subsection{Global properties\label{sub:Global-properties}}

In Table \ref{tab:global_properties}, we have listed the stellar
parameters together with the thermodynamic values fixed for the inflows
at the bottom, i.e. the internal energy $\varepsilon_{\mathrm{bot}}$,
density $\rho_{\mathrm{bot}}$ and entropy $s_{\mathrm{bot}}$, as
well as important global properties for our 3D simulations. Before
we consider the $\left\langle 3\mathrm{D}\right\rangle $ stratifications
in Sect. \ref{sub:The-mean-atmosphere}, we briefly discuss some (temporally
averaged) global properties.

\subsubsection{Stellar parameters\label{sub:Stellar-parameters}}

Surface gravity and metallicity are input parameters for a simulation,
while the effective temperature is a property ensuing from the fixed
entropy of the inflowing material at the bottom $s_{\mathrm{bot}}$.
We calculate the effective temperature from the spatially averaged
emergent radiative energy flux $F_{\mathrm{rad}}$ and the Stefan-Boltzmann
law $T_{\mathrm{eff}}=\left[F_{\mathrm{rad}}/\sigma\right]^{1/4}$,
with $\sigma$ being the Stefan-Boltzmann constant. In Column 1 of
Table \ref{tab:global_properties} we have listed the resulting temporally
averaged $T_{\mathrm{eff}}$ of our final, relaxed simulations. These
differ somewhat from the targeted $T_{\mathrm{eff}}$s, since we do
not know a priori, the relation between $s_{\mathrm{bot}}$ and $T_{\mathrm{eff}}$.
However, the majority of our models ($72\%$) deviate less than 50K,
and the mean deviation for the whole grid is $\overline{\Delta T_{\mathrm{eff}}}\sim32\,\mathrm{K}$.

\subsubsection{Constant entropy of the adiabatic convection zone\label{sub:Entropy-bottom}}

\begin{figure}
\includegraphics[width=88mm]{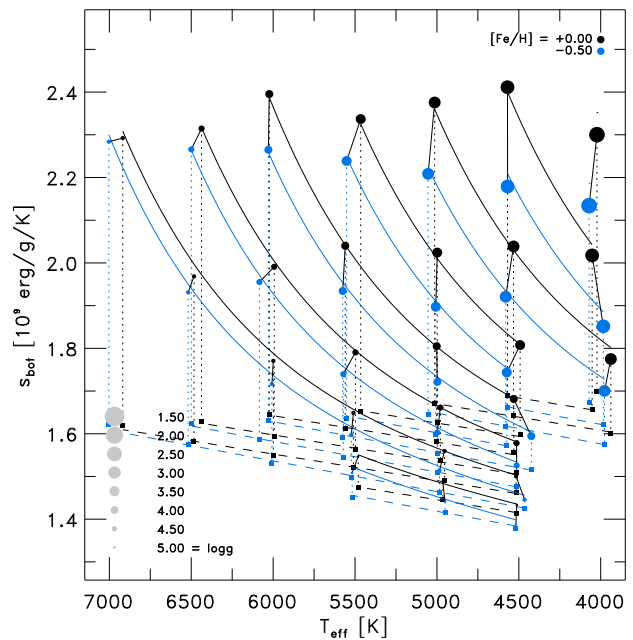}

\caption{Overview of the constant entropy value of the adiabatic convection
zone, which is given by the fixed entropy of the inflowing plasma
at the bottom, $s_{\mathrm{bot}}$, (circles) as well as the atmospheric
entropy minimum, $s_{\mathrm{min}}$, (squares) for two metallicities
($\left[\mathrm{Fe}/\mathrm{H}\right]=-0.5$ and $0.0$, blue and
black respectively) against $\teff$. The jump in entropy, $\Delta s$,
is indicated with vertical dotted lines. Simulations with the same
gravity are connected with functional fits for $s_{\mathrm{bot}}$
and $s_{\mathrm{min}}$ (solid and dashed lines respectively; see
App. \ref{sub:Functional-fits}), while similar simulations with different
$\feh$ are connected with short solid black lines.}
\label{fig:entropy-bottom}
\end{figure}

The main input parameter that has to be adjusted is $s_{\mathrm{bot}}$,
which has the same value as the entropy in the deep convection zone
due to the adiabaticity of convection, i.e. $s_{\mathrm{bot}}=s_{\mathrm{ad}}$
\citep{Steffen:1993ASPC...40..300S}. This is also the reason, why
the results from our rather shallow boxes are valid. We set $s_{\mathrm{bot}}$
by specifying a fixed value for the density and energy per unit mass
for the inflowing material at the bottom, $\rho_{\mathrm{bot}}$ and
$\varepsilon_{\mathrm{bot}}$. The actual values of $\varepsilon_{\mathrm{bot}}$,
$\rho_{\mathrm{bot}}$ and $s_{\mathrm{bot}}$ applied in our simulations
are given in Table \ref{tab:global_properties}. Furthermore, we provide
functional fits for $s_{\mathrm{bot}}$ (see App. \ref{sub:Functional-fits}).
We compute the entropy by integrating%
\footnote{The values for $p_{\mathrm{th}}\left(\rho,\varepsilon\right)$ and
$T\left(\rho,\varepsilon\right)$ are given in the EOS tables in the
covered range of $\log\left(\rho/\left[\mathrm{g}/\mathrm{cm}{}^{3}\right]\right)\in\left[-14,1\right]$
in 57 steps and $\log\left(\varepsilon/\left[\mathrm{erg}/\mathrm{g}\right]\right)\in\left[11,14\right]$
in 300 steps.%
} the first law of thermodynamics in the form 
\begin{equation}
ds=\frac{1}{T}\left(d\varepsilon-p_{th}\frac{d\rho}{\rho^{2}}\right),\label{eq:entropy}
\end{equation}
 adding an arbitrary integration constant in order to shift the zero-point
of the entropy to a similar value as in \citet{Ludwig:1999p7606}.
In Fig. \ref{fig:entropy-bottom}, we show $s_{\mathrm{bot}}$ against
$\teff$ for $\left[\mathrm{Fe}/\mathrm{H}\right]=0.0$ and $-0.5$,
as an example. The value for $s_{\mathrm{bot}}$ increases exponentially
with higher $T_{\mathrm{eff}}$ and with lower $\log g$, and linearly
with metallicity with a moderate slope.\\

In order to increase the effective temperature solely, i.e. the emergent
radiative flux $F_{\mathrm{rad}}$ at the top boundary, the total
energy flux ascending from the convection zone has to increase by
that same amount due to conservation of energy (see Sect. \ref{sub:transport-of-energy}).
On the other hand, when we keep $\teff$ fixed and decrease the surface
gravity, this in turn will cause the density to decrease correspondingly
(see Sect. \ref{sub:total-pressure_and_density}). Therefore, to maintain
the same energy flux, either the transported heat content ($\dss$,
$\varepsilon$) or the mass flux ($\rho$ or $v_{z}$) needs to be
enhanced. When the energy flux is carried by a larger mass flux, then
we speak of an enhancement in \textit{convective efficiency}%
\footnote{In 1D MLT modeling the term convective efficiency is commonly referred
to the mixing-length. The latter is in 3D RHD simulations referred
to the mass mixing-length, which is the inverse gradient of the mass
flux \citep[see][]{Trampedach:2011p5920}.%
} (see Sects. \ref{sub:total-pressure_and_density}, \ref{sub:velocity-field}
and \ref{sub:transport-of-energy}). When one considers $\varepsilon_{\mathrm{bot}}$
with stellar parameters, then it clearly depicts qualitatively the
same characteristic changes as $s_{\mathrm{bot}}$. By inserting the
perfect gas law%
\footnote{$p=kT\rho/\mu m_{u}$, with $k$ being the Boltzmann constant, $\mu$
the mean molecular weight, and $m_{u}$ the atomic mass constant.%
} in Eq. \ref{eq:entropy} one obtains 
\begin{eqnarray}
ds & = & \frac{d\varepsilon}{T}-\frac{k}{\mu m_{u}}\frac{d\rho}{\rho},\label{eq:entropy_dependence}
\end{eqnarray}
from which one can immediately see that the entropy increases with
internal energy, $s\propto\varepsilon$, and also increases with lower
density $s\propto-\ln\rho$. 

On the other hand, an increase in metallicity leads to a higher entropy
of the adiabat and also a larger atmospheric entropy-jump (see Fig.
\ref{fig:entropy-bottom}). Furthermore, we find increased velocities
(and $\dss$) and decreased densities at higher $\feh$ (see Sects.
\ref{sub:velocity-field} and \ref{sub:total-pressure_and_density}),
which in turn affects the convective efficiency. The dependence on
metallicity can be unveiled with the following approximation. The
opacity (and absorption coefficient) increases with higher $\feh$,
since the opacity depends directly on the metallicity. The hydrostatic
equilibrium can be written in terms of optical depth as $dp_{\mathrm{th}}/d\tau_{\mathrm{Ross}}=g/\kappa_{\mathrm{Ross}}$.
From the EOS (also ideal gas law), one can see that the pressure scales
with the density, $p_{\mathrm{th}}\propto\rho$. Therefore, when one
would fix $\teff$ and $\logg$, but increase metallicity (and opacity),
then the hydrostatic balance will be realized at a lower density stratification
(see bottom and middle panel in\ref{fig:density_peak}), which is
also given by 1D MLT models. The lower density stratification will
result in higher $\ssbot$ and $\dss$ (Eq. \ref{eq:entropy_dependence}
and top panel in\ref{fig:density_peak}).

We emphasize that the dependence of both $s_{\mathrm{bot}}$ and $\dss$
with stellar parameters is quite non-trivial, since not only is it
coupled to the changes in the total energy fluxes, but it is also
affected by the differences in the transition from convective to radiative
transport of energy with stellar parameters. In particular, the non-local
radiative transfer depends non-linearly on the conditions present
in stellar atmospheres, especially changes in the opacity and the
EOS will strongly influence the radiative transfer. Additionally,
$s_{\mathrm{bot}}$ will be influenced by changes in the efficiency
of the convective energy transport, that is in the convective fluxes
arising from the hydrodynamics (see Sect. \ref{sub:transport-of-energy}
for more details).\\

\begin{figure}
\includegraphics[width=88mm]{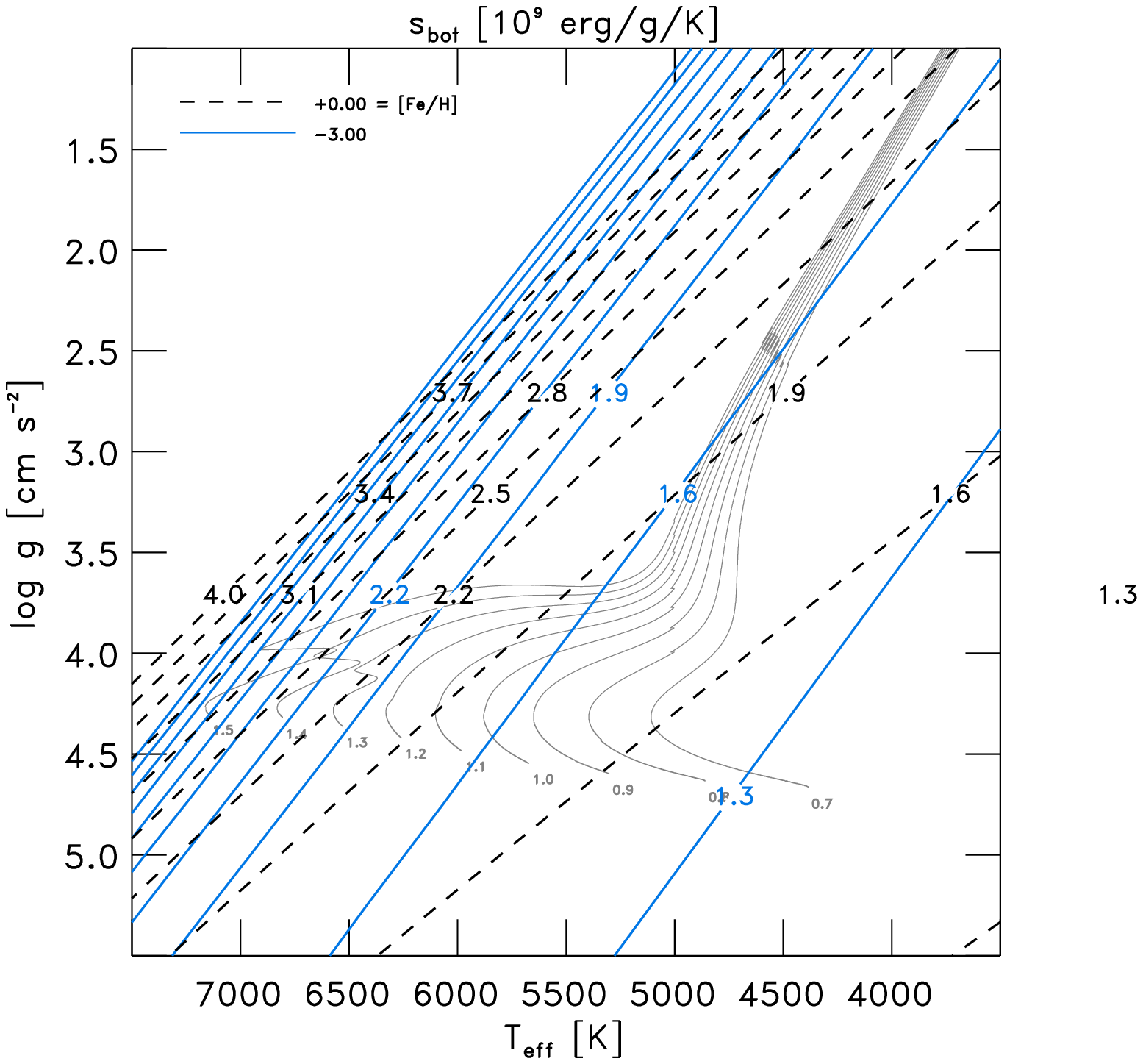}

\caption{Showing lines of constant entropy of the adiabat $s_{\mathrm{bot}}$
from $1.3$ to $4.0\times10^{13}\mathrm{erg}/\mathrm{g}/\mathrm{K}$
in steps of $0.3\times10^{13}\mathrm{erg}/\mathrm{g}/\mathrm{K}$
and for $\left[\mathrm{Fe}/\mathrm{H}\right]=-3.0$ and $0.0$ (blue
solid and black dashed lines respectively) in the Kiel diagram. Additionally,
we show evolutionary tracks for $0.6$ to $1.5\, M_{\odot}$ with
solar metallicity (thin grey lines).}
\label{fig:const-ssbot-lines-1}
\end{figure}

An analytical derivation of $s_{\mathrm{bot}}$ as a function of stellar
parameters is rather difficult, as explained above. Nonetheless, we
can fit $s_{\mathrm{bot}}$ with stellar parameter in a functional
form based on the results from our simulations (see App. \ref{sub:Functional-fits}).
This has been done previously, based on 2D RHD models with solar metallicity
by \citet{Ludwig:1999p7606}. In Fig. \ref{fig:const-ssbot-lines-1},
we show how $s_{\mathrm{bot}}$ varies across the Kiel diagram ($\teff-\logg$
diagram) for $\left[\mathrm{Fe}/\mathrm{H}\right]=0.0$ and $-3.0$.
In the case of $s_{\mathrm{bot}}$, our results with solar metallicity
are qualitatively in good agreement with those of \citet{Ludwig:1999p7606},
despite the inherent differences between 2D and 3D convection simulations,
the adopted EOS, opacities, and radiative transfer treatment. We find
that $s_{\mathrm{bot}}$ (which depicts $s_{\mathrm{ad}}$) lies on
lines of constant entropy in the Kiel diagram, in fact for the solar
metallicity these lines with the same entropy of the adiabat run almost
diagonally. Moreover, towards lower metallicity, we find two significant
differences for the lines of constant $s_{\mathrm{bot}}$, the first
one being that the slopes of the lines steepen, and the second being
that the distances in the $\teff-\logg$ plane between the lines decrease.
The latter implies metal-poor stars feature a broader range in $s_{\mathrm{bot}}$
compared to metal-rich ones. As we will see in Sect. \ref{sub:The-mean-atmosphere},
this has important consequences for the stratifications.

\subsubsection{Entropy jump\label{sub:The-entropy-jump}}

The upflows enter the simulation box at the bottom with the constant
entropy value of the adiabatic convection zone, $s_{\mathrm{bot}}$,
and ascend until they reach the superadiabatic region (SAR) just below
the visible surface, where the convective energy is converted to radiative
energy. In the photosphere, the mean free path for the continuum radiation
grows large enough for the gas to become transparent, and the overturning
upflow at the surface loses its internal energy as photons escape
and carry away entropy. Further above in the nearly isothermal atmosphere
(with constant $\varepsilon$, see Fig. \ref{fig:ee-rho-plane-1})
with an exponentially decreasing density the entropy increases again
due to the EOS (see Eq. \ref{eq:entropy_dependence}). This leads
to a minimum, $s_{\mathrm{min}}=\min\left[\left\langle s\right\rangle \right]$,
just above the surface ($\log\tau_{\mathrm{Ross}}<0.0$) in the temporal
and horizontal averaged entropy (see Fig. \ref{fig:mean_entropy}).
We determined the entropy jump from the difference of the entropy
minimum and the fixed entropy at the bottom, i.e. 
\begin{equation}
\Delta s=s_{\mathrm{bot}}-s_{\mathrm{min}}.
\end{equation}
In order to calculate $s_{\mathrm{min}}$, we used averages of the
entropy on constant Rosseland optical depth%
\footnote{Averages on constant column mass density yield a very similar $s_{\mathrm{min}}$.%
}, since it is the radiation losses that cause the sharp changes in
the thermodynamic state around the optical surface and, therefore,
the optical depth scale offers a better reference frame for comparisons.
The averages on constant geometrical height $\left\langle 3\mathrm{D}\right\rangle _{z}$
smear out and thereby overestimate $s_{\mathrm{min}}$ increasingly
towards higher $T_{\mathrm{eff}}$ due to the increasing level of
corrugation of iso-$s$ surfaces on the geometrical scale (see Fig.
\ref{fig:mean_entropy}). The constant entropy at the bottom $s_{\mathrm{bot}}$,
on the other hand, is a fixed input value for each simulation. It
is worthwhile to mention that the main contribution to the variation
in $\Delta s$ as a function of stellar parameters is due to $s_{\mathrm{bot}}$,
since $s_{\mathrm{min}}$ varies just slightly with stellar parameter
compared to $s_{\mathrm{bot}}$ (see Fig. \ref{fig:entropy-bottom}).\\

In Fig. \ref{fig:entropy-bottom} we show $s_{\mathrm{min}}$ (dashed
lines) as well as $\Delta s$ (dotted lines), and it is obvious that
the minimum in entropy increases just slightly with increasing $T_{\mathrm{eff}}$,
while the jump increases with the constant entropy value of the adiabatic
convection zone quasi-exponentially at higher $T_{\mathrm{eff}}$
and lower $\log g$. This can be concluded more easily from Fig. \ref{fig:entropy_jump}
(top panel), where we display $s_{\mathrm{bot}}$ vs. $T_{\mathrm{eff}}$
(see also Col. 8 of Table \ref{tab:global_properties} and for $\Delta s$
and $s_{\mathrm{min}}$ in App. \ref{sub:Functional-fits}). We note
that the location of the entropy jump essentially represents the boundary
of stars and the jump is to be regarded as physically realistic, which
is a result of 3D RHD simulations. The entropy minimum coincides with
the position of the upper end of the SAR. A similar sharp drop occurs
for most of the thermodynamic quantities of interest ($\varepsilon$,
$T$ and $n_{\mathrm{el}}$), whereas $\rho$ and $p_{\mathrm{tot}}$
display a marked change of gradient. Moreover, the jump in entropy
is an important value, since it is a direct measure of the efficiency
of convective energy transport \citep[see][]{Trampedach:2013arXiv1303.1780T}.
The latter is in 1D modeling set by the four MLT parameters, especially
$\alpha_{\mathrm{MLT}}$, in the framework of MLT \citep[see][]{BohmVitense:1958p4822,Henyey:1965p15592}.
Towards cool dwarfs $\Delta s$ becomes smaller, indicating a higher
convective efficiency, while towards hotter stars the large entropy
jumps reflect a low convective efficiency. We will present a calibration
of $\alpha_{\mathrm{MLT}}$ based on the entropy jump in a subsequent
study, as previously done by using multidimensional convection simulations
\citep[see][]{Ludwig:1999p7606,Trampedach:1999p22663}. While we found
$\ssbot$ to be qualitatively similar to those of \citet{Ludwig:1999p7606},
in the case of $\Delta s$ our findings are also similar, but the
differences are here distinctively larger.\\

\begin{figure}
\includegraphics[width=88mm]{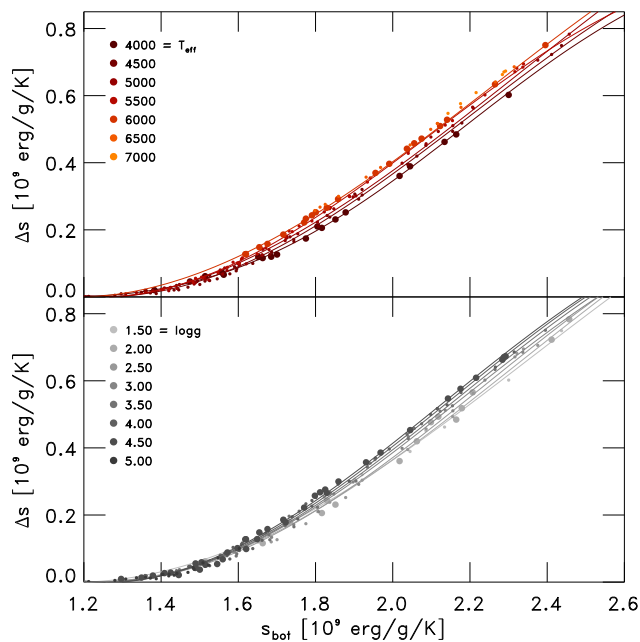}

\caption{We compare the entropy jump $\Delta s$ against the constant entropy
value of the adiabatic convection zone $s_{\mathrm{bot}}$ for all
grid models (filled circles; color-coding explained in the legend).
Furthermore, we show also hyperbolic tangent functional fits (see
Eq. \ref{eq:hyp_fit} and App. \ref{sub:Functional-fits}) between
$T_{\mathrm{eff}}=4000$ and $6000\,\mathrm{K}$ (red lines; top panel)
and between $\log g=1.5$ and $4.5$ (grey lines; bottom panel).}
\label{fig:ssbot_vs_ssj}
\end{figure}
It is quite remarkable how closely the variation of $\Delta s$ resembles
the change of $s_{\mathrm{bot}}$ with stellar parameters (see Fig.
\ref{fig:entropy-bottom}). Motivated by this, we compare directly
$\Delta s$ against $s_{\mathrm{bot}}$ in Fig. \ref{fig:ssbot_vs_ssj}.
We find a nice correlation between $\Delta s$ vs. $s_{\mathrm{bot}}$.
At lower $s_{\mathrm{bot}}$ values, $\Delta s$ seems to converge
towards $0.0$ (a negative jump is not expected, since the atmosphere
is \textit{losing} energy in form of radiation from the photosphere),
while for $s_{\mathrm{bot}}\gtrsim1.7$, $\Delta s$ grows linearly
with $s_{\mathrm{bot}}$ and only a modest level of scatter. In Fig.
\ref{fig:ssbot_vs_ssj}, we color-coded the $\teff$- and $\logg$-values
respectively, to show how the residuals depend systematically on atmospheric
parameters. Models with higher $\teff$ (bright orange dots) and higher
$\logg$ (dark grey dots) settle along higher $\Delta s$ and vice
versa. In order to illustrate this better, we have fitted a set of
hyperbolic tangent functions (see Eq. \ref{eq:hyp_fit}), which we
show also in Fig. \ref{fig:ssbot_vs_ssj}. We included functional
fits between $T_{\mathrm{eff}}=4000$ and $6000\,\mathrm{K}$ (red/orange
lines in top panel) and between $\log g=1.5$ and $4.5$ (grey lines
in bottom panel). Hence, we find hotter dwarfs along lines at larger
$\Delta s$, while cooler giants settle along lines at smaller entropy
jumps.

Interestingly, in the linear part ($s_{\mathrm{bot}}\gtrsim1.7$)
$\Delta s\left(s_{\mathrm{bot}}\right)$ displays a rather universal
slope of $\Delta s/s_{\mathrm{bot}}\sim0.85$, while the actual offset
in the ordinate depends mainly on $T_{\mathrm{eff}}$ and $\log g$.
Another interesting aspect is that $T_{\mathrm{eff}}$ shows a similar
strong influence as $\logg$. The latter, however, is obviously expressed
in logarithmic scale, therefore the influence of $T_{\mathrm{eff}}$
is much stronger. On the other hand, when one performs a similar hyperbolic
tangent functional fit for a fixed value of $\feh$, then $\Delta s$
is dispersed around the functional fit with such a large scatter that
a fit is rather meaningless. Therefore, in contrast to $\teff$ and
$\logg$ we find no systematic trends with metallicity.\\

Based on the strong correlation between the entropy jump $\Delta s$
and $s_{\mathrm{bot}}$, it is of interest to investigate what other
\textit{scaling relations} may be manifested for other stellar properties.
With $\dss$ as an inverse measure of convective efficiency, we expect
that in light of such scaling relations, important quantities depending
on the entropy jump will also similarly scale systematically with
$s_{\mathrm{bot}}$, in particular density and velocity (see Sects.
\ref{sub:velocity-field} and \ref{sub:total-pressure_and_density}),
and therefore also the calibrated mixing-length of a particular MLT
implementation. We note briefly that qualitatively similar relations
can be achieved with 1D MLT models with a fixed mixing length.

\subsubsection{Emergent intensity\label{sub:The-emergent-intensity}}

\begin{figure*}
\hspace{5mm}\includegraphics[width=175mm]{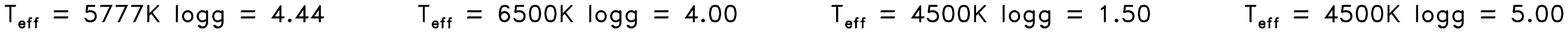}\\
\begin{minipage}[t]{1\columnwidth}%
\includegraphics[height=175mm]{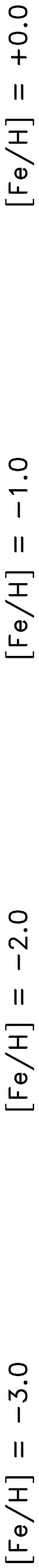}\hspace{1mm}\includegraphics[width=175mm]{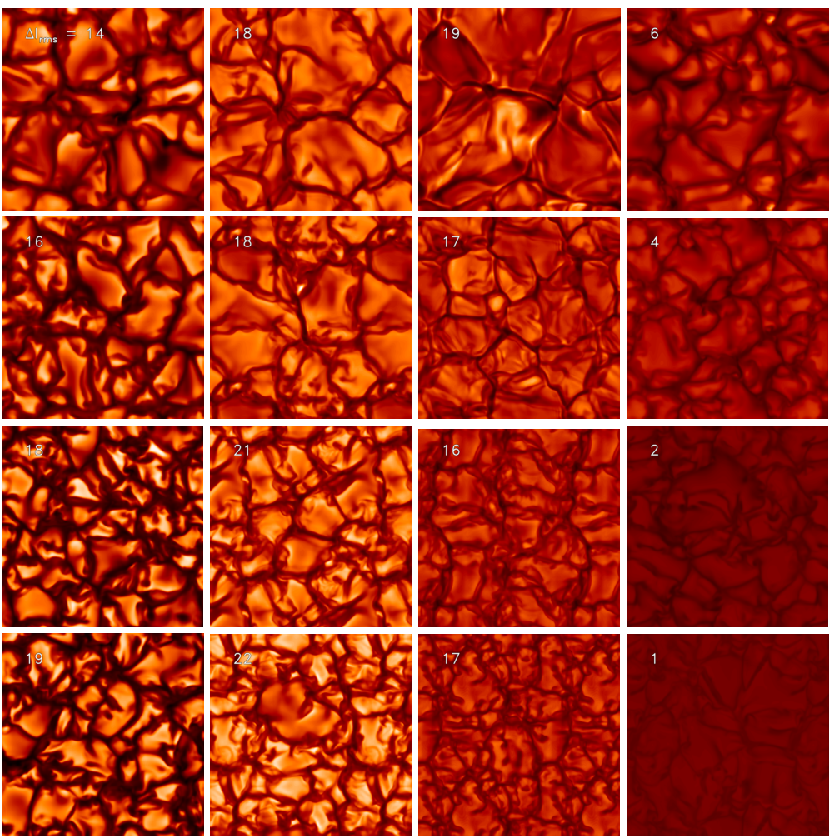}%
\end{minipage}\\
\hspace{5mm}\includegraphics[width=175mm]{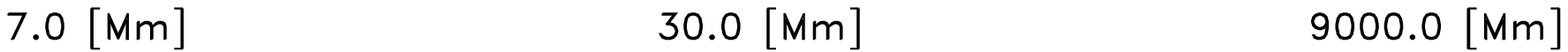}

\caption{We show an overview of the emergent (bolometric) intensity for a selection
of stars, namely main-sequence (MS), turnoff (TO), K-giant K-dwarf
(from left to right, respectively) at a given time instant. For each
star, we show four metallicities $\feh=0.0,-1.0,-2.0$ and $-3.0$
(from top to bottom, respectively). To facilitate comparisons between
the different metallicity of each star, the intensity scale and the
horizontal geometrical size of the metal-poor simulations are identical
to $\feh=0.0$, and the individual intensity contrasts $\left[\%\right]$
are indicated in each box.}
\label{fig:intenisty-map}
\end{figure*}
While classic 1D models are inherently horizontally symmetric, therefore
lacking a visible granulation pattern, the emergent intensity of 3D
models features inhomogeneities exhibiting rich details, which arise
due to the presence of turbulent convective motions. We give an overview
the emergent intensity of our simulations in Fig. \ref{fig:intenisty-map}.
Therein we display a main-sequence (MS) simulation (the Sun), a turnoff
(TO) simulation, a K-giant, and a K-dwarf model, each with four different
metallicities. To facilitate direct comparisons among the four metallicities,
we kept the horizontal sizes and the color scales for the continuum
intensities fixed from $\feh=0.0$ for the individual stellar categories
(we extended the metal-poor simulations by exploiting the periodic
horizontal boundary conditions). The dark regions depict the cold
intergranular lanes, while the brighter areas are the hot granules.
The radiation above the granules originate at higher geometrical heights,
while for downdrafts it comes from much lower heights. This is because
the opacity is highly nonlinear due to the strong temperature sensitivity
of the $\mathrm{H}^{-}$-opacity ($\kappa_{\mathrm{H}^{-}}\sim T^{10}$,
see SN98), which is the by far the dominant continuum opacity source
in the visible for late-type stellar photospheres. Since the temperature
difference between the granules and the intergranular lanes is very
large ($>10^{3}\,\mathrm{K}$), layers of constant optical depth will
be increasingly more corrugated and become largest around the peak
of the SAR. Therefore, the radiation above granules is emerging from
higher geometrical depths, $z_{\mathrm{up}}$, while above downdrafts
it originates from deeper geometrical heights, $z_{\mathrm{dn}}$
(for the Sun the largest difference between the averaged geometrical
heights can amount up to $\left\langle z\right\rangle _{\mathrm{up}}-\left\langle z\right\rangle _{\mathrm{dn}}\simeq-140\,\mathrm{km}$
at $\ltaur=2.0$).\\

An immediate, interesting aspect that leaps to the eye from the overview
presented in Fig. \ref{fig:intenisty-map} is the qualitative self-similarity
of the granulation patterns despite the large variations in size-scales.
The emergent intensity increases towards higher $T_{\mathrm{eff}}$
and decreases for lower surface gravities, as expected. From Fig.
\ref{fig:intenisty-map}, it is also clear that the granule sizes
decrease with metallicity (due to smaller $H_{P}$, see Sect. \ref{sub:Granule-size};
see also \citealp{Collet:2007p5617}). Also apparent is the change
of intensity contrast with stellar parameters, as we will discuss
below.\\

\begin{figure}
\includegraphics[width=88mm]{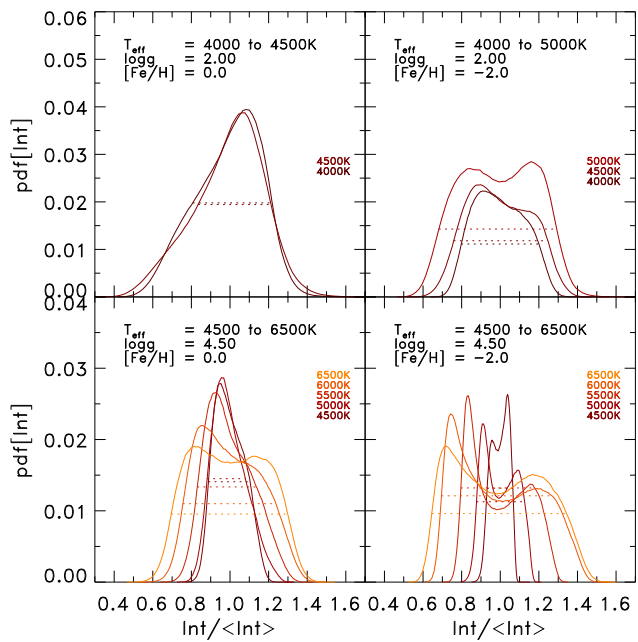}

\caption{We show the temporally averaged histograms of the (bolometric) intensity
(solid lines) for various stellar parameters (the bin-size is 100).
To ease comparisons between the different stellar parameters, we normalized
the individual intensity scales with its mean value. Furthermore,
we have indicated the respective FWHM of the intensity histograms,
$I_{\mathrm{FWHM}}$, (dotted lines), a measure of the intensity contrast.
Note the different ordinate scale in the top panel. The bimodal distributions
are given due to the asymmetry and the large contrast in the up and
downflows. }
\label{fig:intensity-histogram}
\end{figure}
In order to discuss the changes in the intensity, we show in Fig.
\ref{fig:intensity-histogram} the temporally averaged histograms
of the intensity $I$ normalized to their individual mean intensity
$\overline{I}$, thereby enabling a direct comparison between different
stellar parameters. The histograms of the intensity show two components:
a peak at lower (darker, $I/\overline{I}<1.0$) intensities, resulting
from the cool downdrafts, and an often broader component at higher
(brighter, $I/\overline{I}>1.0$) intensities, arising from the upflowing
hot granules. We note that these findings are qualitatively to be
expected \citep[see SN98;][]{Trampedach:2013arXiv1303.1780T}. 

\begin{figure}
\includegraphics[width=88mm]{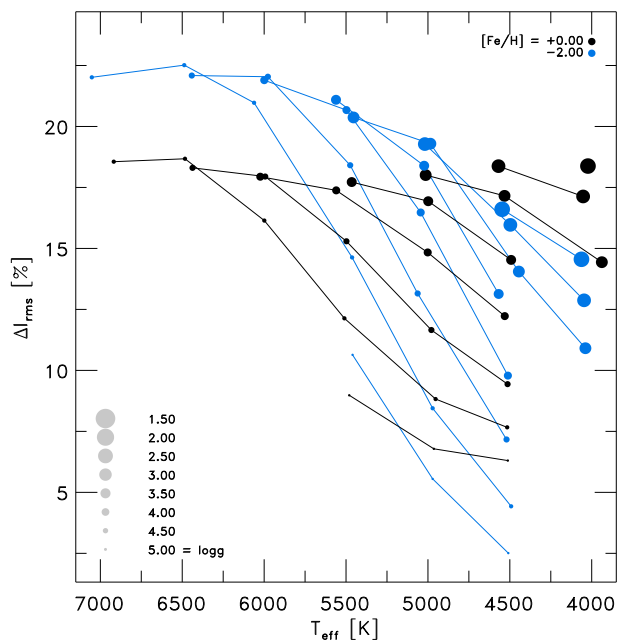}

\caption{Overview of the (bolometric) intensity contrast $\Delta I_{\mathrm{rms}}$
against $\teff$ for $\feh=-2.0$ and $0.0$ (blue and black respectively).
Models with the same gravity, but different $\teff$ are connected.
The \textit{enhanced} naked or hidden granulation at lower metallicity
can be extracted in the larger range of intensity contrasts (see text
for details).}
\label{fig:intensity-contrast}
\end{figure}

As clearly depicted in Fig. \ref{fig:intensity-histogram}, the shapes
of the two components change with stellar parameters, in particular
the amplitudes and widths, thereby changing the overall shape. The
two components can be clearly extracted from histograms at higher
$T_{\mathrm{eff}}$, where the intensity contrast is increasingly
enhanced and eventually produces a distinctly bimodal distribution,
which is a manifestation of the \textit{hidden or naked granulation}
(see Fig. \ref{fig:intensity-contrast} and \citealt{Nordlund:1990p6720}).
In order to better illustrate this, we also included the full width
at half maxima (FWHM) of the individual intensity histograms $I_{\mathrm{FWHM}}$
in Fig. \ref{fig:intensity-contrast}. On the other hand, at lower
$T_{\mathrm{eff}}$ the intensity contrast decreases in general, so
that the two components overlap, leading to a single narrower higher
peak in the histogram, thereby becoming indistinguishable from each
other in the histogram. \citet{Ludwig:2012A&A...547A.118L} found
also an unimodal intensity distribution in the context of a 3D giant
model with solar metallicity. Furthermore, we find that the individual
contribution to the intensity from upflows and downflows is often
asymmetric, meaning that the amplitudes of the two peaks in the bimodal
distribution are unequal (see Fig. \ref{fig:intensity-histogram}).
In general, for dwarfs, we find that the relative importance of downflows
with respect to upflows in terms of the peak contribution to the intensity
distribution increases with increasing $\teff$. However, we also
find exceptions, e.g. at lower metallicity where the behavior at $T_{\mathrm{eff}}=4500\,\mathrm{K}$
is actually the opposite. Also, the balance between upflows and downflows
varies with surface gravity. The intensity histograms for giants are
in general broader (higher contrast) compared to dwarfs of the same
$\teff$ (see $I_{\mathrm{FWHM}}$), hence exhibiting a larger intensity
contrast. For dwarfs at lower metallicity (right bottom panel) the
bimodality is more pronounced and the $I_{\mathrm{FWHM}}$ (contrast)
is broader (higher) towards higher $\teff$, while at lower $\teff$
the $I_{\mathrm{FWHM}}$ (contrast) becomes narrower (lower) compared
to solar metallicity (left bottom panel). The latter hints at an \textit{enhancement}
of the effect of hidden or naked granulation.\\

To illustrate the latter in more detail, we show in Fig. \ref{fig:intensity-contrast}
the rms of the bolometric disk center intensity fluctuations for $\left[\mathrm{Fe}/\mathrm{H}\right]=0.0$
and $-2.0$, which is commonly referred as the intensity contrast
\begin{eqnarray}
\Delta I_{\mathrm{rms}} & = & \left[\sum\left(I_{i}-\overline{I}\right)^{2}/N\overline{I}^{2}\right]^{1/2}\label{eq:intensity-contrast}
\end{eqnarray}
with $\overline{I}$ being the (spatial) mean intensity and $N$ the
number of data points \citep[see][]{Roudier:1986p20337}. We remark
that the shown $\Delta I_{\mathrm{rms}}$ are is temporal averages.
It is essentially defined as the relative standard deviation, hence
it reflects the width of the intensity distribution (see Fig. \ref{fig:intensity-histogram}).
This often measured value is very suitable for quantifying the range
of brightness fluctuations due to granulation. The intensity contrast
increases with higher $T_{\mathrm{eff}}$ and lower $\log g$. For
our solar simulation we get an intensity contrast of $15\,\%$, which
is close to the one found by SN98 with $16\,\%$ (see Col. 10 in Table
\ref{tab:global_properties}).\\

Towards higher $T_{\mathrm{eff}}$, we find that $s_{\mathrm{bot}}$,
$\Delta s$, and the vertical velocity increase, as shown in Sects.
\ref{sub:Entropy-bottom} and \ref{sub:velocity-field}. For increasingly
hotter stars, the top of the convective zone, $z_{\mathrm{top,cz}}$,
penetrates higher and higher above the optical surface due to larger
vertical velocities (see Fig. \ref{fig:rms_vertical_velocity}). Additionally,
at higher $T_{\mathrm{eff}}$ (higher $s_{\mathrm{bot}}$ and $\Delta s$),
the overall temperatures and their fluctuations also increase, implying
that one observes increasingly higher layers, since the dominant $\mathrm{H}^{-}$-opacity,
hence the optical depth, depends sensitively on the temperature. Therefore,
the granulation pattern is enhanced at higher $T_{\mathrm{eff}}$,
while on the contrary for lower $T_{\mathrm{eff}}$ the granulation
becomes less visible, since $z_{\mathrm{top,cz}}$ recedes below the
optical surface in the latter case (see overview in Fig. \ref{fig:intenisty-map}).
This phenomenon has been already described by \citet{Nordlund:1990p6720}
as \textit{naked granulation}.\\

Interestingly, in our simulations, we find that at lower metallicity
the effect of naked and hidden granulation is more pronounced, in
the sense that the range of contrast from cool, low-contrast dwarfs,
to hotter, high-contrast dwarfs, is $61\,\%$ larger (from $10.9$
to $17.6$) for our $\left[\mathrm{Fe}/\mathrm{H}\right]=-2.0$ simulations
than for solar metallicity (see Fig. \ref{fig:intensity-contrast}).
At lower metallicity the major electron donors (metals) are depleted,
therefore the formation of the dominant opacity source $\mathrm{H}^{-}$
depends primarily on the ionization of hydrogen, which is the reason
for the steep increase of intensity contrast with for higher $\teff$
\citep{Nordlund:1990p6720}.\\

The variations in the intensity and in the intensity contrast with
stellar parameters have important ramifications for observations.
At the one hand, the \textit{\emph{enhanced}} naked or hidden granulation
at lower metallicity affects the formation of spectral lines and the
limb darkening. On the other hand, it should also lead to distinct
signatures in the granulation background of asteroseismological observations
and spectro-interferometric imaging. Here, we limit ourselves to the
discussion of the global properties of the emergent intensity patterns,
and a detailed analysis will be performed in subsequent papers.

\subsubsection{Granule size\label{sub:Granule-size}}

\begin{figure}
\includegraphics[width=88mm]{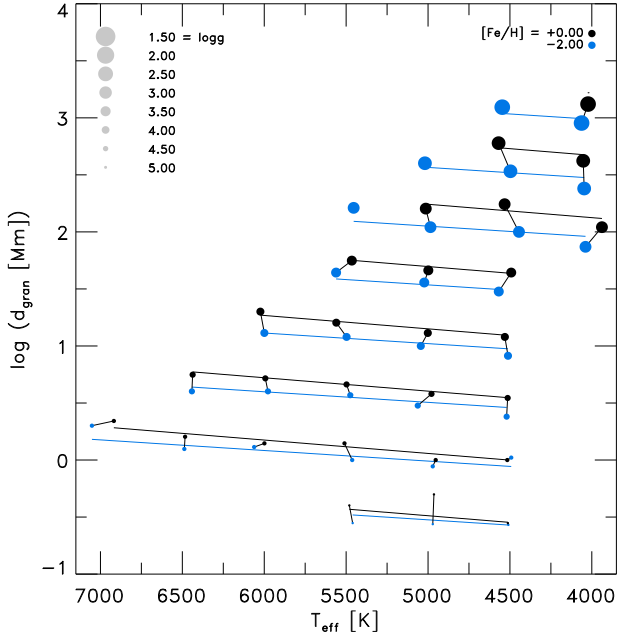}

\caption{Overview of the granule diameter $d_{\mathrm{gran}}$ derived from
the maximum of the mean 2D spatial power spectrum of the bolometric
intensity against $\teff$ for $\feh=-2.0$ and $0.0$ (blue and black
respectively). Models with the same gravity are connected by their
respective functional fits (solid lines; see App. \ref{sub:Functional-fits}).}
\label{fig:granule-size-1}
\end{figure}
The physical dimensions of the simulations boxes ($s_{x},s_{y}$ and
$s_{z}$ in Cols. 11 and 12 of Table \ref{tab:global_properties})
are selected based on the mean diameter of granules (see Sect. \ref{sub:Granule-counting}
and Table \ref{tab:global_properties}) and a target of about 10 granules
in the box. Additionally, we measured the granule sizes by calculating
the 2D spatial power spectrum of the bolometric intensity for the
time series, and determining its maximum from the smoothed time average
\citep[see Fig. 9 in][]{Trampedach:2013arXiv1303.1780T}. This method
is quite robust despite the large variations in gravity. In Fig. \ref{fig:granule-size-1},
we present the measured granule sizes $d_{\mathrm{gran}}$ (given
in Col. 13 in Table \ref{tab:global_properties}; see also App. \ref{sub:Functional-fits}),
showing that they become larger with smaller surface gravity. Also,
the granules of the simulations with fixed $\logg$, the lowest $T_{\mathrm{eff}}$
are typically $\sim50\,\%$ smaller compared to the simulations with
the hottest $T_{\mathrm{eff}}$, while for the models with the lowest
metallicity they are typically $\sim30\,\%$ smaller than for the
metal-rich ones.\\

\begin{figure}
\includegraphics[width=88mm]{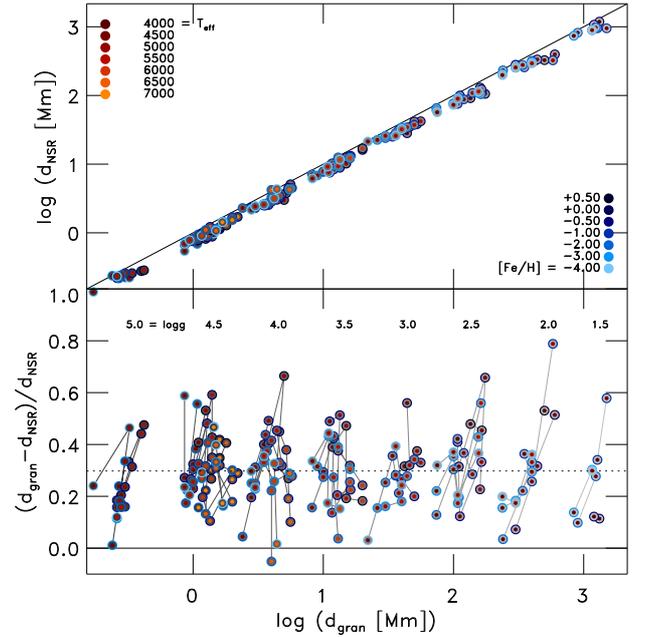}

\caption{In the top panel, we compare the granule size approximated with the
pressure scale height $d_{\mathrm{NIC}}$ (see Eq. \ref{eq:dNIC})
against the mean granule diameter $d_{\mathrm{gran}}$ (same as Fig.
\ref{fig:granule-size-1}) for all models. The individual stellar
parameters are indicated ($\teff$, $\logg$ and $\feh$ with red,
gray, and blue respectively). We indicated the position of $d_{\mathrm{NIC}}=d_{\mathrm{gran}}$
the solid diagonal line. In the bottom panel, we show also the relative
residuals. We indicated also the mean residual (dotted line), which
amounts to $\sim30\,\%$. Here, models with the same gravity are connected
(solid grey lines).}
\label{fig:granule-size-2}
\end{figure}
We find a remarkable validation for the approximation of the maximal
horizontal extent of a granule based on mass-conservation considerations
made by SN98 \citep[see also][]{Nordlund:2009p4109}. Hereafter, we
denote the following relation as the \textit{Nordlund scaling relation}
(NSR). The ascending buoyant plasma inside a cylindrical granule with
radius $r$ gives rise to a vertical mass flux with $j_{z}=[\pi r^{2}]\rho v_{z}$.
This mass flux has to deflect and overturn increasingly towards the
top. Due to conservation of mass, the upflow has to drain off sideways
through the edge of the granule within approximately one pressure
scale height $H_{P}$, hence resulting in a horizontal mass flux $j_{h}=[2\pi rH_{P}]\rho v_{h}$.
The pressure is a quantity that preserves its characteristic shape
with stellar parameters, i.e. the pressure of two different simulations
look rather similar on a uniform depth scale, therefore the pressure
scale height is preferred over the density scale height. Equating
$j_{z}$ and $j_{h}$ we can solve for the (maximal) granular diameter,
$d=2r$: 
\begin{eqnarray}
d_{\mathrm{NSR}} & = & 4\left[v_{h}/v_{z}\right]H_{P}\approx4H_{P}\label{eq:dNIC}
\end{eqnarray}
We show in Fig. \ref{fig:granule-size-2} a comparison of the granular
diameters estimated with $d_{\mathrm{NIC}}$ and from the maximum
of 2D spatial power spectra $d_{\mathrm{gran}}$, which is shown in
Fig. \ref{fig:granule-size-1}. The astonishing tight correlation
can solely be interpreted as clear evidence for the validity of the
NSR. We find that the mean pressure scale height taken at the height
of the maximum vertical rms-velocity below the optical surface ($\log\tau_{\mathrm{Ross}}\sim2.0$,
see Fig. \ref{fig:rms_vertical_velocity}) gives the best match between
$d_{\mathrm{NIC}}$ and the granule sizes. Furthermore, we also confirm
that the relevant scale-height is that of the \textit{total} pressure
scale height, $H_{P}=p_{\mathrm{tot}}/\rho g$, since we find a better
agreement with the latter. The granular diameters found from the peak
of 2D spatial power spectra are about $30\,\%$ larger than the estimate
from Eq. \ref{eq:dNIC}, i.e., $d_{\mathrm{gran}}\sim1.3d_{\mathrm{NIC}}$
(see lower panel of Fig. \ref{fig:granule-size-2}). The variation
of the velocity ratio $v_{h}/v_{z}$ in the convection zone is rather
small ($v_{h}/v_{z}$$\sim1.0$) as both are of the order of the sound
speed, therefore the variation in Eq. \ref{eq:dNIC} stems predominantly
from $H_{P}$. In hydrostatic equilibrium the pressure scale height
is inversely proportional to the surface gravity ($H_{P}\propto1/g$),
which explains the strong correlation between the granular sizes and
$\log g$. On the other hand, with increasing $T_{\mathrm{eff}}$
and $\left[\mathrm{Fe}/\mathrm{H}\right]$, the pressure scale height
increases slightly because of the increase in the ratio of pressure
and density ($H_{P}\propto p_{\mathrm{tot}}/\rho$). The ratio actually
increases even though both values decrease, since the density drops
with height slightly more rapidly than the pressure.\\

Finally, we want to mention our finding on the filling factor for
upflows and downflows, $f_{\mathrm{up}}$ and $f_{\mathrm{dn}}$ respectively.
We derived the filling factor from the sign of the velocity field
in the unaltered simulations on layers of constant geometrical height.
Then we computed the mean filling factor in the convection zone, which
yields on average for all simulations $f_{\mathrm{up}}\simeq0.65$
with a minute deviation of $\sigma=0.014$. Therefore, we find that
the mean filling factor is rather universal, and close to previous
findings by SN98 with $f_{\mathrm{up}}\sim2/3$ and $f_{\mathrm{dn}}\sim1/3$.
In deeper solar simulations, which reach down to $20\,\mathrm{Mm}$
\citep{Stein:2011p6057}, we find very similar values for the filling
factor.

\subsection{The mean atmosphere\label{sub:The-mean-atmosphere}}

In the following, we want to discuss the properties of the mean stratifications
and the temporal and spatial averages of various important quantities.
Unless specified otherwise, the $\left\langle 3\mathrm{D}\right\rangle $
stratifications presented here are averages on surfaces of constant
Rosseland optical depth, i.e. $\left\langle 3\mathrm{D}\right\rangle =\left\langle 3\mathrm{D}\right\rangle _{\mathrm{Ross}}$.
Whenever we employ alternative averages in the text, e.g., on constant
geometrical height $\left\langle 3\mathrm{D}\right\rangle _{z}$,
we indicate that explicitly. We remark briefly that only the averages
on constant geometrical height $\left\langle 3\mathrm{D}\right\rangle _{z}$
strictly fulfill the equations of conservation (Eqs. \ref{eq:mass},
\ref{eq:momentum} and \ref{eq:energy}), therefore also the hydrostatic
equilibrium, while all other averages exhibit slight deviations. This
and the actual methods we used to compute the mean stratifications
are discussed in a separate paper (see Magic et al.\textbf{ }in preparation).
For the sake of clarity, we display here only a subsample of our grid
models including MS and RGB stars ($\log g=4.5$ and $2.0$, which
we refer to as dwarfs and giants, respectively) with solar and sub-solar
metallicity ($\left[\mathrm{Fe}/\mathrm{H}\right]=0.0$ and $-2.0$,
solar and metal-poor, respectively). Whenever possible, we compare
with corresponding 1D models that are obtained with our 1D code (see
App. \ref{sec:Stagger-grid-1D-atmospheres}).\\

Before continuing our discussion, we would like to point out the importance
of the superadiabatic region (SAR), as it will be referred repeatedly
in the following. It is the region, where the transport of energy
changes character, from convective to radiative. The top of the SAR,
where $\nsad=0$, marks the top the convection zone, since it is the
uppermost point, where the Schwarzschild criterion is fulfilled. At
the location of the peak of the superadiabatic gradient, one also
finds the largest fluctuations and inhomogeneities in the thermodynamic
variables due to the non-adiabatic transition to the photosphere.
Furthermore, it is here in the SAR, where the entropy jump and the
peak in the vertical velocity occur. In fact, the SAR effectively
represents the physical outer boundary of the convective envelope.
It is the most dynamic part in the interior of late-type stars, where
the largest fluctuations are found. This is the reason why hydrostatic
1D modeling has the greatest challenges in this rather small region.

\subsubsection{Temperature stratification\label{sub:temperature-stratification}}

\begin{figure}
\includegraphics[width=88mm]{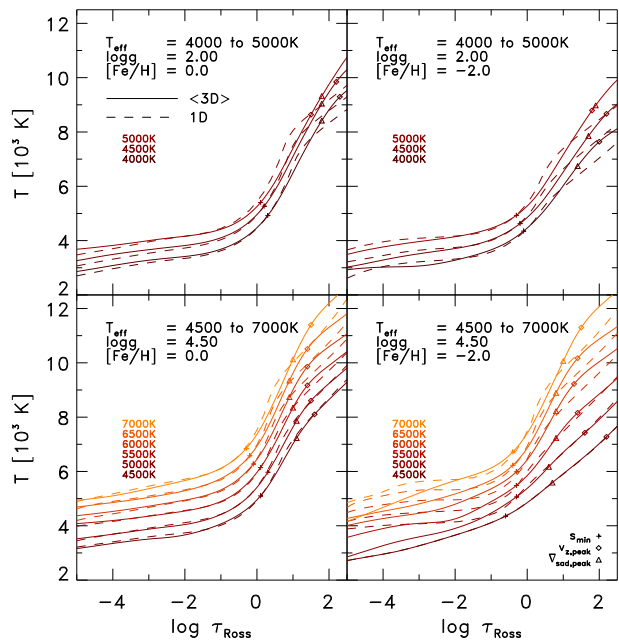}

\caption{$\left\langle 3\mathrm{D}\right\rangle $ stratifications of the temperature
vs. optical depth for various stellar parameters (solid lines) and
1D models with $\alpha_{\mathrm{MLT}}=1.5$ (dashed lines). Furthermore,
we have marked the positions of entropy minimum (plus), vertical peak
velocity (diamond), and maximum in $\nabla_{\mathrm{sad}}=\nabla-\nabla_{\mathrm{ad}}$
(triangle).}
\label{fig:mean_temperature-1}
\end{figure}
We first consider the temperature stratifications in optical depth,
which we show in Fig. \ref{fig:mean_temperature-1}. We also show
the corresponding stratifications of 1D theoretical model atmosphere
with $\alpha_{\mathrm{MLT}}=1.5$ based on our 1D code (dotted lines)
with identical EOS and opacity tables as the 3D models. In the continuum
forming layers around the optical surface ($-1.0<\log\tau_{\mathrm{Ross}}<0.5$),
the differences between $\left\langle 3\mathrm{D}\right\rangle $
models with different $\teff$s, but same $\logg$ and $\feh$, are
rather small besides the shift in the temperature stratification corresponding
to the difference in effective temperature $\Delta T_{\mathrm{eff}}$,
which is to be expected since $T_{\mathrm{eff}}\approx T\left(\tau=2/3\right)$.
Well above and below the optical surface, on the other hand, we find
significant differences between the $\hav$ models depending on the
stellar parameters.\\

In the upper layers ($\log\tau_{\mathrm{Ross}}<-2.0$) of atmospheres
with solar metallicity, we find that the behavior in mean temperature
is similar between 3D and 1D models. On the other hand, the metal-poor
$\hav$ models exhibit significantly cooler temperature stratifications
compared to $\hav$ models with solar metallicity ($\Delta T/T\left(\log\tau_{\mathrm{Ross}}=-0.5\right)\sim-1\%$
and $\sim-14\%$ for $\left[\mathrm{Fe}/\mathrm{H}\right]=-1.0$ and
$-3.0$ respectively), in particular for dwarfs ($\log g=4.5$). The
temperature stratification in the upper photospheres of solar-metallicity
models is largely controlled by radiative equilibrium, while for low-metallicity
models this is not generally the case: for metal-poor models, the
absorption features become considerably weaker, therefore, the radiative
heating by spectral line re-absorption ($q_{\mathrm{rad}}$) is dominated
by the adiabatic cooling due to expansion of the ascending gas ($-p_{\mathrm{th}}\vec{\nabla}\cdot\vec{v}$)
in the energy balance (Eq. \ref{eq:energy}), leading to an equilibrium
structure at cooler temperatures \citep{Asplund:1999p11771}. For
cool, metal-poor giants (e.g., $T_{\mathrm{eff}}=4000\,\mathrm{K}$,
$\log g=2.0$), we recognize the effects of molecule formation on
the structure of the high atmosphere. At sufficiently low temperatures,
molecules start to form, which contribute with a large line opacity,
shifting the balance from adiabatic to radiative heating and cooling,
resulting in a stratification closer to the radiative equilibrium
one \citep[see][]{Gustafsson:2008p3814}. On the other hand, for giants
with solar metallicity the radiative equilibrium is even more dominating,
since these exhibit hotter stratifications than 1D models in the upper
layers. \citet{Ludwig:2012A&A...547A.118L} find the same but on a
much milder level. These effects are rather non-linear, and we find
no simple systematic trends within our grid models.\\

\begin{figure}
\includegraphics[width=88mm]{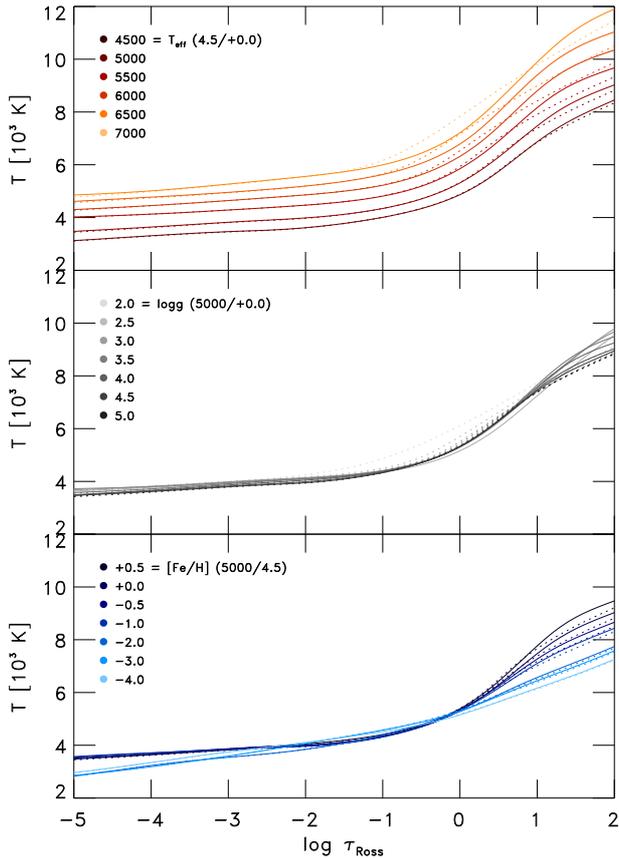}

\caption{$\left\langle 3\mathrm{D}\right\rangle $ temperature stratifications
with the variation of one stellar parameter at a time, while the two
others are fixed ($T_{\mathrm{eff}}$, $\log g$, and $\left[\mathrm{Fe}/\mathrm{H}\right]$,
from top to bottom, respectively). We show our standard averages on
constant optical depth $\left\langle 3\mathrm{D}\right\rangle $ (solid
line) and on constant geometrical depth $\left\langle 3\mathrm{D}\right\rangle _{z}$
(dotted line).}
\label{fig:mean_temperature-2}
\end{figure}
We would now like to examine the influence of individual stellar parameters
on the temperature stratification. Therefore, we show in Fig. \ref{fig:mean_temperature-2}
the temperature stratifications of models where we separately vary
one at the time ($T_{\mathrm{eff}}$, $\log g$, and $\left[\mathrm{Fe}/\mathrm{H}\right]$),
while keeping the other two parameters constant. Figure \ref{fig:mean_temperature-2}
(top panel) shows, as expected, that with increasing $T_{\mathrm{eff}}$,
the temperature stratification becomes overall hotter above, but also
below the optical surface, in order to provide the required total
energy flux (higher enthalpy, Eq. \ref{eq:fconv_approx}). We find
in our simulations (both 1D and 3D) that the increased $T_{\mathrm{eff}}$s
with hotter stratifications are accompanied by lower densities and
higher vertical velocities below the surface (see $p_{\mathrm{tot}}\left(\log\tau_{\mathrm{Ross}}=2.0\right)$
in Fig. \ref{fig:mean_total-pressure} representatively for the $\rho$).
The net effect on the convective flows are lower mass fluxes for higher
$T_{\mathrm{eff}}$s, since the decrease in density is predominating
the increase in velocity, therefore resulting in a more inefficient
convection. This is compensated with higher entropy jumps (see Fig.
\ref{fig:entropy_jump} with $\dss$ as an inverse measure for convective
efficiency), hence higher temperatures and steeper temperature gradients.
On the other hand, the temperatures in the upper, radiative layers
increase less with increasing $T_{\mathrm{eff}}$ than in the deeper,
convective ones. We find with decreasing surface gravity (middle panel
in Fig. \ref{fig:mean_temperature-2}) the same correlations as with
increasing $\teff$s before, the temperature stratifications become
hotter below the photosphere, and due to lower densities we find a
more inefficient convection, while the upper atmosphere is less affected.
For lower metallicities (bottom panel), the temperature stratifications
are significantly cooler, both above and below the optical surface
($\Delta T/T\left(\log\tau_{\mathrm{Ross}}=2.0\right)\sim-5\%$ and
$-15\%$ for $\left[\mathrm{Fe}/\mathrm{H}\right]=-1.0$ and $-3.0$
respectively). At the top the stratifications are cooler at lower
$\feh$ due to the dominance of adiabatic cooling over radiative heating.
Below the optical surface, we find higher densities with lower velocities
and entropy jumps (while the mass flux is increasing), therefore,
leading to an efficient convection with shallow temperature gradients
at lower metallicities. We find cooler models that fall below the
opacity edge, which we describe below (compare $\teff=5000\,\mathrm{K}$
in Fig. \ref{fig:ee-rho-plane-1}), follow an adiabatic temperature
stratification even in the atmosphere, which coincides with the rather
sudden change between $\left[\mathrm{Fe}/\mathrm{H}\right]=-1.0$
and $-2.0$ in Fig. \ref{fig:mean_temperature-2} (bottom panel).
Besides our standard averages on constant Rosseland optical depth,
we show also the averages on constant geometrical depth scale $\left\langle 3\mathrm{D}\right\rangle _{z}$
(here is $z$ fixed and $\taur=\left\langle \taur\right\rangle _{z}$),
which are systematically different, in particular below the optical
surface, but behave qualitatively in a similar way with stellar parameters.\\

\begin{figure}
\includegraphics[width=88mm]{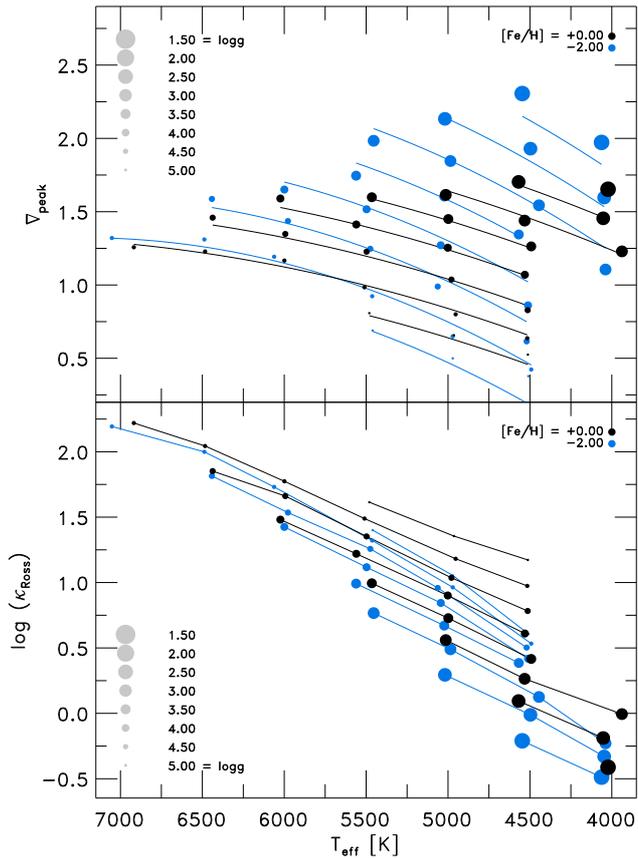}

\caption{Overview of the maximum temperature gradient $\nabla_{\mathrm{peak}}$
(top panel) and Rosseland opacity $\kappa_{\mathrm{Ross}}$ taken
at the height $\taur\approx3.0$ (bottom panel) against $\teff$ for
$\feh=-2.0$ and $0.0$ (blue and black respectively). Models with
the same surface gravity are connected by their respective functional
fits in the top panel (solid lines; see App. \ref{sub:Functional-fits}).}
\label{fig:nabla_peak}\label{fig:kapr}
\end{figure}
In the sub-photospheric region ($\log\tau_{\mathrm{Ross}}>0.5$),
where convection dominates, the temperature gradients $\nabla=d\ln T/d\ln p_{\mathrm{tot}}$%
\footnote{$\nabla$ increases, if only the thermodynamic pressure is included,
neglecting the turbulent component.%
} become increasingly steeper with higher $T_{\mathrm{eff}}$, reflecting
the hotter interior stratifications. This can be illustrated with
the maximum in temperature gradient, $\nabla_{\mathrm{peak}}=\max\nabla$,
which we show in Fig. \ref{fig:nabla_peak} (functional fits are also
given in App. \ref{sub:Functional-fits}). The increase of $\nabla_{\mathrm{peak}}$
with $T_{\mathrm{eff}}$ is close to linear, but it seems to saturate
at higher $T_{\mathrm{eff}}$ (see $T_{\mathrm{eff}}\ge6500\,\mathrm{K}$).
We find that the maximum in temperature gradient $\nabla_{\mathrm{peak}}$
reproduces qualitatively a similar behavior as the intensity contrast
$\Delta I_{\mathrm{rms}}$ with stellar parameter (compare Figs. \ref{fig:intensity-contrast}
and \ref{fig:nabla_peak}), which is consistent with the strong temperature
sensitivity of the $\mathrm{H}^{-}$ opacity. Furthermore, our metal-poor
simulations exhibit a larger range of $\nabla_{\mathrm{peak}}$-values
than their solar metallicity counterparts, and $\nabla_{\mathrm{peak}}$
is similarly \textit{enhanced} at lower metallicity (compare also
$T\left(\log\tau_{\mathrm{Ross}}=2.0\right)$ for dwarfs in Fig. \ref{fig:mean_temperature-1}),
as the intensity contrast (see Sect. \ref{sub:The-emergent-intensity}).
Our cool metal-poor simulations have flatter and hot metal-poor simulations
have steeper temperature stratifications than the metal-rich part
of our grid (see Fig. \ref{fig:mean_temperature-2}). Curiously, $\nabla_{\mathrm{peak}}$
is close to constant with metallicity for the solar $T_{\mathrm{eff}}$
and $\logg$.\\

We identify three main reasons for the given variations in temperature
gradients with stellar parameters in the SAR, which are rooted in
the hydrodynamics and the radiative transfer: velocity field, convective
efficiency, and radiative back-warming.
\begin{enumerate}
\item As we discussed above (see Sect. \ref{sub:The-entropy-jump}), the
entropy jump $\Delta s$ increases with effective temperature according
to a power law (see Fig. \ref{fig:entropy_jump}). This behavior arises
due to the variations in the radiative losses (see Sect. \ref{sub:transport-of-energy}),
which is accompanied by changes in internal energy and density (see
Fig. \ref{fig:ee-rho-plane-1}). The velocities rise rapidly, as exhibited
by the growth of $v_{z,\mathrm{rms}}^{\mathrm{peak}}$ and also $p_{\mathrm{turb}}^{\mathrm{peak}}$
with $T_{\mathrm{eff}}$ (see Figs. \ref{fig:rms_vertical_velocity}
and \ref{fig:mean_turbulent-pressure-fraction} respectively). Similar
to $\nabla_{\mathrm{peak}}$, both $v_{z,\mathrm{rms}}^{\mathrm{peak}}$
and $p_{\mathrm{turb}}^{\mathrm{peak}}$ occur in the SAR, and both
increase towards higher $T_{\mathrm{eff}}$ and lower $\log g$. 
\item The \emph{mass mixing length} $\alpha_{m}$ changes with stellar parameters
\citep{Trampedach:2011p5920}. The latter is evaluated as the inverse
gradient of the vertical mass flux, separately in the up- or downflows,
hence $\alpha_{m}^{-1}=d\ln j_{z,\mathrm{up}}/d\ln p$, with $j_{z,\mathrm{up}}$
being the vertical mass flux in the upflows. Therefore, the mass mixing
length is composed of the gradients of the density and the vertical
velocity, i.e. $\alpha_{m}\propto1/d\ln\rho+1/d\ln v_{z}$. We find
that $\alpha_{m}$ increases for lower $\Delta s$, $v_{z,\mathrm{rms}}^{\mathrm{peak}}$
and $\nabla_{\mathrm{peak}}$. We will publish our findings on the
mass mixing length in a separate paper.
\item In the lower photospheric layers, where the continuum forms, radiation
is absorbed (\emph{blocked}) by spectral lines; this implies that
less radiative flux can be transported at the wavelengths corresponding
to spectral lines and, conversely, that more flux has to be pushed
through continuum windows, an effect commonly referred to as \textit{line-blanketing}.
This in turn leads to a steepening of the temperature gradient and
to additional heating of the sub-surface layers, also known as \textit{back-warming}
\citep[see][]{Mihalas:1970p21310,Nordlund:2009p4109}. In 1D models
it is straightforward to quantify the effect of back-warming, as done
for example by \citet{Gustafsson:2008p3814}, who found it to contribute
a slight increase in temperatures below the surface ($\Delta T/T\left(\tau_{\mathrm{Ross}}=10\right)\simeq5\%$
for solar metallicity stars with $\teff\approx5000\,\mathrm{K}$ and
$\logg=3.0$). In our 3D RHD atmosphere models, line-blanketing and
back-warming effects are also naturally included through our opacity-binning
method. Isolating the radiative back-warming effect in our 3D simulations
is, however, a little more involved than in 1D and we defer the analysis
of this mechanism to a future paper in this series.
\end{enumerate}
The three mentioned effects are nonlinearly coupled and compete with
each other, making it difficult to disentangle the individual contributions.
A quantitative analysis will be presented in a later paper.\\

\begin{figure*}
\includegraphics[width=176mm]{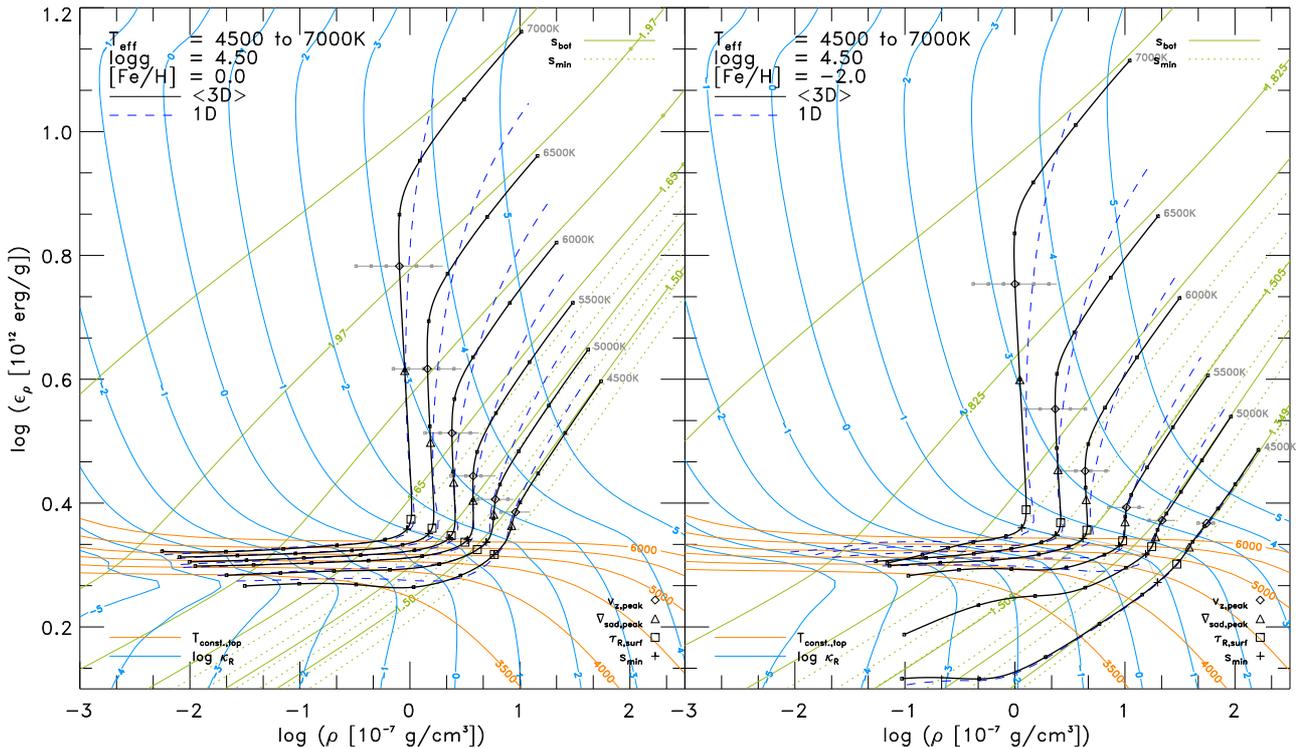}

\caption{We show the mean internal energy against mean density for dwarf models
($\log g=4.5$) with $\left[\mathrm{Fe}/\mathrm{H}\right]=0.0$ and
$-2.0$ (left and right respectively). The specific isocontours for
the entropy $\ssbot$ (green solid) and $s_{\mathrm{min}}$ (green
dotted), Rosseland opacity per volume, $\rho\kappa_{\mathrm{Ross}}$,
(blue) and temperature $T$ (orange) are underlayed. Moreover, the
positions of entropy minimum $\ssmin$ (plus), optical surface (large
square), vertical peak velocity $\vzrmsp$ (diamonds) and maximum
in $\nabla_{\mathrm{sad}}$ (triangle) are marked respectively. The
amplitude of $v_{z,\mathrm{rms}}^{\mathrm{peak}}$ is indicated with
horizontal bars with markings in 1 km/s. We also included the 1D models
with $\alpha_{\mathrm{MLT}}=1.5$ (blue dashed lines). The range in
optical depth is shown from $\ltaur=-5.0$ to $+5.0$ for each dex
(small squares). However, we note that our simulations boxes are much
deeper ($\left\langle \ltaur\right\rangle \approx+7.5$).}
\label{fig:ee-rho-plane-1}
\end{figure*}
We would like now to examine more closely a sample of $\left\langle 3\mathrm{D}\right\rangle $
models in the $\varepsilon-\rho$-plane, as shown in Fig. \ref{fig:ee-rho-plane-1},
in order to better illustrate the variations with stellar parameters.
One can clearly distinguish three different regimes: the adiabatic
convection zone, the photospheric transition, and the almost isothermal
upper atmosphere.

At the bottom boundary, sufficiently deep in the convection zone where
entropy fluctuations become small, $\left\langle \delta s\right\rangle =0.3\,\%$,
the models follow closely the associated adiabats with $s=s_{\mathrm{bot}}$
(green lines). They deviate increasingly from their adiabats, as the
top of the convection zone is approached. This is due to the entropy
deficient downdrafts (cooled in the photosphere) becoming less diluted
by the entropic upflows, as the optical surface is approached. For
the 1D models (blue dashed lines), the value of the entropy at the
bottom of the stratifications is evidently overestimated, particularly
at higher $\teff$, but this is because we haven't calibrated the
$\alpha_{\mathrm{MLT}}$ parameter here, and we have used a value
of $1.5$ for all models. The transition of energy transport from
fully convective to fully radiative is clearly visible, since, at
the optical surface, one finds a sharp isochoric ($\Delta\rho\sim0.0$)
drop in internal energy (this is basically the enthalpy-jump $\Delta h$
in Eq. \ref{eq:fconv_approx}). The $\varepsilon$-jump coincides
with the sub-photospheric region ($0.0<\log\tau_{\mathrm{Ross}}<2.0$),
where the atmosphere starts to become transparent. The transition
zone ends eventually at the optical surface ($\log\tau_{\mathrm{Ross}}\simeq0.0$,
marked with big squares). Above the optical surface ($\log\tau_{\mathrm{Ross}}<0.0$)
the atmosphere is almost isothermal (compare with the orange isotherms
in Fig. \ref{fig:ee-rho-plane-1}), with exponentially decreasing
density and almost constant internal energy ($\Delta\varepsilon\sim0.0$). 

The entropy $s_{\mathrm{bot}}$ at the bottom grows exponentially
with increasing $T_{\mathrm{eff}}$ and decreasing $\log g$. We showed
above that the entropy jump increases in a similar way (see Fig. \ref{fig:ssbot_vs_ssj}).
Here we find a similar behavior for the jump in internal energy ($\Delta\varepsilon$,
hence $\Delta h$) in the photosphere. Moreover, we show in Fig. \ref{fig:ee-rho-plane-1}
the positions of $\nabla_{\mathrm{sad}}^{\mathrm{peak}}$ and $v_{z,\mathrm{rms}}^{\mathrm{peak}}$
located in the $\varepsilon$-jump, and, again, we find both $v_{z,\mathrm{rms}}^{\mathrm{peak}}$
and $\nabla_{\mathrm{sad}}^{\mathrm{peak}}$ to scale exponentially
with $T_{\mathrm{eff}}$. For $v_{z,\mathrm{rms}}^{\mathrm{peak}}$
we have indicated the amplitudes as well, which also increases exponentially
with $T_{\mathrm{eff}}$. All of the aforementioned positions are
distributed rather regularly in the $\varepsilon-\rho$-plane, while
they are less so on the $\log\tau_{\mathrm{Ross}}$-scale (see Fig.
\ref{fig:mean_temperature-1}). The position of $s_{\mathrm{min}}$
is close to the optical surface and shows little variation in optical
depth.

At lower energies and densities in Fig. \ref{fig:ee-rho-plane-1}
($\log\varepsilon\sim0.3$ and $\log\rho\sim3.0$ to $0.0$) we notice
the effect of \ion{H}{i} and \ion{He}{i} opacity in form of an edge
in the opacity contours ($\log\kappa_{\mathrm{Ross}}\sim-5.0$ to
$-1.0$), since the bound-free absorption increases (more excited
states) towards higher energy below the ionization energy, and they
fade away again above it. Models that fall below this edge exhibit
a rather different stratification. In particular, towards cool metal-poor
dwarfs, i.e. lower $\teff$, higher $\logg$, and lower $\feh$, the
models more closely follow adiabats than isotherms, in the atmosphere.
This effect of the competition between radiative and dynamic heating
(see beginning of this Sect.) above the convection zone becomes particularly
evident at lower metallicity (for $\left[\mathrm{Fe}/\mathrm{H}\right]\leq-2.0$).
However, for the 1D models (blue dashed lines), this is obviously
not the case, since these always follow isotherms due to the enforcement
of radiative equilibrium. Furthermore, the cool metal-poor $\left\langle 3\mathrm{D}\right\rangle $
models also display higher densities at the optical surface, thereby
spanning a larger $\rho$-range for different $T_{\mathrm{eff}}$s.
The stratifications of simulations of hotter dwarfs, on the other
hand, depend little on metallicity. For the simulations, we have only
plotted the range $\ltaur\in\left[-5.0,5.0\right]$, and the top of
this is reached at much higher densities for the metal-poor dwarfs
than for the solar metallicity dwarfs. Therefore, the density ranges
covered above the optical surface by the individual atmospheres is
small for metal-poor models ($\min\left[\log\rho\right]\sim-1.0$
and $-2.0$, respectively; see Fig. \ref{fig:ee-rho-plane-1}).

\subsubsection{Velocity field\label{sub:velocity-field}}

\begin{figure}
\includegraphics[width=88mm]{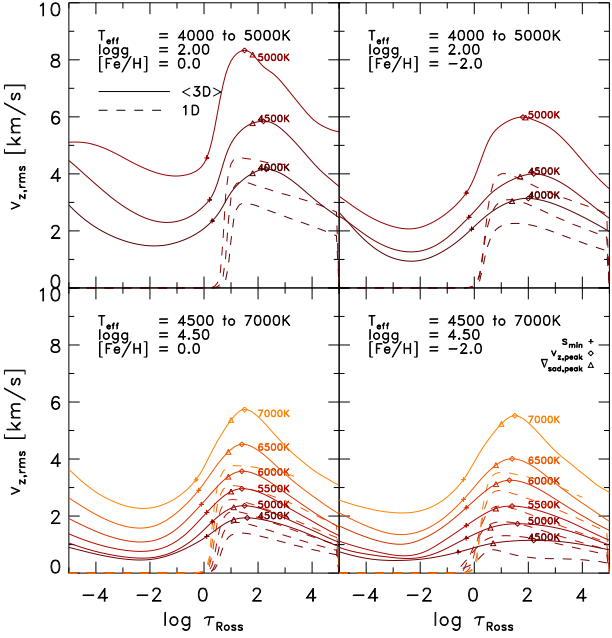}

\caption{Vertical rms-velocity, $v_{z,\mathrm{rms}}$, from the $\left\langle 3\mathrm{D}\right\rangle $
stratifications (solid lines) and convective velocity $v_{\mathrm{MLT}}$
from the corresponding 1D MLT models (dotted lines) as function of
optical depth $\ltaur$ for various stellar parameters.}
\label{fig:rms_vertical_velocity}
\end{figure}
Next, we consider the velocity field in our simulations, which arise
self-consistently from the solution of the hydrodynamic equations.
In Fig. \ref{fig:rms_vertical_velocity} we show the rms-velocity
of the vertical component $v_{z,\mathrm{rms}}$, being the flux carrying
component of the convective flows, and being the broadening component
of spectral lines at disk center. 

\begin{figure}
\includegraphics[width=88mm]{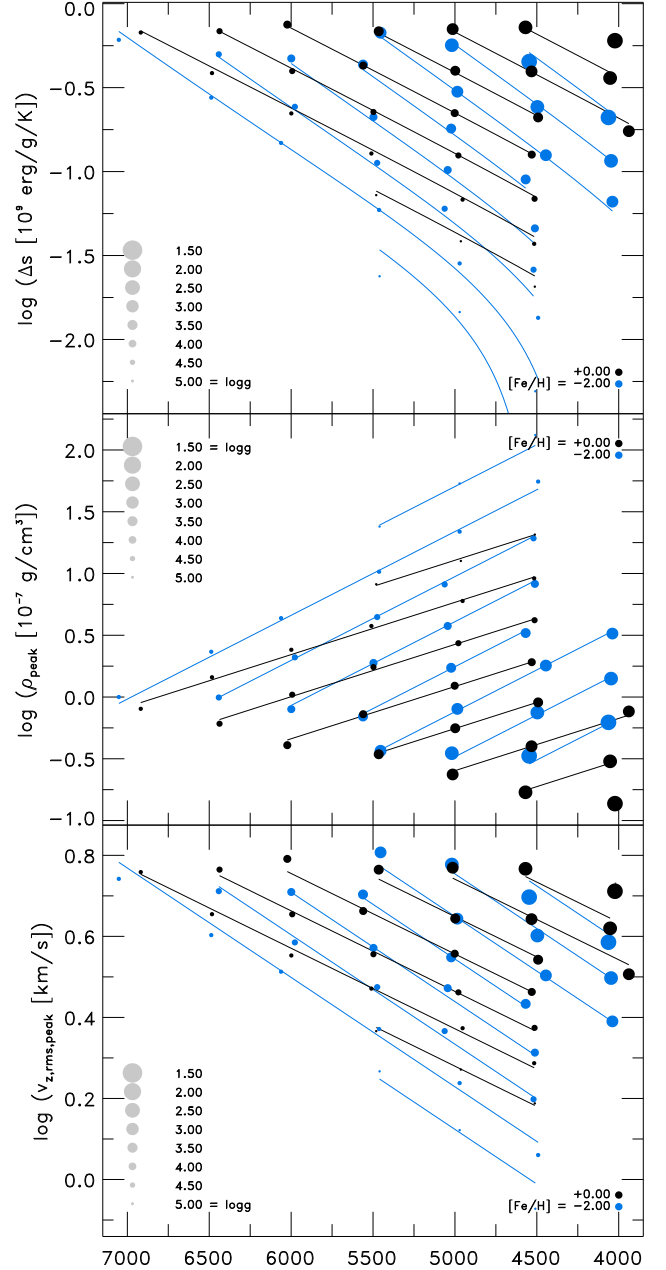}

\caption{Overview of the entropy jump (top panel), maximal vertical rms-velocity
below the surface $v_{z,\mathrm{rms}}^{\mathrm{peak}}$ (middle panel)
and the density at the same height $\rho_{\mathrm{peak}}$ (bottom
panel) vs. $\teff$ for $\feh=-2.0$ and $0.0$ (blue and black respectively).
Models with the same gravity are connected with their respective functional
fits (solid lines; see App. \ref{sub:Functional-fits}). Note the
inverse correlation between density and velocity.}
\label{fig:entropy_jump}\label{fig:velocity_peak}\label{fig:density_peak}
\end{figure}

The buoyant uprising plasma will experience increasingly a decrease
in density towards the photosphere, hence a strong density gradient,
and due to mass conservation, the convective motions will eventually
overturn. Therefore, $v_{z,\mathrm{rms}}$ peaks in the SAR around
$\log\tau_{\mathrm{Ross}}\sim1.5$ for dwarfs and $\sim2.3$ for giants.
Furthermore, since in the SAR, the transition region from convective
to radiative transport of energy takes place due to decrease in opacity
and the subsequent radiative losses, here we find the strongest turbulent
motions concomitant with the greatest fluctuations in all thermodynamical
quantities (in particular entropy, temperature and density, see Figs.
\ref{fig:mean_entropy}, \ref{fig:mean_temperature-1} and \ref{fig:density_peak}).
Further towards the interior, $v_{z,\mathrm{rms}}$ drops as the density
increases. From the slightly sub-photospheric maximum, velocities
fall off to a minimum above the optical surface, then increases again
in the higher atmosphere (see Fig. \ref{fig:rms_vertical_velocity}).
Towards upper layers, $v_{z,\mathrm{rms}}$ increases again due to
p-modes, excited in the SAR but leaking out of the acoustic cavity
as they have frequencies above the acoustic cut-off. The metal poor
simulations show a slightly smaller increase in $v_{z,\mathrm{rms}}$,
since their density gradients are shallower due to steeper $T$-gradients.
The declining velocity above the surface is due to the fact that the
convective motions overshoot well above the top of the convection
zone. We find the velocity minimum to occur between $\log\tau_{\mathrm{Ross}}\sim-2.3$
for dwarfs and $-1.5$ for giants.

As expected, the magnitude of the velocity field is enhanced towards
higher $T_{\mathrm{eff}}$, lower $\log g$, and higher $\left[\mathrm{Fe}/\mathrm{H}\right]$,
similar as $\dss$. The symmetry of the velocity profile changes with
$\log g$ and metallicity, while it is little affected by $T_{\mathrm{eff}}$.
For lower $\log g$, the peak in the velocity field is increasingly
shifted to optically deeper layers (e.g. at solar metallicity the
average peak position for dwarfs is $\left\langle \log\tau_{\mathrm{Ross}}\right\rangle \sim1.5$,
while for giants it is $\sim2.0$). The coolest metal-poor simulations
display a flatter profile, and the position of the minimum is increasingly
shifted towards higher layers, especially for extreme metal-poor dwarfs
($\left[\mathrm{Fe}/\mathrm{H}\right]<-3.0$), and the profile is
therefore stretched and skewed.

For comparison, we also show in Fig. \ref{fig:rms_vertical_velocity}
the convective velocity $v_{\mathrm{MLT}}$ of our 1D models determined
by MLT. It is apparent that the general trends of increasing velocities
with increasing $\teff$ and $\feh$ and decreasing $\logg$, are
common between the simulations and the 1D MLT models, although much
less pronounced in 1D. Furthermore, $v_{\mathrm{MLT}}$ drops rather
sharply at the top of the convection zone (as given by the Schwarzschild
criterion), as no overshooting is allowed for in our implementation
of MLT. Several non-local variants of MLT exists, and they allow for
overshooting, but none of them produce velocity profiles close to
that of our simulations. We also want to mention the large asymmetry
in velocities of the up- and downflows (SN98): in 3D simulations,
the latter are much faster than the former \textbf{(}up to \textbf{$\sim2$}
faster, in particular for cool dwarfs\textbf{)}, contrary to what
is normally assumed in 1D descriptions of convection.\\

The peak vertical rms-velocity, $v_{z,\mathrm{rms}}^{\mathrm{peak}}$
(see bottom panel of Fig. \ref{fig:velocity_peak}), is a good measure
of the global magnitude of velocities in the simulations. It also
serves as a measure of the amount of turbulence present in the simulations.
The actual values are also given in Col. 9 in Table \ref{tab:global_properties},
together with a functional fit in App. \ref{sub:Functional-fits}.
The variation of $v_{z,\mathrm{rms}}^{\mathrm{peak}}$ with stellar
parameters resembles that of the entropy jump $\Delta s$ (compare
top and bottom panel in Fig. \ref{fig:entropy_jump}), namely it also
increases exponentially with higher $T_{\mathrm{eff}}$ and lower
$\log g$ and linearly with $\left[\mathrm{Fe}/\mathrm{H}\right]$.
An interesting aspect is the increase of $v_{z,\mathrm{rms}}^{\mathrm{peak}}$
and $\Delta s$ with $T_{\mathrm{eff}}$, which are close to exponential,
indicating a correlation between the two. The characteristic variations
of $v_{z,\mathrm{rms}}^{\mathrm{peak}}$ correspond to the inverse
variations of the density taken at the same heights as the peaks in
$v_{z,\mathrm{rms}}$ (see middle panel in Fig. \ref{fig:density_peak}).
This behavior arises due to conservation of mass (Eq. \ref{eq:mass}),
which can be expressed as
\begin{equation}
\left\langle \partial_{z}\ln\rho\right\rangle =-\left\langle \partial_{z}\ln v_{z}\right\rangle \label{eq:dens_velo_relation}
\end{equation}
for a stationary flow ($\partial_{t}\rho=0$). Of course, under this
assumption, this equation is strictly speaking valid only locally,
while we compare here averaged values. Despite that, we find that
this relation is, in general, qualitatively fulfilled across the whole
depth of our simulations. Towards the optical surface, the density
decreases, which has to be compensated by faster velocities, in order
to fulfill conservation of mass as well as sustain the energy flux.
The velocity field profile results ultimately from the interplay between
the vertical and horizontal acceleration due to buoyancy and overturning
respectively. The latter in turn is set by the radiative losses that
arises from the prevailing opacity conditions according to the thermodynamic
state of the plasma (see Sect. \ref{sub:transport-of-energy}). Furthermore,
one can also reason that at a higher effective temperature, hence
hotter temperature stratification, the density will be lower (ideal
gas gives $T\propto1/\rho$; see also middle panel in Fig. \ref{fig:density_peak}),
however, at the same time, more energy (enthalpy) has to be carried
to the surface, which necessitates a faster flow (as is given in Eq.
\ref{eq:ftot_ssj_rho_uy}). The entropy jump, density, and velocity
are coupled intimately with each other (the vertical mass flux is
$j_{z}=\rho v_{z}$). Therefore, changes in one quantity imply corresponding
variations in the values of the other quantities, and vice versa.
The radiative energy losses at the photospheric transition generate
the entropy fluctuations according to the prevailing opacity and the
irradiation-duration, hence it sets the amplitude of the entropy jump
$\Delta s$. On the other hand, the entropy deficient plasma with
its density excess determines the buoyancy force, $f_{B}\sim\Delta\rho$,
and therefore the vertical velocities $v_{z}$ of the downdrafts.
The downdrafts in turn will settle the upflows in order to deliver
the required convective energy flux. The subtle details in the chain
of causalities are non-trivial and beyond the scope of the present
paper.\\

\begin{figure}
\includegraphics[width=88mm]{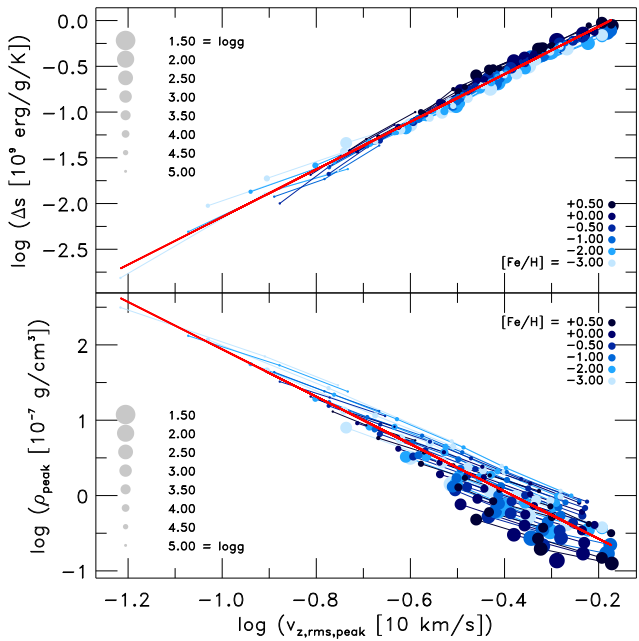}

\caption{Comparison of the entropy jump $\Delta s$ (top panel) and the density
at the optical depth of the maximal vertical velocity below the optical
surface $\rho_{\mathrm{peak}}$ (bottom panel) vs. maximal vertical
rms-velocity $v_{z,\mathrm{rms}}^{\mathrm{peak}}$ for all models.
A linear fit is also indicated in both panels (red lines).}
\label{fig:velocity_peak-vs-entropy-jump}
\end{figure}
Similar to our finding in Sect. \ref{sub:The-entropy-jump} of a scaling
relation between the entropy jump and the constant entropy value of
the adiabatic convection zone $\Delta s\left(s_{\mathrm{bot}}\right)$,
we find here again another interesting, tight scaling relations between
$\rho_{\mathrm{peak}}$, $v_{z,\mathrm{rms}}^{\mathrm{peak}}$ and
$\Delta s$, which we show in Fig. \ref{fig:velocity_peak-vs-entropy-jump}.
The values are plotted on a double logarithmic scale, to more clearly
illustrate the power-law character of the relations. From the above
discussion, it follows that the vertical velocity is also correlated
with the constant entropy value $s_{\mathrm{bot}}$ of the adiabatic
convection zone and the density. We also show linear fits of the density
$\rho_{\mathrm{peak}}$ and entropy jump $\dss$ as a function of
$v_{z,\mathrm{rms}}^{\mathrm{peak}}$ in log-log scale (red lines
in Fig. \ref{fig:velocity_peak-vs-entropy-jump}), exhibiting the
slopes of $\log\Delta s/\log v_{z,\mathrm{rms}}^{\mathrm{peak}}\sim0.46$
and $\log\rho_{\mathrm{peak}}/\log v_{z,\mathrm{rms}}^{\mathrm{peak}}\sim-1.20$,
hence a scaling with the respective slopes.\\

In 3D RHD simulations, the non-thermal, macroscopic velocity fields
arising from convective instabilities are computed self-consistently
from first principles and therefore have an immediate physical meaning.
They represent the buoyant motions associated with convection and
its turbulent features, and their statistical properties carry equally
important physical information as the mean temperature or density
stratifications. By contrast, in 1D atmosphere modeling, a free-parameter-dependent
velocity field $v_{\mathrm{MLT}}$ is derived for the convective flux
in MLT. Also, for radiative transfer and spectral line formation calculations,
two ad-hoc Gaussian velocity distributions -- the so-called micro-
and macroturbulence ($\xi_{\mathrm{turb}}$ and $\chi_{\mathrm{turb}}$,
respectively) -- are usually introduced to model Doppler broadening
of spectral lines associated with non-thermal (e.g., convective, turbulent,
oscillatory, etc.) gas motions in stellar atmospheres. The values
of the micro- and macroturbulence parameters are determined by comparing
synthetic and observed spectral line profiles and line strengths.
Usually, a depth-independent value of the microturbulence $\xi_{\mathrm{turb}}$
and one global value of the macroturbulence $\chi_{\mathrm{turb}}$
are applied in theoretical spectrum syntheses with 1D model atmospheres.
Full-3D line formation calculations using 3D models similar to those
described here, have demonstrated that in late-type stars the required
non-thermal Doppler line broadening is indeed primarily the result
of Doppler shifts from the convective motions and to a lesser extent
oscillations in the atmosphere \citep{Asplund:2000p20875}. As such
this non-thermal velocity field is clearly depth-dependent, while
micro- and macroturbulence are almost always assumed to be non-varying
with depth. Furthermore, $v_{\mathrm{MLT}}$ is solely assigned to
satisfy the necessary amount of convective flux by the individual
prescription of MLT. While, interestingly, $v_{\mathrm{MLT}}$ mimics
to a certain extent the run of $v_{z,\mathrm{rms}}$ in the interior
for cooler dwarfs. We remark that this interpretation is however not
physically consistently motivated. Moreover, the convective velocity
varies depending on its actual implementation \citep[e.g. ][]{BohmVitense:1958p4822,Henyey:1965p15592}
and, as such, $v_{\mathrm{MLT}}$ should be interpreted and used with
caution. We point out that one important motivation for conducting
3D RHD atmosphere models is the fact that the before mentioned spurious
inconsistent velocities become redundant. The hydrodynamical simulations
account consistently for only one unique velocity field.\\

\begin{figure}
\includegraphics[width=88mm]{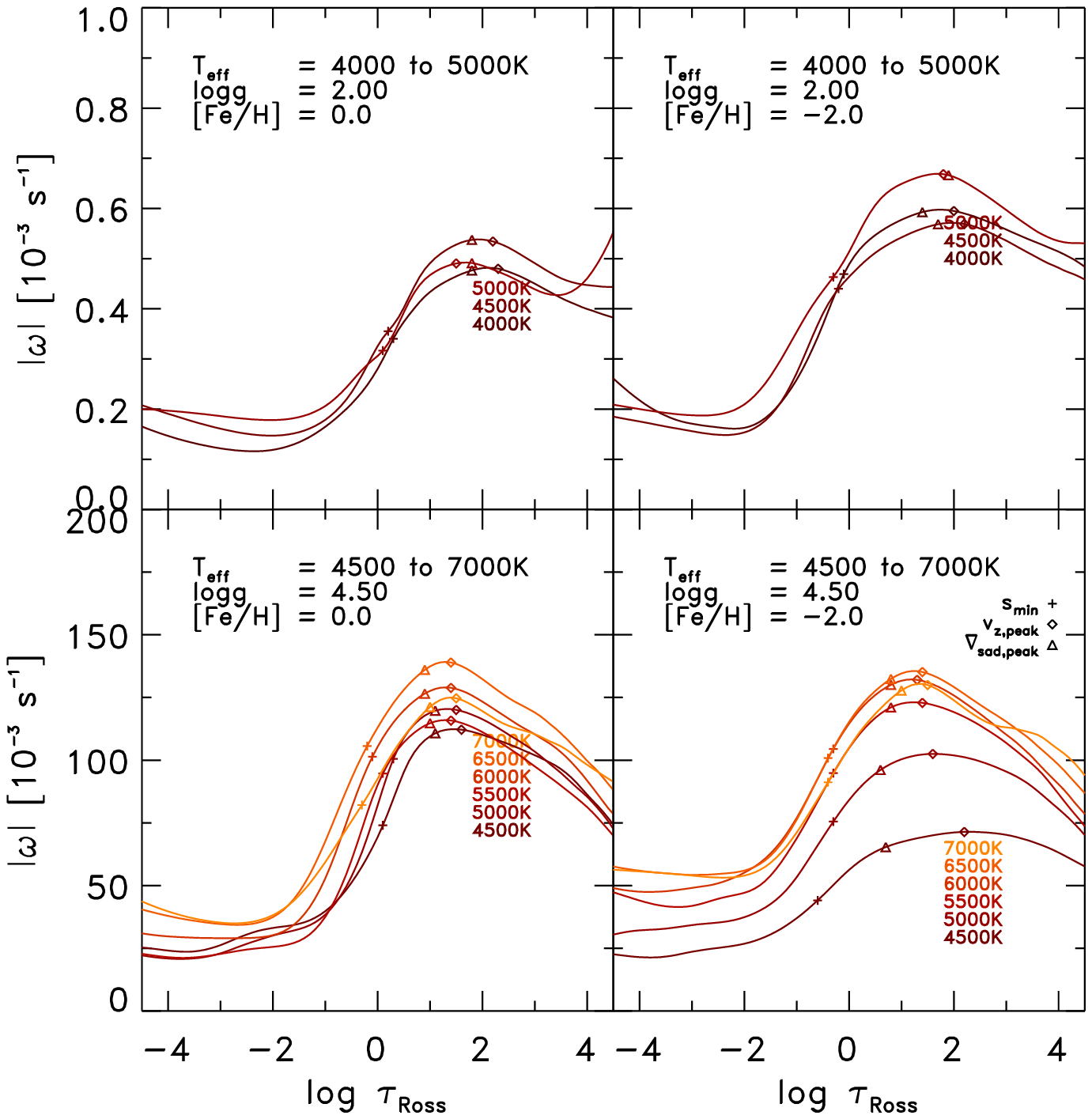}

\caption{$\left\langle 3\mathrm{D}\right\rangle $ stratification of the absolute
value of the vorticity $\left|\vec{\omega}\right|$ vs. $\teff$ is
shown for various stellar parameters. }
\label{fig:mean_vorticity}
\end{figure}
Another good measure for the turbulence of a velocity field is the
absolute value of the vorticity $\left|\vec{\omega}\right|=\left|\vec{\nabla}\times\vec{v}\right|$,
which is shown in Fig. \ref{fig:mean_vorticity}. The vorticity arises
below the surface in SAR due the overturning of the upflows and the
turbulent downdrafts experiencing the density gradient. The peak in
$\vec{\omega}$ is associated with pronounced shear flows, which arise
due deflection of the horizontal flows into downdrafts of the overturning
plasma (see SN98). The vorticity is concentrated in tube-like structures
in the intergranular lanes around the edges of granules. The run of
the vorticity follows closely that of $v_{z,\mathrm{rms}}$ (see Fig.
\ref{fig:rms_vertical_velocity}).

\subsubsection{Turbulent pressure\label{sub:turbulent-pressure}}

\begin{figure}
\includegraphics[width=88mm]{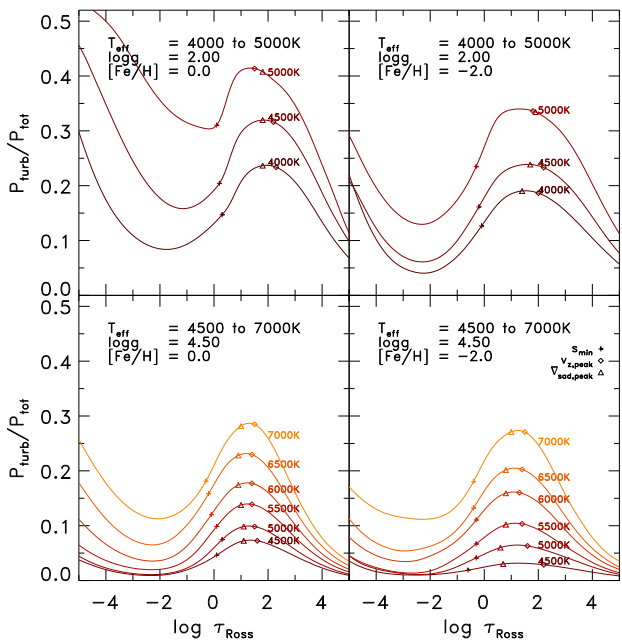}

\caption{The fraction of turbulent pressure to total pressure $p_{\mathrm{turb}}/p_{\mathrm{tot}}$
vs. optical depth $\ltaur$ is displayed for various stellar parameters.}
\label{fig:mean_turbulent-pressure-fraction}
\end{figure}
The turbulent pressure, $p_{\mathrm{turb}}=\rho v_{z}^{2}$, is an
additional (dynamical) pressure that arises from the (macroscopic)
vertical bulk flows due to the convective motions. It appears when
considering the horizontal averages of the momentum equation (Eq.
\ref{eq:momentum}), more specifically of the advection term therein.
The ratio of turbulent to total pressure, $p_{\mathrm{turb}}/p_{\mathrm{tot}}$,
shown in Fig. \ref{fig:mean_turbulent-pressure-fraction}, follows
qualitatively very closely the run of $v_{z,\mathrm{rms}}$ (compare
with Fig. \ref{fig:rms_vertical_velocity}), namely, it peaks in the
SAR ($\log\tau_{\mathrm{Ross}}\sim1.5$), reaches a minimum around
$\log\tau_{\mathrm{Ross}}\sim-2.0$, and increases in the upper layers
(a functional fit for $\left[p_{\mathrm{turb}}/p_{\mathrm{tot}}\right]_{\mathrm{peak}}$
is given in App. \ref{sub:Functional-fits}). In the SAR, the shape
of the $p_{\mathrm{turb}}/p_{\mathrm{tot}}$ profile with optical
depth looks similar to a Gaussian function, however, towards lower
$T_{\mathrm{eff}}$ and metallicity, it becomes increasingly skewed.
Averages on constant geometrical depth $\havz$ are similar, only
the peak and the upper layers are slightly lower at higher $\teff$s.

For hotter stars, in particular metal-rich giants, the turbulent pressure
becomes comparable to the gas pressure ($p_{\mathrm{turb}}/p_{\mathrm{tot}}\sim0.4$)
in the SAR, and the atmosphere is increasingly supported by $p_{\mathrm{turb}}$.
This means that neglecting the turbulent pressure, as is usually done
in 1D models, would significantly overestimate the gas pressure. The
consequence of this is a faulty, inconsistent stratification, since
the overestimation in gas pressure comes at the cost of altering other
physical quantities like the density, even when the temperature stratification
looks similar compared to a $\left\langle 3\mathrm{D}\right\rangle $
model. 

We find that $v_{z,\mathrm{rms}}$ is very close to $v_{\mathrm{turb}}=\left[\left\langle p_{\mathrm{turb}}\right\rangle /\left\langle \rho\right\rangle \right]^{1/2}$,
since the latter is basically the density-weighted analog of the former.
In 1D stellar structure models that include turbulent pressure, the
convective velocity from MLT is considered, i.e. $p_{\mathrm{turb,1D}}=\beta\rho v_{\mathrm{MLT}}^{2}$.
However, as shown in Fig. \ref{fig:rms_vertical_velocity}, the convective
velocities, $v_{\mathrm{MLT}}$, are underestimating $v_{\mathrm{turb}}$
systematically towards higher $\teff$ and lower $\logg$. Therefore,
the 1D models can be improved by using $v_{z,\mathrm{rms}}$ resulting
from 3D simulations, or one can fit the scaling factor $\beta$ to
match $v_{z,\mathrm{rms}}$, which would clearly reduce the error.

As shown by \citet{Wende:2009p5826}, we also expect $v_{z,\mathrm{rms}}$
to correlate with the \textit{microturbulence}%
\footnote{This is not necessarily the case for the macroturbulence $\chi_{\mathrm{turb}}$,
which compensates for the missing large-scale motions that alter the
shape of the emergent spectral line profile, but not its strength
(equivalent width).%
} $\xi_{\mathrm{turb}}$, since $v_{z,\mathrm{rms}}$ is a horizontal
average of the velocity field. In 1D line formation calculations,
a depth-independent $\xi_{\mathrm{turb}}$ is introduced in order
to compensate for missing Doppler broadening in the line extinction
profile. The actual correlation of the velocity field with $\xi_{\mathrm{turb}}$
is nontrivial due to the non-locality of the radiation field, which
is additionally impeded by non-linear atomic physics. Therefore, this
correlation has to be determined empirically by comparing the results
of 3D line-formation calculations with their $\left\langle 3\mathrm{D}\right\rangle $
counterparts \citep[see also][]{Steffen:2009p12544}. We intend to
perform such calculations in an upcoming work.

Finally, \citet{Chiavassa:2011p12542} showed that using a realistic
turbulent pressure contribution to the hydrostatic equilibrium in
1D red supergiant atmospheres, greatly improves the derived surface
gravity in these stars. This extra pressure component also leads to
an expansion of the atmosphere compared to a 1D model stratification
without turbulent pressure. This is referred to as \textit{atmospheric
levitation} \citep[see][]{Trampedach:2001p5597}. This will affect
p-modes by affording them a larger cavity, and hence lowering their
frequencies. This is part of the seismic near-surface effect which
has plagued helio- and asteroseismology.

\subsubsection{Total pressure and density \label{sub:total-pressure_and_density}}

\begin{figure}
\includegraphics[width=88mm]{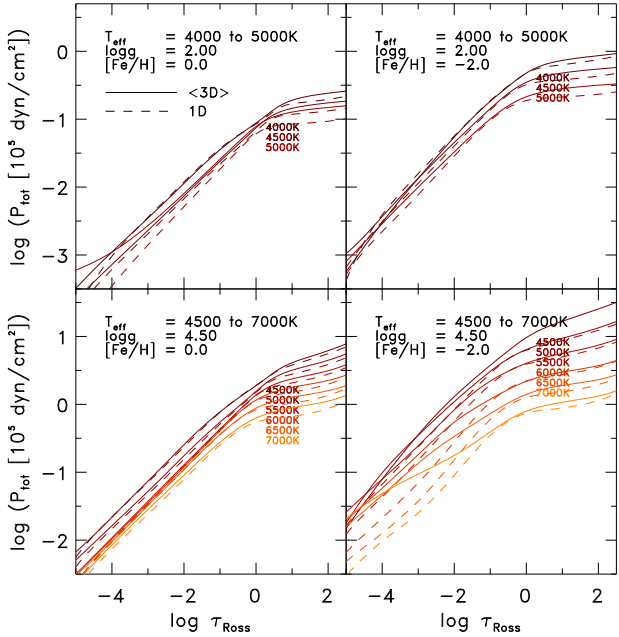}

\caption{We display the $\left\langle 3\mathrm{D}\right\rangle $ total pressure
$p_{\mathrm{tot}}$ against the optical depth $\ltaur$ for various
stellar parameters (solid lines). For comparison we also plot the
1D models with dashed lines. Note the different ordinate scale in
the top panel.}
\label{fig:mean_total-pressure}
\end{figure}
The total pressure is defined as the sum of thermodynamic and turbulent
pressure, $p_{\mathrm{tot}}=p_{\mathrm{th}}+p_{\mathrm{turb}}$, and
the former consists of gas and radiation pressure, $p_{\mathrm{th}}=p_{\mathrm{gas}}+p_{\mathrm{rad}}$.
In Fig. \ref{fig:mean_total-pressure}, we show the total pressure
for various stellar parameters. In contrast to the previous quantities,
$p_{\mathrm{tot}}$ decreases with higher $T_{\mathrm{eff}}$, lower
$\log g$, and higher metallicity. From the three stellar parameters,
the influence of the metallicity is the strongest. We find the highest
pressures (and densities) in the coolest metal-poor dwarfs and the
lowest pressures in the hottest metal-rich giants. In the upper layers
of hot metal-poor dwarfs, we find pressures systematically increased
with respect to their 1D counterparts, which is accompanied by similar
behavior in $\rho$, $p_{\mathrm{gas}}$ and $p_{\mathrm{turb}}$.
As we showed above, a significant fraction of the total pressure is
contributed by turbulent pressure in the SAR and in the upper layers
(see Fig. \ref{fig:mean_turbulent-pressure-fraction}), in particular
towards higher $\teff$ and lower $\logg$. Moreover, we note that
the temporal and horizontal $\havz$-averages from our relaxed simulations
are very close to hydrostatic equilibrium, and the turbulent pressure
contributes significantly to this equilibrium.\\

Since the mean density stratifications look qualitatively similar
to the total pressure ones, we refrain from showing them. Instead
we prefer to show, in Fig. \ref{fig:density_peak}, the peak density%
\footnote{The total pressure would lead to a very similar plot.%
} $\rho_{\mathrm{peak}}$, which is the density at the height of the
maximum rms-vertical velocity (see Fig. \ref{fig:rms_vertical_velocity}).
The density $\rho_{\mathrm{peak}}$ increases with lower $\teff$,
higher $\logg$, and lower $\feh$. These variations with stellar
parameters arise due to the radiative transfer, since the cooling
and heating rates (see Eq. \ref{eq:radiative_heating}) depend on
density $\rho$ and opacity $\kappa$. We showed in Sect. \ref{sub:Entropy-bottom}
that with higher metallicity and opacity, the hydrostatic stratification
is set at lower $\rho$.

\subsubsection{Electron number density\label{sub:electron-density}}

\begin{figure}
\includegraphics[width=88mm]{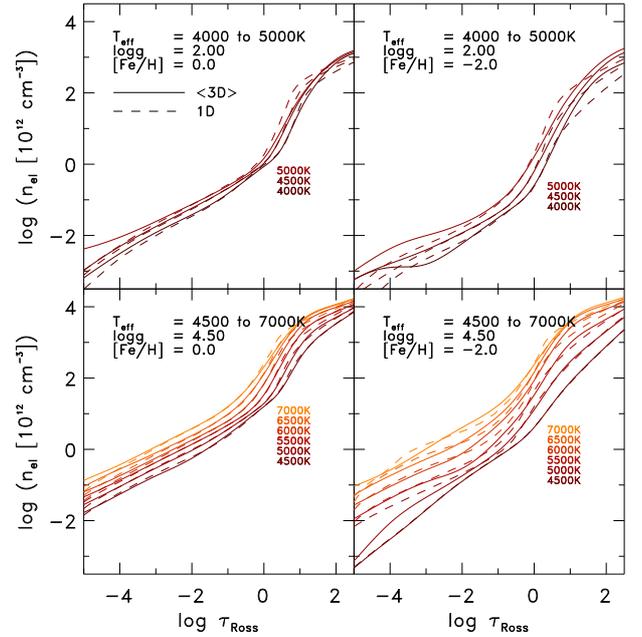}

\caption{We present the $\left\langle 3\mathrm{D}\right\rangle $ stratifications
of the local electron number density $n_{\mathrm{el}}$ against optical
depth $\ltaur$ for various stellar parameters (solid lines). For
comparison, we also show the corresponding stratifications from 1D
models (dashed lines). }
\label{fig:mean_electron-density}
\end{figure}
Next, we discuss the properties of the electron number density $n_{\mathrm{el}}$
(Fig. \ref{fig:mean_electron-density}), which is the temporal and
spatial average of the local electron density on layers of constant
Rosseland optical depth. The electron number density drops by about
$\sim2\,\mathrm{dex}$ at the transition from the interior to the
photosphere. This is due to the fact that the density itself decreases
here, and due to the recombination of hydrogen at the photospheric
transition. The convective flux consists to $\sim1/3$ of $F_{\mathrm{ion}}$
(see Sect. \ref{sub:transport-of-energy}), therefore, as the hot
ionized plasma reaches the surface, it radiates away energy, recombines,
and overturns into downdrafts, thereby reducing the number of free
electrons. The electron density increases with higher $T_{\mathrm{eff}}$,
lower $\log g$, and higher $\left[\mathrm{Fe}/\mathrm{H}\right]$.
The electron pressure $p_{\mathrm{el}}=n_{\mathrm{el}}k_{\mathrm{B}}T$
follows similar trends as the electron density in terms of variations
with stellar parameters and depth.

\subsubsection{Entropy \label{sub:entropy}}

\begin{figure}
\includegraphics[width=88mm]{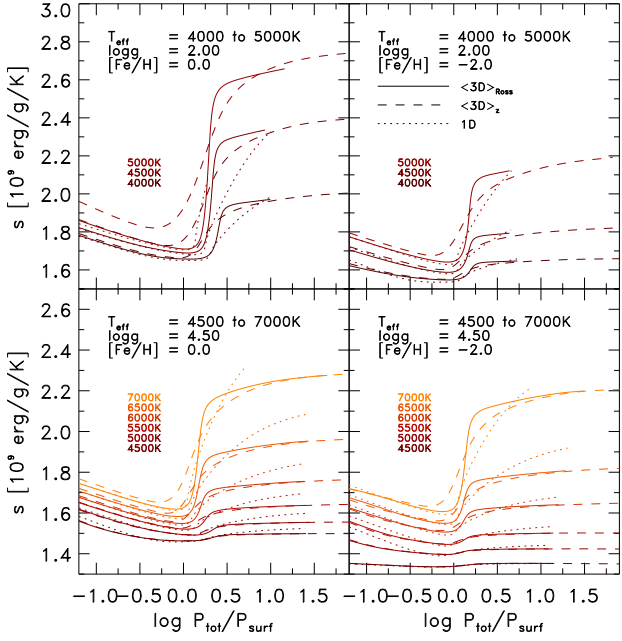}

\caption{$\left\langle 3\mathrm{D}\right\rangle $ and $\left\langle 3\mathrm{D}\right\rangle _{z}$
mean stratifications (solid and dashed respectively) of the entropy
$s$ vs. the total pressure normalized to the pressure at the optical
surface $\log p_{\mathrm{tot}}/p_{\mathrm{surf}}$ for various stellar
parameters. We show also the $s$-stratifications of the 1D models
(dotted lines). Note the different ordinate scale in the top panel.}
\label{fig:mean_entropy}
\end{figure}
Local, box-in-a-star, 3D RHD atmosphere models have well defined boundary
conditions at the bottom boundary because of the adiabaticity of the
convection zone, even though they are relatively shallow and comprise
only a small fraction of the convection zone. Indeed, the specific
entropy per unit mass of the plasma stays constant across most of
the convective zone, in particular for the upflows. In Fig. \ref{fig:mean_entropy},
we show the average entropy. Below the optical surface, the entropy
converges asymptotically against $s_{\mathrm{bot}}$ into deeper layers,
especially the averages on constant geometrical height $\left\langle 3\mathrm{D}\right\rangle _{z}$
(dashed lines). As the hot plasma in the granules reaches the optical
surface, it becomes transparent, thereby a large fraction of the energy
is radiated away. This results in a decrease in entropy, until it
reaches a minimum at the top of the convection zone ($\log\tau_{\mathrm{Ross}}\sim0.0$).
Further up, the entropy then increases again due to the decoupling
of the radiation and matter above the photosphere, which results in
an almost isothermal atmosphere. The 1D models (dotted lines) exhibit
larger entropy stratifications in the deeper convection zone, in particular
for higher $\teff$, thereby overestimating the entropy jump increasingly
due to the fixed mixing-length parameter $\alpha_{\mathrm{MLT}}$
with 1.5 for all stellar parameters.

\subsubsection{Superadiabatic temperature gradient \label{sub:temperature-gradient}}

\begin{figure}
\includegraphics[width=88mm]{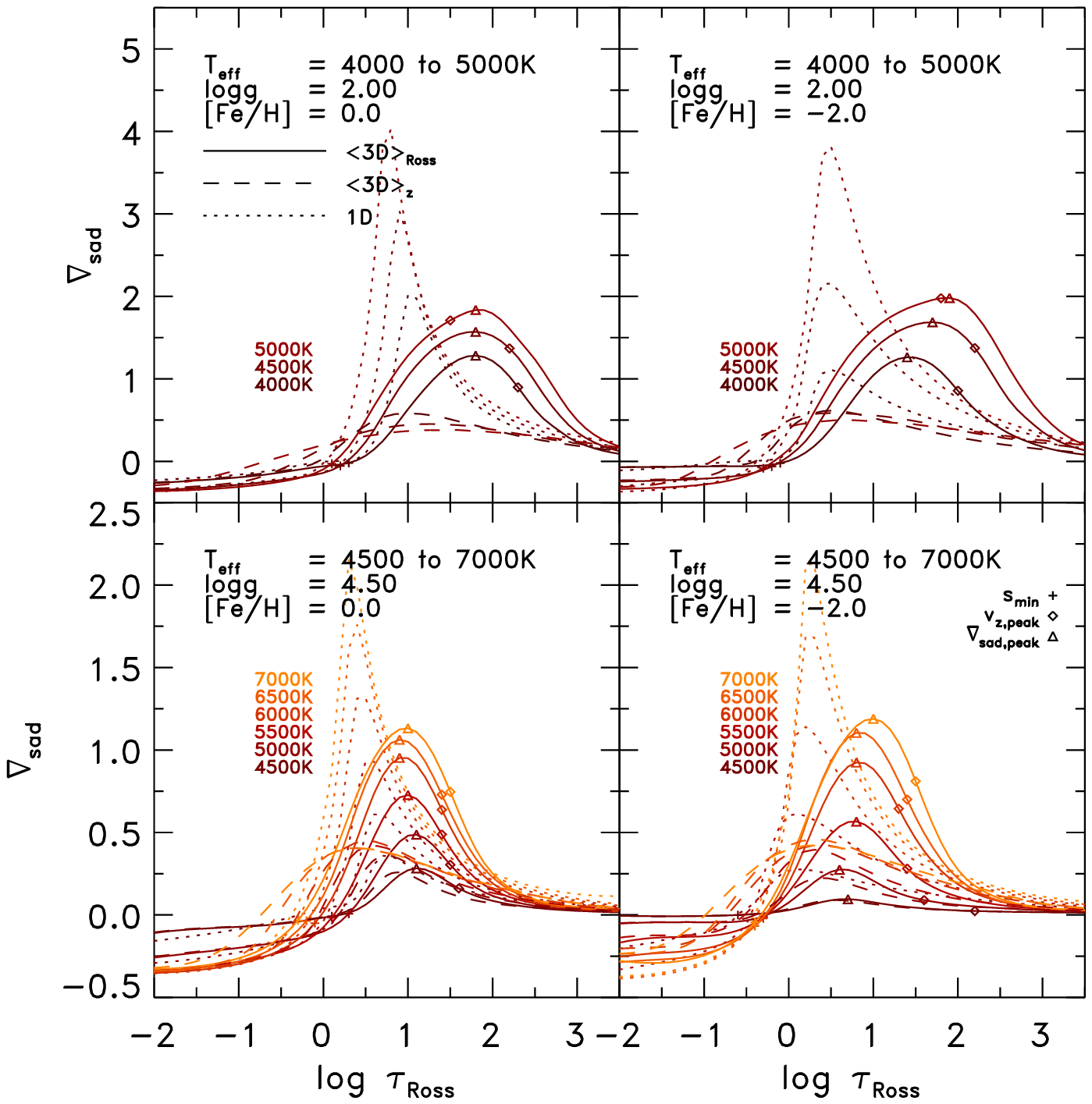}

\caption{We show the $\havr$ (solid lines) and $\havz$ (dashed lines) superadiabatic
gradient $\nabla_{\mathrm{sad}}$ vs. optical depth $\ltaur$ for
various stellar parameters. Furthermore, for comparison, we also show
the corresponding values from 1D models (dotted lines). Note the different
ordinate scale in the top panels.}
\label{fig:nabla_sad}
\end{figure}
We limit ourselves to show only the superadiabatic gradient $\nabla_{\mathrm{sad}}=\nabla-\nabla_{\mathrm{ad}}$,
since it combines the important properties of both the total and the
adiabatic temperature gradient ($\nabla$ and $\nabla_{\mathrm{ad}}$
respectively). In Fig. \ref{fig:nabla_sad}, we show $\nabla_{\mathrm{sad}}$
averaged over constant geometrical height or Rosseland optical depth.
The peak in $\nabla_{\mathrm{sad}}$, which looks like a skewed Gaussian
function, arises solely from the temperature gradient $\nabla$. The
superadiabatic gradient peaks around $\log\tau_{\mathrm{Ross}}\sim1.0-2.0$,
and becomes sub-adiabatic, i.e. $\nabla_{\mathrm{sad}}<0.0$, above
the optical surface at $\log\tau_{\mathrm{Ross}}<0.0-0.5$. The entropy
jump correlates directly with the superadiabatic gradient, since $\nabla_{\mathrm{sad}}=1/c_{p}\left[\partial s/\partial\ln p_{\mathrm{tot}}\right]$
and one can show that $\partial s/dz=c_{P}/H_{P}\left(\nabla-\nabla_{\mathrm{ad}}\right)$.
Hence, it is no surprise that they exhibit similarity in the peak
amplitude and position. In particular, the peak amplitude increases
with increasing $T_{\mathrm{eff}}$ and $\log g$ (see $\Delta s$
in Fig. \ref{fig:entropy_jump}; a functional fit for $\nabla_{\mathrm{sad}}^{\mathrm{peak}}$
is given in App. \ref{sub:Functional-fits}). The position of $\nabla_{\mathrm{sad}}^{\mathrm{peak}}$
on the optical depth scale $\havr$ (triangles), hence the position
of the steepest temperature gradient, changes slightly with stellar
parameters. However, similar to the position of $v_{z,\mathrm{rms}}^{\mathrm{peak}}$,
in the $\varepsilon-\rho$-plane, the distribution of $\nabla_{\mathrm{sad}}^{\mathrm{peak}}$
is regular (see Fig. \ref{fig:ee-rho-plane-1}), namely it shifts
systematically towards higher $\varepsilon$ and lower $\rho$ with
increasing $T_{\mathrm{eff}}$.

As it is clear in Fig. \ref{fig:nabla_sad}, one finds substantial
differences in $\nabla_{\mathrm{sad}}$ when comparing the two $\left\langle 3\mathrm{D}\right\rangle $
stratifications with their 1D counterparts, namely, the 1D gradients
exhibit distinctively larger amplitudes. These differences arise partly
due the missing turbulent pressure in the 1D case, but do not resolve
the discrepancies. Furthermore, we find an asymmetrically skewed shape
towards the optical surface in the 1D gradients, which is shared by
the geometrical averages $\havz$, but is not the case for the averages
on constant Rosseland optical depth $\havr$. A main reason for the
shown differences between $\havr$ and 1D comes from the averaging
over layers of constant $\taur$. The underlying $\nabla_{\mathrm{ad}}$s
are rather insensitive to the deviations between the $\havr$ and
1D stratifications, so the differences arise mainly due to $\nabla$.
Between $\havz$ and 1D the adiabatic gradients differ in the sub-photospheric
gradient.

\subsubsection{Transport of energy \label{sub:transport-of-energy}}

\begin{figure}
\includegraphics[width=88mm]{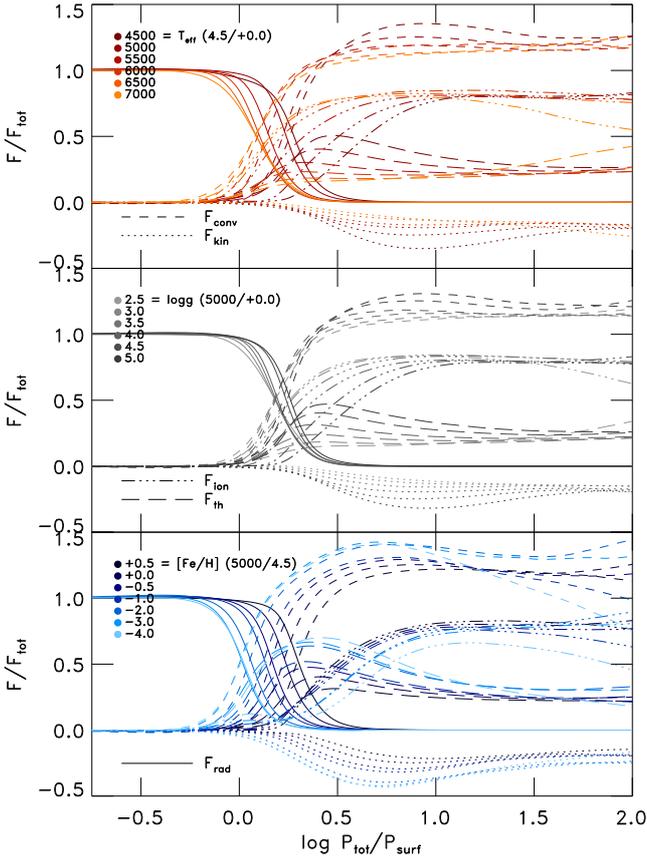}

\caption{Behavior of the normalized energy fluxes $F/F_{\mathrm{tot}}$ against
the total pressure normalized to the pressure at the optical surface
$\log p_{\mathrm{tot}}/p_{\mathrm{surf}}$ as a function of variations
in the individual stellar parameters ($T_{\mathrm{eff}}$, $\log g$,
and $\left[\mathrm{Fe}/\mathrm{H}\right]$, from top to bottom, respectively
). In each panel, the various curves are shown varying one of the
parameters while keeping the other two fixed. The individual normalized
components of $F_{\mathrm{tot}}$ are $F_{\mathrm{enth}}$ (dashed),
$F_{\mathrm{kin}}$ (dotted), $F_{\mathrm{ion}}$ (dash-triple-dotted),
$F_{\mathrm{th}}$ (long dashed) and $F_{\mathrm{rad}}$ (solid).
Averages are evaluated at constant geometrical height.}

\label{fig:fluxes} 
\end{figure}
The individual energy fluxes are quantities worthy of further consideration.
The energy flux is conserved only on averages of constant geometrical
height $\havz$, therefore, we show and discuss the latter here. The
total energy flux $F_{\mathrm{tot}}=F_{\mathrm{rad}}+F_{\mathrm{conv}}+F_{\mathrm{visc}}$
emerges from the photosphere solely in the form of radiative energy
flux. The total energy flux is supplied from the convection zone by
the convective energy flux, which is the sum of the enthalpy flux
\begin{equation}
F_{\mathrm{enth}}=\left[\varepsilon+\frac{p_{\mathrm{th}}}{\rho}\right]\delta j_{z},\qquad\mathrm{with}\,\delta j_{z}=\rho v_{z}-\left\langle \rho v_{z}\right\rangle \label{eq:fconv}
\end{equation}
($\delta j_{z}$ being the horizontal fluctuations of the vertical
mass flux; the average vertical mass flux vanishes) carried in the
upflows and the kinetic energy flux 
\begin{equation}
F_{\mathrm{kin}}=\left[\frac{1}{2}\rho\vec{v}^{2}\right]\delta j_{z}\label{eq:fkin}
\end{equation}
arising from the downdrafts (see SN98 and \citealt{Nordlund:2009p4109}).
Since the mean kinetic energy flux $F_{\mathrm{kin}}$ is negative,
the positive enthalpy flux $F_{\mathrm{enth}}$ is the major component
of the convective energy flux $F_{\mathrm{conv}}$. The enthalpy flux
in turn consists of the energy fluxes due to ionization $F_{\mathrm{ion}}=\left[\varepsilon-\frac{3}{2}\frac{p_{\mathrm{th}}}{\rho}\right]\delta j_{z}$,
thermal heat $F_{\mathrm{th}}=\frac{3}{2}\frac{p_{\mathrm{th}}}{\rho}\delta j_{z}$
and acoustic (sound) waves $F_{\mathrm{acous}}=\left\langle p_{\mathrm{th}}v_{z}\right\rangle -\left\langle p_{\mathrm{th}}\right\rangle \left\langle v_{z}\right\rangle $.
In Fig. \ref{fig:fluxes}, we show the energy fluxes $F_{\mathrm{rad}}$,
$F_{\mathrm{enth}}$, $F_{\mathrm{kin}}$, $F_{\mathrm{ion}}$, and
$F_{\mathrm{th}}$ normalized to the total emergent energy flux $\sigma T_{\mathrm{eff}}^{4}$
(for clarity, we refrain from showing $F_{\mathrm{visc}}$ and $F_{\mathrm{acous}}$,
since their contribution to $F_{\mathrm{tot}}$ is very small). We
vary one stellar parameter at a time, while the other two are fixed
($\teff$, $\logg$, and $\feh$, from top to bottom in Fig. \ref{fig:fluxes},
respectively). Just below the optical surface ($0.5<\log p_{\mathrm{tot}}/p_{\mathrm{surf}}<1.0$),
both $F_{\mathrm{kin}}$ (solid lines) and $F_{\mathrm{enth}}$ (dashed
lines) increase towards cool metal-poor dwarfs, i.e. lower $T_{\mathrm{eff}}$,
$\feh$ and higher $\log g$, due to higher densities and velocities.
The increased reduction of the total flux by $F_{\mathrm{kin}}$ ($<0.0$)
is compensated by a simultaneously higher $F_{\mathrm{enth}}$ ($>1.0$).
On the other hand, in deeper layers ($\log p_{\mathrm{tot}}/p_{\mathrm{surf}}>1.5)$,
both converge to similar fractions for all stellar parameters ($-0.17$
and $1.14$ for $F_{\mathrm{kin}}$ and $F_{\mathrm{enth}}$ respectively).
This convergence to very similar values is rather remarkable. The
convective motions seem to follow an exact guideline, which might
be correlated to the universal filling factor, however, we will study
this more carefully in a future work. We remark that in deeper solar
simulations%
\footnote{Our shallow solar simulation is $2.8\,\mathrm{Mm}$ deep.%
} ($20\,\mathrm{Mm}$) that $F_{\mathrm{enth}}$ and $F_{\mathrm{kin}}$
increase with depth, while their sum remains constant \citep[see ][]{Stein:2009p6085}.

The majority of the total energy flux $F_{\mathrm{tot}}$ in the convection
zone is carried in form of ionized hydrogen%
\footnote{The given fractions are averages of all grid models.%
} with $F_{\mathrm{ion}}\simeq0.67$, while thermal heat is the second
most important component with $F_{\mathrm{th}}\simeq0.29$. The acoustic
energy constitutes only a small fraction with $F_{\mathrm{acous}}\simeq0.04$.
SN98 found similar fractions with $F_{\mathrm{kin}}\sim-0.10$ to
$-0.15$, $F_{\mathrm{ion}}\sim2/3$ and $F_{\mathrm{th}}\sim1/3$
for the Sun. The $F_{\mathrm{ion}}$ and $F_{\mathrm{th}}$ fractions,
which are the major constituents of the enthalpy flux, undergo a significant
change below the surface, as we show in Fig. \ref{fig:fluxes} for
models with different stellar parameters. In particular, the fraction
of thermal heat $F_{\mathrm{th}}$ becomes more significant at the
cost of $F_{\mathrm{ion}}$ towards cool metal-poor dwarfs. The thermal
flux $F_{\mathrm{th}}$ reaches a maximum (up to $F_{\mathrm{th,max}}\simeq0.5$)
just below the surface, but eventually converges close to the above
mentioned fractions in deeper layers (long dashed lines).\\

In 1D MLT models, the convective flux is assumed to consist of the
enthalpy flux only, $F_{\mathrm{conv,1D}}$ (see Appendix \ref{sub:Basic-equations}).
This is a result of the MLT assumption of symmetric flows which means
that the kinetic energy fluxes in the up- and downflows cancel exactly.
As remarked by \citet{Henyey:1965p15592}, the details on $F_{\mathrm{kin}}$,
$F_{\mathrm{ion}}$, $F_{\mathrm{th}}$ and $F_{\mathrm{acous}}$
are not at hand due to the lack of a self-consistent velocity field.
The energy fluxes from 3D RHD simulations, on the other hand, arise
self-consistently from solving the coupled equations of radiative
hydrodynamics, without further assumptions.\\

As mentioned above, the emergent total energy flux $F_{\mathrm{tot}}$
is carried in the convection zone mainly by the positive enthalpy
flux $F_{\mathrm{enth}}$ (Eq. \ref{eq:fconv}). Therefore, one can
approximate the convective energy flux with the mean jump in enthalpy%
\footnote{For example, $\Delta h$ can be determined at the top and bottom of
the photospheric transition region (see Fig. \ref{fig:ee-rho-plane-1}).%
} $\Delta h$ times the mean vertical mass flux of the upflows below
the optical surface, hence 
\begin{equation}
F_{\mathrm{conv}}\approx\left\langle \Delta h\right\rangle \left\langle \rho v_{z}\right\rangle .\label{eq:fconv_approx}
\end{equation}
At the transition region, the enthalpy jump $\Delta h$ is primarily
caused by the strong drop in internal energy $\varepsilon$, hence
entropy $s$, and the thermodynamic pressure work is rather small
(note the change of $p_{\mathrm{tot}}$ below the surface $\ltaur>0.0$
in Fig. \ref{fig:mean_total-pressure}), i.e. $\Delta h\approx T\Delta s$,
where $T$ is the temperature at the surface. By approximating $T\simeq T_{\mathrm{eff}}$,
one can expect the total energy flux $F_{\mathrm{tot}}=\sigma T_{\mathrm{eff}}^{4}$
to depend to first order on the mean entropy jump%
\footnote{Here, we prefer to use $\Delta s$ instead of directly $\Delta h$
or $\Delta\varepsilon$ due to the adiabaticity of convection.%
}, density, and vertical velocity:

\begin{equation}
\sigma T_{\mathrm{eff}}^{3}\propto\left\langle \Delta s\right\rangle \left\langle \rho\right\rangle \left\langle v_{z}\right\rangle .\label{eq:ftot_ssj_rho_uy}
\end{equation}
This approximation can already be retrieved on dimensional grounds,
however, we derived the latter in order to explain the systematic
variations of $\Delta s$, $\vzrmsp$ and $\rho_{\mathrm{peak}}$
with stellar parameters, which we have observed above (see Figs. \ref{fig:entropy-bottom},
\ref{fig:velocity_peak} and \ref{fig:density_peak}, respectively).
The emergent radiative energy flux is correlated with $\Delta s$,
$\rho$ and $v_{z}$, and the respective composition resulting from
the individual contributions varies with stellar parameters. 

The interplay between the radiative heating and cooling rates $q_{\mathrm{rad}}$
(Eq. \ref{eq:radiative_heating}) and hydrostatic equilibrium, require
a different density stratification for different stellar parameters
due to the dependence of opacity on thermodynamic variables, as we
showed in Sect. \ref{sub:Entropy-bottom}. The resulting density variations
will induce adjustments in the vertical velocity and entropy jump.
Furthermore, we find with increasing $\logg$ and $\feh$ at a fixed
$\teff$, the density increases, which is compensated by higher $\Delta s$
and $v_{z}$ ($\Delta s\propto\Delta\rho^{-1}$, see Eq. \ref{eq:entropy_dependence}).
We would like also to emphasize the remarkably important (non-local)
influence of the rather thin photospheric transition region on basically
the whole convection zone, since the entropy deficiency of the turbulent
downdrafts are generated mainly here. The latter sets the entropy
jump and the convective driving \citep[see][]{Nordlund:2009p4109}.\\

\begin{figure}
\includegraphics[width=88mm]{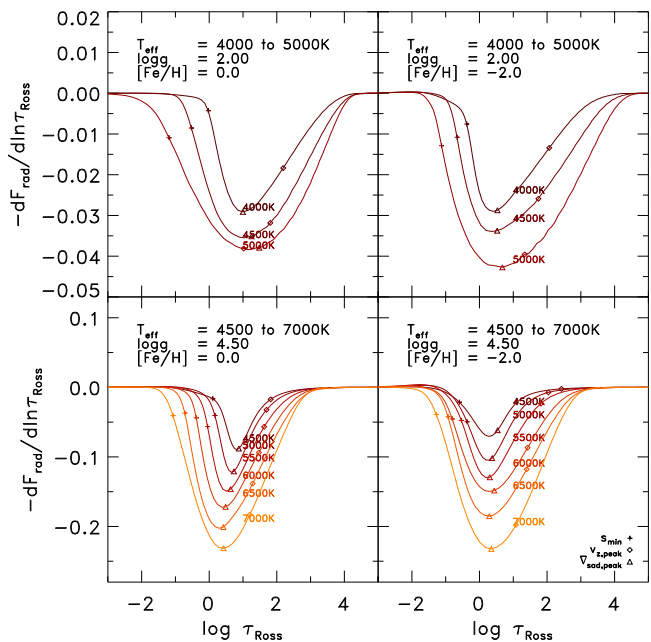}\caption{We show the normalized cooling and heating rates $q_{\mathrm{rad}}^{\mathrm{norm}}$
vs. optical depth $\ltaur$ for various stellar parameters. The shown
averages are retrieved on constant geometrical height $\left\langle 3\mathrm{D}\right\rangle _{z}$.
Note the different ordinate scale in the top panel.}

\label{fig:qrad_ov} 
\end{figure}
The radiative heating and cooling rates $q_{\mathrm{rad}}$ (Eq. \ref{eq:radiative_heating})
due to radiative losses enter the hydrodynamic equations as a source
and sink term in the energy equation (Eq. \ref{eq:energy}). It is
the divergence of the radiative flux $q_{\mathrm{rad}}=-\vec{\nabla}\cdot\vec{F}_{\mathrm{rad}}$,
and a large, negative $q_{\mathrm{rad}}$, the \emph{cooling peak},
marks the transition of energy transport from fully convective below
the optical surface to fully radiative close to the photosphere. To
better illustrate the depth dependence of $q_{\mathrm{rad}}$ and
the comparison among different models, in Fig. \ref{fig:qrad_ov},
we show the normalized cooling and heating rates $q_{\mathrm{rad}}^{\mathrm{norm}}=-dF_{\mathrm{rad}}/d\ln\taur$.
One can see that the amplitude of $q_{\mathrm{rad}}^{\mathrm{norm}}$
increases with higher $\teff$, accompanied by an increase in the
width of the cooling peak. The position of the maximum absolute amplitude
coincides with the position of $\nabla_{\mathrm{sad}}^{\mathrm{peak}}$,
since the cooling rate (radiative loss) is setting the entropy fluctuations,
hence the superadiabatic gradient (see Sect. \ref{sub:temperature-gradient}).
Furthermore, this location moves into upper layers for higher $\teff$
(from $\ltaur\simeq1.0$ up to $0.2$ for $\teff=4000$ to $7000\,\mathrm{K}$
respectively). On the other hand, the width of the photospheric transition
region $\Delta_{\mathrm{ph}}=\Delta\ltaur\left(q_{\mathrm{rad}}<0\right)$
clearly widens for hotter $\teff$, but also, in particular, for metal-poor
giants (see top right panel in Fig. \ref{fig:qrad_ov}). While for
cool dwarfs the width is typically $\Delta_{\mathrm{ph}}\approx3.0\,\mathrm{dex}$,
for hot metal-poor giants, it reaches $\Delta_{\mathrm{ph}}\approx5.0\,\mathrm{dex}$
(see, e.g., model with $5000\,\mathrm{K}$ in right top panel of Fig.
\ref{fig:qrad_ov}).

\subsection{Comparison with 1D models\label{sub:Comparison-with-1D}}

\subsubsection{1D models\label{sub:1D-models}}

\begin{figure}
\includegraphics[width=88mm]{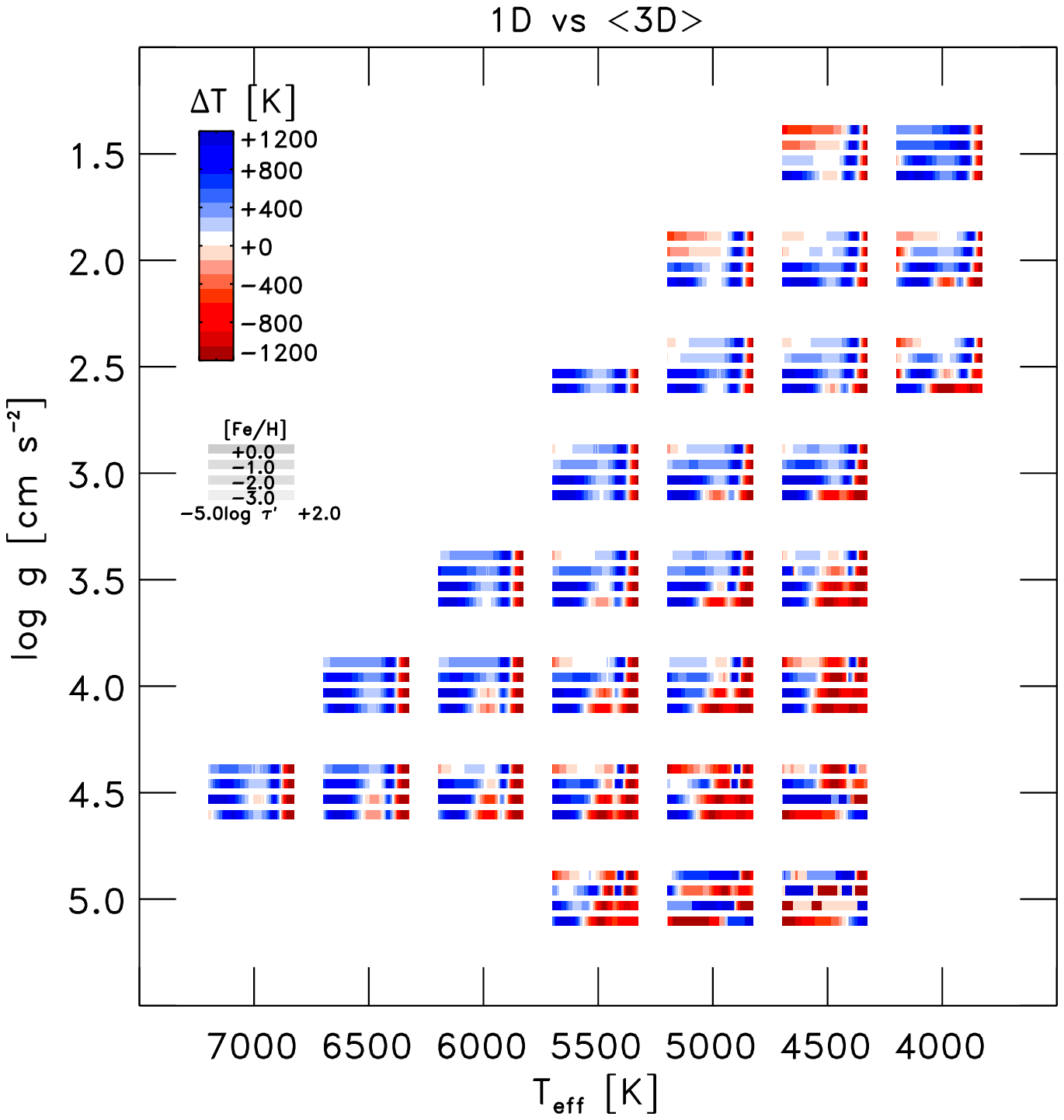}

\caption{Comparison of the temperature stratification for the $\left\langle 3\mathrm{D}\right\rangle $
and the 1D models by showing the difference $\left\langle 1\mathrm{D}\right\rangle -\left\langle 3\mathrm{D}\right\rangle $
(colored bars) between $\ltaur=-5.0$ and $2.0$ in the Kiel-diagram.
We present four different metallicity ($\feh=-3.0,-2.0,-1.0$ and
$0.0$).}
\label{fig:comp_1D}
\end{figure}
A differential comparison between 1D and 3D in terms of approaches
in the modeling of stellar atmospheres is of obvious relevance here.
Therefore, we developed a plane-parallel, hydrostatic, 1D atmosphere
code (hereafter simply referred to as the \emph{1D code}) that is
based on a similar physical treatment as the MARCS code with a few
simplifications (see Appendix \ref{sec:Stagger-grid-1D-atmospheres}
and \citealt{Gustafsson:2008p3814} for more details). We employ exactly
the same EOS and opacities as in the individual 3D models, thereby
excluding differences due to dissimilar input physics. Also, we applied
the $\hav$ models as initial stratifications for the 1D models. These
mean $\hav$ stratifications are defined on an equidistant optical
depth scale from $\ltaur=-5.0$ to $+5.0$ in steps of $0.1$. The
well-resolved optical depth scale reduces discretization errors in
the 1D atmosphere calculations, thereby making the 1D-$\hav$ comparison
more reliable. 

In Fig. \ref{fig:comp_1D}, we show a comparison of the 1D and $\left\langle 3\mathrm{D}\right\rangle $
temperature stratifications. One can immediately extract that the
upper layers of the atmospheres are systematically overestimated in
the 1D models by up to $\sim1000\,\mathrm{K}$, in particular for
metal-poor stars $\feh\leq-2.0$ (for solar models the maximal difference
is $\sim500\,\mathrm{K}$). In the optically thin layers of 1D models,
stable against convection, radiative equilibrium is enforced. However,
in the upper layers of the metal-poor $\hav$ models, the effect of
the non-vanishing adiabatic cooling rate is to shift the balance with
radiative heating to lower temperatures due to a scarcity and weakness
of spectral lines at lower metallicities \citep{Asplund:1999p11775,Collet:2007p5617}.
Interesting are also the hotter temperature stratifications for a
few giants ($\teff/\logg=4500\,\mathrm{K}/1.5$ and $5000\,\mathrm{K}/2.0$)
towards higher metallicity ($\feh>-2.0$), which results from the
radiative equilibrium at higher temperatures. On the other hand, with
the 1D models, we find systematically cooler temperatures below the
photosphere $\ltaur\simeq2.0$ with up to $\sim1000\,\mathrm{K}$
(here there is no difference with different metallicities). Therefore,
one has to keep in mind that, with 1D atmosphere models, and for metal-poor
stars in particular, these severe effects on the stratifications can
lead to large systematic errors in spectroscopic abundance determinations,
up to 0.5 dex or more in logarithmic abundance, depending on the formation
height of the individual spectral lines used in the analysis \citep[e.g.][]{Asplund:1999p11775,Asplund:2001p21515,Collet:2006p10854,Collet:2007p5617,Caffau:2008p13377,Caffau:2010p4631,GonzalezHernandez:2010p21570,Kucinskas:2013A&A...549A..14K}.
We will return to this issue using our new grid of 3D stellar models
in subsequent investigations.

In the 1D model calculations the mixing-length parameter is kept constant
with $\alpha_{\mathrm{MLT}}=1.5$, which is the commonly applied value
\citep[see][]{Gustafsson:2008p3814}. However, it is well-known that
$\alpha_{\mathrm{MLT}}$ varies with stellar parameters (see \citealt{Ludwig:1999p7606};
\citealt{Bonaca:2012p19742}, Magic et al. in preparation). Therefore,
we caution that a single fixed value will lead to severe differences
in atmospheric stratification. The systematic deviations beneath the
optical surface in the temperature stratification between 1D and $\hav$
towards cool dwarfs can be interpreted as the manifestation of the
wrong $\alpha_{\mathrm{MLT}}$ (see Fig. \ref{fig:comp_1D}). Furthermore,
$p_{\mathrm{turb}}$ is neglected in the 1D code, which affects the
stratification by reducing the gas pressure (see Sect. \ref{sub:turbulent-pressure}).

\subsubsection{\textsc{MARCS} and ATLAS models\label{sub:MARCS-models}\label{sub:ATLAS-models}}

\begin{figure}
\includegraphics[width=88mm]{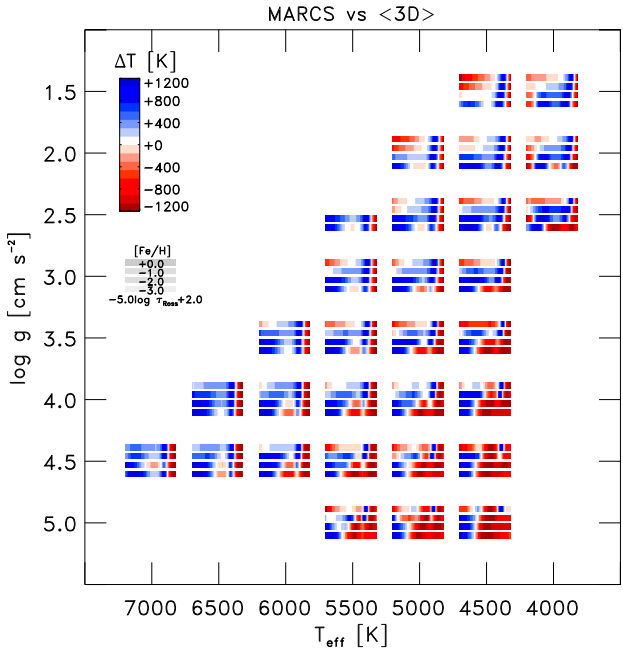}

\caption{Similar as Fig. \ref{fig:comp_1D}, however here we compare the $\left\langle 3\mathrm{D}\right\rangle $
with the MARCS models for $\feh=-3.0,-2.0,-1.0$ and $0.0$. For a
better comparison, we applied the same temperature range as given
Fig. \ref{fig:comp_1D}.}
\label{fig:comp_marcs}
\end{figure}
Last, we would also like to briefly compare our $\left\langle 3\mathrm{D}\right\rangle $
stratifications with the currently widely applied MARCS and ATLAS
models (see Fig. \ref{fig:comp_marcs}, we show only the comparison
with MARCS modes, since the ATLAS models look qualitatively rather
similar). We find qualitatively similar deviations as with the 1D
models above. At the same time, here we also have additional differences
due to the different input physics (EOS and opacities). The largest
differences between the $\left\langle 3\mathrm{D}\right\rangle $
and 1D MARCS stratifications of metal-poor stellar atmospheres are
slightly higher, with $\sim1300\,\mathrm{K}$ at $\feh=-3.0$, while
for solar metallicity the temperatures are underestimated in 1D by
$\sim500\,\mathrm{K}$ at mosts. Below the surface, the differences
amount to $\sim1000\,\mathrm{K}$. The ATLAS models are up to $\sim850\,\mathrm{K}$
hotter at the top and $\sim1000\,\mathrm{K}$ cooler below the surface.
In both cases, the deviations at the top increase towards lower $\feh$.

\section{Conclusions\label{sec:Conclusions}}

We presented here a comprehensive grid of realistic, state-of-the-art,
three-dimensional (3D), time-dependent, radiative-hydrodynamic (RHD)
stellar atmosphere models for late-type stars, covering a substantial
portion of stellar parameter space, and provided a detailed description
of the approach we followed for the construction of models. With the
aid of our realistic 3D RHD simulations, we are able to access and
render details of stellar atmospheres and subsurface convection, that
are out of reach for 1D models and also inaccessible by observations.
We presented and discussed a number of important global physical properties
of the simulations as well as the mean stratifications resulting from
the relatively large amount of data. \\

The constant entropy value of the adiabatic convection zone has a
profound influence on several aspects and properties of the 3D RHD
simulations. In particular, we find systematic correlations among
the constant entropy value of the adiabatic convection zone, the entropy
jump, and the vertical velocity, which we interpreted as \textit{scaling
relations}. In addition, we find that the variation in intensity contrast
is enhanced at lower metallicity. Also, we determined that the granule
size scales basically with the pressure scale height close to the
surface, which can be explained in the picture of what we refer to
as Nordlund scaling relation (NSR).\\

We discussed in great detail the depth-dependent temporal and spatial
averages of various important physical quantities. In particular,
we determined and examined various systematic trends in the variations
of the entropy jump, the density, and the vertical velocity with stellar
parameters. The latter can be discussed by regarding the changes in
the transition of energy transport from convective to radiative at
the photosphere. Namely, for different stellar parameters, the coupling
between radiation and matter through the radiative transfer necessitates
specific physical conditions due to changes in the opacity, which
in turn alters the density. These variations in the density on the
other hand require adjustments in the entropy jump and the vertical
velocity. This can be illustrated under consideration of the total
energy flux and conservation of energy. The named important values
are coupled with each other, and these also set basically the general
physical framework of the stellar atmosphere. The actual particular
connections of these correlations have to be studied carefully in
more detail, thus possibly leading to an improved understanding of
the physical mechanisms operating in subsurface convection, hence
stellar atmospheres.\\

We compared our 3D models and their mean stratifications to 1D models
employing the same input physics, thereby revealing important systematic
differences between the two kinds of models due to the incomplete
treatment of convection by the 1D mixing-length theory (MLT) and the
assumption of radiative equilibrium. The latter leads to an overestimation
of the temperature stratification in metal-poor stars. While below
the optical surface, we find that the temperatures are typically underestimated
due to a fixed mixing length ($\alpha_{\mathrm{MLT}}=1.5$), in particular
for higher $\teff$ and lower $\logg$. Also, we find that MLT fails
to render a realistic vertical velocity field. The often neglected
turbulent pressure has towards giants a non-negligible contribution
on the total pressure, thereby, indicating that the thermal gas pressure
is also overestimated significantly. We also quantified the differences
with widely used 1D atmosphere models, in particular ATLAS and MARCS.
For a number of important values we provide functional fits with stellar
parameters, so that these can be accessed immediately. Thereby, one
can easily scale new 3D models based on these informations.\\

The present work is meant to be an introduction to a series of papers
on the \textsc{Stagger-}grid. The material discussed here is in fact
just a small fraction of the actual information contained in the complete
simulation data set. On the other hand, the list of potential applications
for 3D models is also long. Because of space constraints, we will
discuss in more detail the different methods we have applied for computing
temporal and spatial averages of our 3D RHD models in a separate paper.
Therein, we will also discuss our routines for the interpolation of
$\left\langle 3\mathrm{D}\right\rangle $ atmosphere models, and explore
the differences between $\left\langle 3\mathrm{D}\right\rangle $
and 3D line formation. We have also compiled details on statistical
properties, which we will present individually. These can be later
on utilized for a so-called $1.5\mathrm{D}$ line formation \citep[see][]{Ayres:2006p21937}.
An obvious plan on the agenda is also to analyze carefully the detailed
properties of granulation and the intergranular lane for different
stellar parameters across the HR-diagram in the future. \\

As the major purpose of theoretical atmosphere models, we are computing
full 3D synthetic spectra with OPTIM3D \citep[see][]{Chiavassa:2009p22491}
for all of our models (Chiavassa et al. in preparation). These will
be made publicly accessible and can be used for various applications,
e.g. spectroscopic parameter determination. Furthermore, we will derive
new limb-darkening coefficients based on the full 3D synthetic spectra
(Magic et al. in preparation), which are vital for applications such
as the characterization of transiting exoplanets \citep{Hayek:2012p21944}.
We intend to calibrate photometric colors and radial velocities from
the spectra. For abundance determinations, we will construct an extensive
library of synthetic high-resolution spectral lines. These will serve
for deriving 3D effects on line-shifts, line-asymmetries and bisectors.\\

Also an important application is the calibration of free parameters
in 1D models by employing our 3D simulations, in particular MLT and
micro- and macroturbulence. We will incorporate our $\left\langle 3\mathrm{D}\right\rangle $
models into stellar evolutionary models, which will result in improved
stellar structures useful for asteroseismology. And with 3D line formation
calculations we will calibrate the microturbulence.\\

Despite the enormous success and the \emph{ab-initio} nature\emph{
}of 3D atmosphere modeling, as last we want to mention the weaknesses.
In order to keep the computation costs reasonable, the radiative transfer
is simplified with the opacity binning method, which may influence
the outcome. Also, the numerical resolutions of these so called large-eddy
simulations are not resolving the microscopic viscous dissipation
length scales, hence, the need to introduce numerical diffusion. However,
these do not affect the main properties of the macroscopic flows and
of the physical stratification. Also, we minimized the diffusion coefficients
once under the constraint of numerical stability, and then applied
for all the simulations. These issues can be solved, and the easiest
rectification will be enhancement of the numerical resolution, in
particular for giant. Furthermore, we will conduct improvements in
the radiative transfer by employing more bins and possibly more more
angles in the future. By inserting magnetic fields with different
field strength, we will explore the influence of the latter on convection
for different type stars. Accounting for non-LTE effects is extremely
expensive for 3D simulations, therefore, these are usually neglected,
however, even these will be included eventually in the future.
\begin{acknowledgements}
We acknowledge access to computing facilities at Rechen Zentrum Garching
(RZG) through Max-Planck Institute for Astrophysics (MPA) and the
National Computational Infrastructure (NCI) of Australia, through
Australian National University (ANU), where the simulations were carried
out. We are grateful to W. D{\"a}ppen for access to the code and
data tables for the EOS. And we thank B. Plez and B. Edvardsson for
providing the MARCS line-opacities. Also, we acknowledge the Action
de Recherche Concert{\'e}e (ARC) grant provided by the Direction
g{\'e}n{\'e}rale de l’Enseignement non obligatoire et de la Recherche
scientifique – Direction de la Recherche scientifique – Communaut{\'e}
française de Belgique, and the F.R.S.- FNRS. Remo Collet is the recipient
of an Australian Research Council Discovery Early Career Researcher
Award (project number DE120102940). We thank the referee for the helpful
comments.
\end{acknowledgements}
\bibliographystyle{aa}
\bibliography{papers}
\appendix

\section{The \textsc{Stagger}-grid 1D atmospheres\label{sec:Stagger-grid-1D-atmospheres}}

The following discussion concerns solely 1D atmosphere models and
MLT, therefore, similar quantities as discussed above may deviate
(e.g. $F_{\mathrm{conv}}$). The numerical code that we used for computing
1D atmospheres for the \textsc{Stagger}-grid models solves the coupled
equations of hydrostatic equilibrium and energy flux conservation
in 1D plane-parallel geometry. The 1D models use the same EOS and
opacity package in order to allow consistent 3D-1D comparisons. The
set of equations and numerical methods employed for their solution
are similar to those of the \textsc{MARCS} code \citep{Gustafsson:2008p3814}
with a few changes and simplifications that will be outlined in the
following. The resulting model atmospheres yet maintain very good
agreement with \textsc{MARCS} models (see Sect. \ref{sub:MARCS-models}).

\subsection{Basic equations\label{sub:Basic-equations}}

Assuming 1D plane-parallel geometry with horizontal homogeneity and
dominance of hydrostatic equilibrium over all vertical flow simplifies
the equation of motion (Eq. \ref{eq:momentum}) to the hydrostatic
equilibrium equation 
\begin{equation}
\frac{d}{d\tau_{\mathrm{std}}}\left(p_{\mathrm{gas}}+p_{\mathrm{turb}}\right)-\frac{\rho g}{\kappa_{\mathrm{std}}}=0,
\end{equation}
where $\kappa_{\mathrm{std}}$ and $\tau_{\mathrm{std}}$ are a standard
opacity and corresponding optical depth (e.g. the Rosseland mean),
$p_{\mathrm{gas}}$ and $p_{\mathrm{turb}}$ denote gas pressure and
turbulent pressure, $\rho$ is the gas density, and $g$ is the surface
gravity. Radiation pressure is neglected, as in the 3D simulations.
Turbulent pressure is estimated using the expression 
\begin{equation}
p_{\mathrm{turb}}=\beta\rho v_{\mathrm{turb}}^{2},
\end{equation}
with the scaling parameter $\beta$ that corrects for asymmetries
in the velocity distribution and the mean turbulent velocity $v_{\mathrm{turb}}$
that is used as a free, independent parameter.\\

The depth-integral of the energy equation (Eq. \ref{eq:energy}) reduces
to the flux conservation equation, 
\begin{equation}
F_{\mathrm{rad}}+F_{\mathrm{conv}}-\sigma T_{\mathrm{eff}}^{4}=0,\label{eq:fluxconv}
\end{equation}
where $F_{\mathrm{rad}}$ is the radiative energy flux, $F_{\mathrm{conv}}$
is the convective energy flux, $\sigma$ is the Stefan-Boltzmann constant
and $T_{\mathrm{eff}}$ is the stellar effective temperature. Contrary
to the 3D case, effective temperature now appears as a boundary value
and is thus a free parameter. Owing to numerical instabilities of
the formulation, Eq. \ref{eq:fluxconv} is replaced in the higher
atmosphere ($\tau_{\mathrm{Ross}}\lesssim10^{-2}$) with the radiative
equilibrium condition 
\begin{equation}
q_{\mathrm{rad}}=4\pi\rho\int_{\lambda}\kappa_{\lambda}(J_{\lambda}-S_{\lambda})d\lambda\equiv0,
\end{equation}
where $J_{\lambda}$ and $S_{\lambda}$ are the mean intensity and
the source function, similar to Eq. (\ref{eq:radiative_transfer}).
In the 3D case, $q_{\mathrm{rad}}$ is explicitly calculated and is
nonzero in general. Enforcing the condition of radiative equilibrium
$q_{\mathrm{rad}}\equiv0$ in 1D leads to an atmospheric stratification
where an exact balance of radiative heating and cooling in each layer
is achieved, ignoring the effects of gas motion.

The mean intensity and the radiative energy flux at each depth are
obtained by solving the radiative transfer equation, 
\begin{equation}
\mu\frac{dI_{\lambda}}{d\tau_{\lambda}}=I_{\lambda}-S_{\lambda},\label{eq:1DRT}
\end{equation}
where $\mu=\cos\theta$ with the polar angle $\theta$ off the vertical,
$I_{\lambda}$ is the specific intensity at wavelength $\lambda$,
and $\tau_{\lambda}$ is the vertical monochromatic optical depth
(with $\tau_{\lambda}=0$ above the top of the atmosphere). A Planck
source function $S_{\lambda}=B_{\lambda}$ is assumed. The monochromatic
mean intensity and radiative flux are then delivered by the integrals
\begin{eqnarray}
J_{\lambda} & = & \frac{1}{2}\int_{-1}^{1}I_{\lambda}d\mu\\
F_{\mathrm{rad},\lambda} & = & 2\pi\int_{-1}^{1}I_{\lambda}\mu d\mu.
\end{eqnarray}

In the absence of an explicit convection treatment, convective energy
transfer is estimated using a variant of the mixing length recipe
described in \citet{Henyey:1965p15592}. The convective flux is given
by the expression 
\begin{equation}
F_{\mathrm{conv}}=\frac{1}{2}\alpha_{\mathrm{MLT}}\delta\Delta c_{p}T\rho v_{\mathrm{MLT}},
\end{equation}
where $\rho$ is the gas density, $c_{p}$ is the specific heat capacity,
$T$ is the temperature, and $v_{\mathrm{MLT}}$ is the convective
velocity. The well-known free mixing length parameter $\alpha_{\mathrm{MLT}}=l_{m}/H_{p}$
sets the distance $l_{m}$ in units of the local pressure scale height
$H_{p}$ over which energy is transported convectively. See \citet{Gustafsson:2008p3814}
for details of the expressions used to obtain the convective velocity
$v_{\mathrm{MLT}}$ and the factor $\delta\Delta=\Gamma/\left(1+\Gamma\right)\nabla_{\mathrm{sad}}$,
which scales super-adiabaticity $\nabla_{\mathrm{sad}}=\nabla-\nabla_{\mathrm{ad}}$
of the atmospheric stratification (see also Sect.~\ref{sub:temperature-gradient}),
by a convective efficiency factor $\Gamma=c_{P}\rho v_{\mathrm{MLT}}\tau_{e}(y+\tau_{e}^{-2})/(8\sigma T^{3})$
 with the optical thickness $\tau_{e}=\kappa_{\mathrm{Ross}}l_{m}$.
We adopt the same parameters $y=0.076$ and $\nu=8$ as \citet{Gustafsson:2008p3814}
for the radiative heat loss term and turbulent viscosity that enter
the above quantities.

\subsection{Numerical methods\label{sub:Numerical-methods}}

The system of equations is solved using a modified Newton-Raphson
method with an initial stratification of temperature $T$ and gas
pressure $p_{\mathrm{gas}}$ on a fixed Rosseland optical depth grid.
Discretized and linearized versions of the hydrostatic equation and
the energy flux equation (or radiative equilibrium condition, respectively)
provide the inhomogeneous term and the elements of the Jacobian matrix
for the system of $2N$ linear equations, where $N$ is the number
of depth layers. The radiation field is computed for each Newton-Raphson
iteration using the integral method, based on a second-order discretization
of the fundamental solution of the radiative transfer equation (Eq.
\ref{eq:1DRT}).

The corrections $\Delta T$ and $\Delta p_{\mathrm{gas}}$ derived
from the system of linear equations are multiplied by a variable factor
$<1$ that is automatically regulated by the code to aid convergence.
Convergence is assumed when the (relative) residuals of the $2N$
equations decrease beneath a preset threshold. Note that, contrary
to the 3D simulations, the effective temperature is now an adjustable
parameter; the requirement of minimal residuals automatically leads
to an atmospheric stratification with correct $T_{\mathrm{eff}}$
through the energy flux equation.

In order to obtain a 1D model, a given $\left\langle 3\mathrm{D}\right\rangle $
stratification provides the initial input for the Newton-Raphson iterations,
along with the targeted effective temperature and surface gravity.
The same EOS tables that were used for the 3D simulation provide gas
density, specific heat capacity, and adiabatic gradient as a function
of $T$ and $p_{\mathrm{gas}}$. Likewise, the tables containing group
mean opacities and the Rosseland mean opacity provide the required
microphysics for solving the radiative transfer equation, ensuring
maximal consistency with the 3D simulations.

Once convergence has been achieved for the 1D stratification, the
mixing length parameter $\alpha_{\mathrm{MLT}}$ can be calibrated
to obtain a better approximation to the $\left\langle 3\mathrm{D}\right\rangle $
stratification in the convection zone beneath the stellar surface.

\section{Functional fits\label{sub:Functional-fits}}

The resulting amount of data from our numerical simulations is enormous.
A convenient way to provide important key values is in form of functional
fits, which can be easily utilized elsewhere (e.g. for analytical
considerations). In the present paper we have frequently discussed
various important global properties that are reduced to scalars. Some
of them are global scalar values and some are determined at a specific
height from the $\left\langle 3\mathrm{D}\right\rangle $ stratifications,
i.e. temporal and spatial averages on layers of constant Rosseland
optical depth. We fitted these scalars with stellar parameters for
individual suitable functions, thereby enforcing a smooth rendering.
However, we would like to warn against extrapolating these fits outside
their range of validity, i.e. outside the confines of our grid. Also,
one should consider that possible small irregularities between the
grid steps might be neglected, which arise due to non-linear response
of the EOS and the opacity. On the other hand, we provide also most
of the actual shown values in Table \ref{tab:global_properties}.\\

We use three different functional bases for our fits and we perform
the least-squares minimization with an automated Levenberg-Marquardt
method. Instead of the actual stellar parameters, we employ the following
transformed coordinates: $x=(\teff-5777)/1000$, $y=\logg-4.44$ and
$z=\feh$. Furthermore, we find that in order to accomplish an optimal
fit with three independent variables, $f_{i}\left(x,y,z\right)$,
simultaneously, the metallicity should be best included implicitly
as nested functions in the form of second degree polynomial $\zeta_{a}\left(z\right)=\sum_{i=0}^{2}a_{i}z^{i}$,
each resulting in three independent coefficients $a_{i}$. The linear
function 
\begin{equation}
f_{1}\left(x,y,z\right)=\zeta_{a}\left(z\right)+x\zeta_{b}\left(z\right)+y\zeta_{c}\left(z\right)\label{eq:lin_fit}
\end{equation}
is applied for the following quantities: $s_{\mathrm{min}}$ (Fig.
\ref{fig:entropy-bottom}), $\log\rho_{\mathrm{peak}}$ (Fig. \ref{fig:density_peak}),
$\log v_{z,\mathrm{rms}}^{\mathrm{peak}}$ (Fig. \ref{fig:velocity_peak}),
$\log d_{\mathrm{gran}}$ (Fig. \ref{fig:intensity-contrast}), $\log\Delta t_{\mathrm{gran}}$
and $f_{u}^{\mathrm{peak}}$. The resulting coefficients are given
in Table \ref{tab:lin_fit}. On the other hand, we considered the
exponential function 
\begin{equation}
f_{2}\left(x,y,z\right)=f_{1}\left(x,y,z\right)+\zeta_{d}\left(z\right)\exp\left[x\zeta_{e}\left(z\right)+y\zeta_{f}\left(z\right)\right]\label{eq:exp_fit}
\end{equation}
for $s_{\mathrm{bot}}$, $\Delta s$ (Figs. \ref{fig:entropy-bottom}
and \ref{fig:entropy_jump}) and $\left[p_{\mathrm{turb}}/p_{\mathrm{tot}}\right]_{\mathrm{peak}}$
(Fig. \ref{fig:mean_turbulent-pressure-fraction}). For $\vec{\nabla}_{\mathrm{peak}}$
and $\vec{\nabla}_{\mathrm{sad}}^{\mathrm{peak}}$\ref{fig:nabla_peak}
we applied the following function 
\begin{equation}
f_{3}\left(x,y,z\right)=f_{1}\left(x,y,z\right)+x^{2}\zeta_{d}\left(z\right),\label{eq:par_fit}
\end{equation}
with coefficients for $f_{2}$ and $f_{3}$ listed in Table \ref{tab:exp_fit}.
Finally, we showed in Fig. \ref{fig:ssbot_vs_ssj} the entropy jump
$\Delta s$ as a function of $s_{\mathrm{bot}}$, which we fitted
with 
\begin{equation}
f_{4}\left(x\right)=a_{0}+a_{1}x+a_{3}\tanh\left[a_{4}+a_{5}x\right].\label{eq:hyp_fit}
\end{equation}
The resulting coefficients are listed in Table \ref{tab:hyp_fit}.
\begin{table}
\caption{\label{tab:lin_fit}The coefficients $a_{i}$ of the linear function
$f_{1}$ (Eq. \ref{eq:lin_fit}) for $s_{\mathrm{min}}$ $[10^{11}\mathrm{erg}/\mathrm{g}/\mathrm{K}]$,
$\log\rho_{\mathrm{peak}}$ $[10^{-7}\mathrm{g}/\mathrm{cm}^{3}]$,
$\log v_{z,\mathrm{rms}}^{\mathrm{peak}}$ $[10\mathrm{km}/\mathrm{s}]$,
$\log d_{\mathrm{gran}}$ $[\mathrm{Mm}]$ and $\log\Delta t_{\mathrm{gran}}$
$[10^{2}\mathrm{s}]$. In the last two rows, we listed the root-mean-square
and maximal deviation of the fits.}

\begin{tabular}{cccccc}\hline\hline
 $a_i$  & $s_{\mathrm{min}}$ & $\lg\rho_{\mathrm{peak}}$ & $\lg v_{z,\mathrm{rms}}^{\mathrm{peak}}$ & $\lg d_{\mathrm{gran}}$ & $\lg\Delta t_{\mathrm{gran}}$ \\
\hline
 $a_0$ &     1.5440 &     0.3968 &    -0.4626 &     0.2146 &    -0.7325 \\
 $a_1$ &     0.0387 &    -0.2549 &     0.0568 &     0.0666 &    -0.0054 \\
 $a_2$ &     0.0046 &    -0.0344 &     0.0068 &     0.0108 &    -0.0016 \\
 $b_0$ &     0.0621 &    -0.4232 &     0.1988 &     0.1174 &     0.0410 \\
 $b_1$ &    -0.0189 &     0.1260 &    -0.0255 &     0.0187 &     0.0046 \\
 $b_2$ &     0.0013 &    -0.0007 &     0.0050 &     0.0033 &     0.0000 \\
 $c_0$ &    -0.0898 &     0.6814 &    -0.1845 &    -1.0922 &    -0.9970 \\
 $c_1$ &     0.0038 &    -0.0282 &     0.0116 &    -0.0462 &    -0.0038 \\
 $c_2$ &    -0.0004 &    -0.0021 &    -0.0006 &    -0.0075 &    -0.0006 \\
 $\mathrm{rms} \Delta$ &     0.0711 &    20.3286 &     1.0018 &   483.4921 &    47.6416 \\
 $\max \Delta$ &     0.1843 &   138.7171 &     1.3365 &  2697.2449 &   144.2399 \\
\hline\end{tabular}
\end{table}
\begin{table}
\caption{\label{tab:exp_fit}The coefficients $a_{i}$ of the functional bases
$f_{2}$ and $f_{3}$ (Eqs. \ref{eq:exp_fit} and \ref{eq:par_fit})
for $s_{\mathrm{bot}}$ $[10^{11}\mathrm{erg}/\mathrm{g}/\mathrm{K}]$,
$\Delta s$ $[10^{11}\mathrm{erg}/\mathrm{g}/\mathrm{K}]$, $\left[p_{\mathrm{turb}}/p_{\mathrm{tot}}\right]_{\mathrm{peak}}$,$\vec{\nabla}_{\mathrm{peak}}$
and $\vec{\nabla}_{\mathrm{sad}}^{\mathrm{peak}}$. In the last two
rows, we listed the root-mean-square and maximal deviation of the
fits.}

\begin{tabular}{cccccc}\hline\hline
 $a_i$  & $s_{\mathrm{bot}}$ & $\Delta s$ & $p_{\mathrm{turb/tot}}^{\mathrm{peak}}$ & $\vec{\nabla}_{\mathrm{peak}}$ & $\vec{\nabla}_{\mathrm{sad}}^{\mathrm{peak}}$ \\
\hline
 $a_0$ &     1.5789 &    -0.0006 &     0.0321 &     1.0941 &     0.8713 \\
 $a_1$ &     0.0455 &     0.0043 &     0.0459 &    -0.0089 &     0.0338 \\
 $a_2$ &     0.0111 &     0.0064 &     0.0111 &     0.0000 &     0.0076 \\
 $b_0$ &     0.0784 &     0.0017 &     0.0138 &     0.2498 &     0.3401 \\
 $b_1$ &    -0.0183 &     0.0049 &     0.0007 &    -0.0532 &    -0.0717 \\
 $b_2$ &     0.0071 &     0.0060 &     0.0019 &    -0.0050 &    -0.0091 \\
 $c_0$ &    -0.1076 &     0.0028 &    -0.0260 &    -0.4004 &    -0.4847 \\
 $c_1$ &    -0.0028 &    -0.0029 &    -0.0087 &     0.1052 &     0.0990 \\
 $c_2$ &    -0.0042 &    -0.0032 &    -0.0016 &     0.0142 &     0.0155 \\
 $d_0$ &     0.1602 &     0.1979 &     0.1335 &    -0.0600 &    -0.0622 \\
 $d_1$ &     0.0618 &     0.0675 &    -0.0257 &     0.0016 &    -0.0006 \\
 $d_2$ &     0.0062 &     0.0059 &    -0.0081 &    -0.0133 &    -0.0128 \\
 $e_0$ &     1.2867 &     1.1479 &     0.5894 & -- & -- \\
 $e_1$ &    -0.0824 &    -0.0866 &     0.1141 & -- & -- \\
 $e_2$ &     0.0970 &     0.0788 &     0.0337 & -- & -- \\
 $f_0$ &    -1.2136 &    -1.0996 &    -0.5330 & -- & -- \\
 $f_1$ &    -0.0338 &    -0.0316 &    -0.0864 & -- & -- \\
 $f_2$ &    -0.0764 &    -0.0614 &    -0.0249 & -- & -- \\
 $\mathrm{rms} \Delta$ &     0.2555 &     0.2047 &     0.0758 &     0.4533 &     0.5016 \\
 $\max \Delta$ &     0.7268 &     0.6602 &     0.1841 &     1.1245 &     1.2550 \\
\hline\end{tabular}
\end{table}
\begin{table}
\caption{\label{tab:hyp_fit}The coefficients $a_{i}$ of the hyperbolic tangent
function $f_{4}$ (Eq. \ref{eq:hyp_fit}) for fitting $\Delta s$
as function of $s_{\mathrm{bot}}$.}

\begin{tabular}{ccccccccccccc}\hline\hline
 $\teff$ & $a_0$ & $a_1$ & $a_2$ & $a_3$ & $a_4$ \\
\hline
    4000.0 &     1.2910 &    -0.3559 &     1.0367 &    -2.6408 &     1.2059 \\
    4500.0 &     5.1768 &    -2.1859 &     4.5280 &    -1.4475 &     0.6756 \\
    5000.0 &     7.0730 &    -3.0946 &     6.8382 &    -1.2330 &     0.5799 \\
    5500.0 &     7.6382 &    -3.4144 &     7.5981 &    -1.1812 &     0.5636 \\
    6000.0 &     6.8963 &    -2.9796 &     6.9907 &    -1.1769 &     0.5504 \\
\hline
 $\logg$ & $a_0$ & $a_1$ & $a_2$ & $a_3$ & $a_4$ \\
\hline
       1.5 &     5.3693 &    -2.0610 &     5.3770 &    -1.2576 &     0.5461 \\
       2.0 &     1.1012 &    -0.2599 &     0.9218 &    -2.8316 &     1.2958 \\
       2.5 &     1.5805 &    -0.5023 &     1.2081 &    -2.5467 &     1.1888 \\
       3.0 &     5.2106 &    -2.1433 &     4.6691 &    -1.4254 &     0.6548 \\
       3.5 &     4.9821 &    -2.0989 &     4.2522 &    -1.5136 &     0.7111 \\
       4.0 &     8.0957 &    -3.5548 &     7.9721 &    -1.1979 &     0.5625 \\
       4.5 &    14.1757 &    -6.3782 &    17.4802 &    -0.8936 &     0.4180 \\
\hline\end{tabular}
\end{table}

\section{Tables\label{sec:Tables}}

In Table \ref{tab:global_properties} we have listed important global
properties with the stellar parameters. Due to the lack of space,
we show only an excerpt with solar metallicity ($\feh=0.0$). The
complete table is available at CDS \url{http://cds.u-strasbg.fr}.\\
\begin{table*}
\caption{\label{tab:global_properties}The stellar parameters: effective temperature,
surface gravity (Cols. 1 and 2 in $\left[\mathrm{K}\right]$ and $\left[\mathrm{dex}\right]$).
The main input variables: the density $\rho_{\mathrm{bot}}$, internal
energy per unit mass $\varepsilon_{\mathrm{bot}}$ (Cols. 3 and 4
in $\left[10^{-7}\mathrm{g}/\mathrm{cm}^{3}\right],\left[10^{5}\mathrm{erg}/\mathrm{g}\right]$).
We also added the temperature $T_{\mathrm{bot}}$, thermodynamic pressure
$p_{\mathrm{th}}^{\mathrm{bot}}$ and entropy values $s_{\mathrm{bot}}$
at the bottom (Cols. 5, 6 and 7 in $\left[\mathrm{K}\right],\left[10^{5}\mathrm{dyne}/\mathrm{cm}^{2}\right],\left[10^{9}\mathrm{erg}/\mathrm{g}/\mathrm{K}\right]$).
Furthermore, the jump in entropy $\Delta s$, the maximal vertical
rms-velocity $v_{z,\mathrm{rms}}^{\mathrm{peak}}$ and intensity contrast
$\Delta I_{\mathrm{rms}}$ values are given (Cols. 8, 9, 10 in $\left[10^{9}\mathrm{erg}/\mathrm{g}/\mathrm{K}\right],\left[\mathrm{km}/\mathrm{s}\right],\left[\%\right]$).
Finally, we display the horizontal $s_{x,y}$ and vertical box size
$s_{z}$ (Cols. 11 and 12 in $\left[\mathrm{Mm}\right],\left[\mathrm{Mm}\right]$),
the mean granule diameter $d_{\mathrm{gran}}$ (Col. 13 in $\left[\mathrm{Mm}\right]$),
the time step $\Delta t$ and total time $t$ (Cols. 14 and 15 in
$\left[10^{2}\mathrm{s}\right],\left[10^{2}\mathrm{s}\right]$).}
\begin{tabular}{llllllllllllllll}
\hline\hline
$T_{\rm{eff}}$ & $\log g$ & $\lg \rho_{\mathrm{bot}}$ & $\lg \varepsilon_{\mathrm{bot}}$ & $\lg T_{\mathrm{bot}}$ &  $\lg p_{\mathrm{th}}^{\mathrm{bot}}$ & $s_{\mathrm{bot}}$ & $\Delta{s}$ & $v_{z,\mathrm{rms}}^{\mathrm{peak}}$ & $\Delta I_{\mathrm{rms}}$ & $\lg s_{x,y}$ & $\lg s_{z}$ & $\lg d_{\mathrm{gran}}$ & $\lg \Delta t$ & $\lg t$\\
\hline
      4023 &    1.50 &      0.717 &      1.124 &      4.272 &      1.061 &      2.300 &      0.602 &      5.145 &       18.4 &      3.820 &      3.490 &      3.121 &      2.176 &      4.352\\
      4052 &    2.00 &      1.125 &      1.004 &      4.233 &      1.368 &      2.018 &      0.361 &      4.167 &       17.1 &      3.322 &      2.971 &      2.623 &      1.695 &      3.871\\
      3938 &    2.50 &      1.691 &      0.908 &      4.239 &      1.889 &      1.775 &      0.174 &      3.210 &       14.4 &      2.740 &      2.446 &      2.041 &      1.188 &      3.364\\
      4569 &    2.00 &      0.679 &      1.187 &      4.342 &      1.120 &      2.411 &      0.723 &      5.845 &       18.4 &      3.380 &      3.069 &      2.778 &      1.740 &      4.041\\
      4532 &    2.50 &      1.357 &      1.060 &      4.279 &      1.669 &      2.039 &      0.395 &      4.391 &       17.2 &      2.845 &      2.517 &      2.243 &      1.241 &      3.282\\
      4492 &    3.00 &      1.785 &      0.955 &      4.266 &      2.029 &      1.808 &      0.210 &      3.486 &       14.5 &      2.342 &      1.966 &      1.643 &      0.692 &      2.692\\
      4530 &    3.50 &      2.103 &      0.900 &      4.269 &      2.322 &      1.682 &      0.126 &      2.903 &       12.2 &      1.778 &      1.442 &      1.079 &      0.188 &      2.364\\
      4513 &    4.00 &      2.419 &      0.858 &      4.277 &      2.625 &      1.578 &      0.069 &      2.367 &        9.4 &      1.146 &      0.895 &      0.544 &     -0.319 &      1.681\\
      4516 &    4.50 &      2.721 &      0.835 &      4.292 &      2.927 &      1.500 &      0.037 &      1.937 &        7.7 &      0.602 &      0.399 &     -0.000 &     -0.824 &      1.276\\
      4512 &    5.00 &      3.013 &      0.819 &      4.308 &      3.226 &      1.434 &      0.021 &      1.541 &        6.3 &      0.146 &     -0.102 &     -0.553 &     -1.301 &      0.875\\
      4932 &    2.00 &      0.042 &      1.291 &      4.535 &      0.700 &      2.757 &      1.047 &      8.331 &       50.4 &      3.544 &      3.127 &      3.544 &      1.876 &      4.052\\
      5013 &    2.50 &      0.883 &      1.202 &      4.374 &      1.358 &      2.376 &      0.706 &      5.880 &       18.0 &      2.903 &      2.586 &      2.204 &      1.287 &      3.463\\
      4998 &    3.00 &      1.534 &      1.082 &      4.308 &      1.882 &      2.024 &      0.399 &      4.407 &       16.9 &      2.362 &      2.055 &      1.663 &      0.765 &      2.765\\
      5001 &    3.50 &      1.960 &      0.987 &      4.295 &      2.243 &      1.805 &      0.223 &      3.608 &       14.8 &      1.813 &      1.496 &      1.114 &      0.220 &      2.317\\
      4978 &    4.00 &      2.292 &      0.919 &      4.293 &      2.538 &      1.661 &      0.125 &      2.896 &       11.7 &      1.279 &      0.952 &      0.580 &     -0.292 &      1.749\\
      4953 &    4.50 &      2.604 &      0.877 &      4.301 &      2.837 &      1.560 &      0.068 &      2.363 &        8.8 &      0.699 &      0.455 &      0.000 &     -0.824 &      1.217\\
      4963 &    5.00 &      2.885 &      0.854 &      4.314 &      3.118 &      1.485 &      0.038 &      1.868 &        6.8 &      0.176 &     -0.048 &     -0.301 &     -1.301 &      0.796\\
      5465 &    3.00 &      1.084 &      1.215 &      4.403 &      1.589 &      2.337 &      0.685 &      5.815 &       17.7 &      2.447 &      2.125 &      1.748 &      0.819 &      2.995\\
      5560 &    3.50 &      1.663 &      1.119 &      4.345 &      2.062 &      2.040 &      0.428 &      4.598 &       17.4 &      1.903 &      1.572 &      1.204 &      0.318 &      2.415\\
      5497 &    4.00 &      2.139 &      1.010 &      4.322 &      2.456 &      1.791 &      0.226 &      3.597 &       15.3 &      1.362 &      1.023 &      0.663 &     -0.284 &      1.892\\
      5510 &    4.50 &      2.486 &      0.947 &      4.322 &      2.769 &      1.649 &      0.128 &      2.959 &       12.1 &      0.845 &      0.503 &      0.146 &     -0.770 &      1.230\\
      5480 &    5.00 &      2.791 &      0.905 &      4.330 &      3.060 &      1.547 &      0.072 &      2.323 &        9.0 &      0.301 &      0.001 &     -0.398 &     -1.301 &      0.699\\
      5768 &    4.44 &      2.367 &      0.997 &      4.336 &      2.688 &      1.725 &      0.186 &      3.293 &       14.6 &      0.903 &      0.601 &      0.204 &     -0.678 &      1.419\\
      6023 &    3.50 &      1.130 &      1.266 &      4.493 &      1.737 &      2.395 &      0.751 &      6.183 &       17.9 &      1.903 &      1.703 &      1.301 &      0.467 &      2.564\\
      5993 &    4.00 &      1.865 &      1.122 &      4.364 &      2.281 &      1.991 &      0.397 &      4.514 &       17.9 &      1.415 &      1.095 &      0.716 &     -0.155 &      1.942\\
      5998 &    4.50 &      2.301 &      1.026 &      4.344 &      2.644 &      1.771 &      0.222 &      3.572 &       16.1 &      0.845 &      0.552 &      0.146 &     -0.721 &      1.279\\
      6437 &    4.00 &      1.384 &      1.263 &      4.495 &      1.989 &      2.315 &      0.686 &      5.818 &       18.3 &      1.447 &      1.221 &      0.748 &     -0.081 &      2.016\\
      6483 &    4.50 &      2.008 &      1.134 &      4.386 &      2.448 &      1.969 &      0.386 &      4.516 &       18.7 &      0.903 &      0.624 &      0.204 &     -0.638 &      1.403\\
      6918 &    4.50 &      1.545 &      1.283 &      4.543 &      2.201 &      2.292 &      0.673 &      5.737 &       18.6 &      1.041 &      0.781 &      0.342 &     -0.638 &      1.362\\
\hline
\end{tabular}
\end{table*}

\end{document}